# Exchange interactions and intermolecular hybridization in a spin-1/2 nanographene dimer

N. Krane[1*], E. Turco[1*], A. Bernhardt[2*], D. Jacob[3,4*], G. Gandus[5], D. Passerone[1], M. Luisier[5], M. Juríček[2#], R. Fasel[1,7], J. Fernández-Rossier[6#], and P. Ruffieux[1#]

[1] *nanotech@surfaces Laboratory, Empa - Swiss Federal Laboratories for Materials Science and Technology, Dübendorf, Switzerland*

[2] *Department of Chemistry, University of Zurich, Zurich, Switzerland*

[3] *Departamento de Polímeros y Materiales Avanzados: Física, Química y Tecnología, Universidad del País Vasco UPV/EHU, Av. Tolosa 72, E-20018 San Sebastián, Spain*

[4] *IKERBASQUE, Basque Foundation for Science, Plaza Euskadi 5, E-48009 Bilbao, Spain*

[5] *Integrated Systems Laboratory, ETH Zürich, Switzerland*

[6] *International Iberian Nanotechnology Laboratory (INL), 4715-330 Braga, Portugal*

[7] *Department of Chemistry, Biochemistry and Pharmaceutical Sciences, University of Bern, 3012 Bern, Switzerland*

*These authors contributed equally to this work.

#Corresponding authors:
michal.juricek@chem.uzh.ch
joaquin.fernandez-rossier@inl.int
pascal.ruffieux@empa.ch

**Phenalenyl is a radical nanographene with triangular shape that hosts an unpaired electron with spin *S* = ½. The open-shell nature of phenalenyl is expected to be retained in covalently bonded networks. Here, we study a first step in that direction and report the synthesis of the phenalenyl dimer by combining in-solution synthesis and on-surface activation and its characterization both on Au(111) and on a monolayer of NaCl on top of Au(111) by means of inelastic electron tunneling spectroscopy (IETS). IETS shows inelastic steps that, together with a thorough theoretical analysis, are identified as the singlet–triplet excitation arising from interphenalenyl exchange. Two prominent features of our data permit to shed light on the nature of spin interactions in this system. First, the excitation energies with and without the NaCl decoupling layer are 48 and 41 meV, respectively, indicating a significant** **renormalization of the spin excitation energies due to exchange with the Au(111) electrons. Second, a position-dependent bias-asymmetry of the height of the inelastic steps is accounted for by an interphenalenyl hybridization of the singly occupied phenalenyl orbitals that is only possible via third neighbor hopping. This hybridization is also essential to activate kinetic interphenalenyl exchange. Our results set the stage for future work on the bottom-up synthesis of spin *S* = ½ spin lattices with large exchange interaction.**

Phenalenyl is a polycyclic conjugated hydrocarbon that possesses a magnetic ground state with spin *S* = ½, whereby an unpaired electron is hosted in a non-bonding $\pi$-molecular orbital. Its experimental study has been hampered by its high chemical reactivity, typical of radical species[1]. Recent progress in the on-surface synthesis and manipulation under ultra-high vacuum conditions has now made it possible to overcome this problem and study the electronic properties of unsubstituted phenalenyl on a gold surface[2]. Inelastic electron tunneling spectroscopy (IETS)[3] based on scanning tunneling microscopy (STM) shows a prominent zero bias peak that is associated with the formation of a Kondo singlet between the unpaired electron of the phenalenyl and the conduction electron of the surface, endorsing the picture of phenalenyl as a *S* = ½ molecule.

Two possible applications can be foreseen for this type of molecules. First, as in the case of the larger *S* = 1 triangulene[4–9] and other open-shell nanographenes[10–14], phenalenyl could serve as a building block for covalent nanographene arrays that behave as chains or lattices of *S* = ½ interacting spins[15]. Both the low spin and the small magnetic anisotropy make these artificial spin networks suitable to explore quantum magnetism, and perhaps to be used for engineering electronic phases with topological order that can be used for robust quantum information processing. Second, *S* = ½ objects provide a natural realization of spin qubits.

Progress in these two directions requires the understanding, and eventually the control, of exchange interactions of phenalenyl, both with the substrate and with other molecules when integrated within covalent arrays. Here, we undertake the first steps in this direction, both by exploring the electronic properties of a phenalenyl dimer (hereafter referred to as diphenalenyl) on a conducting surface as well as by including a decoupling layer between diphenalenyl and the conducting substrate.

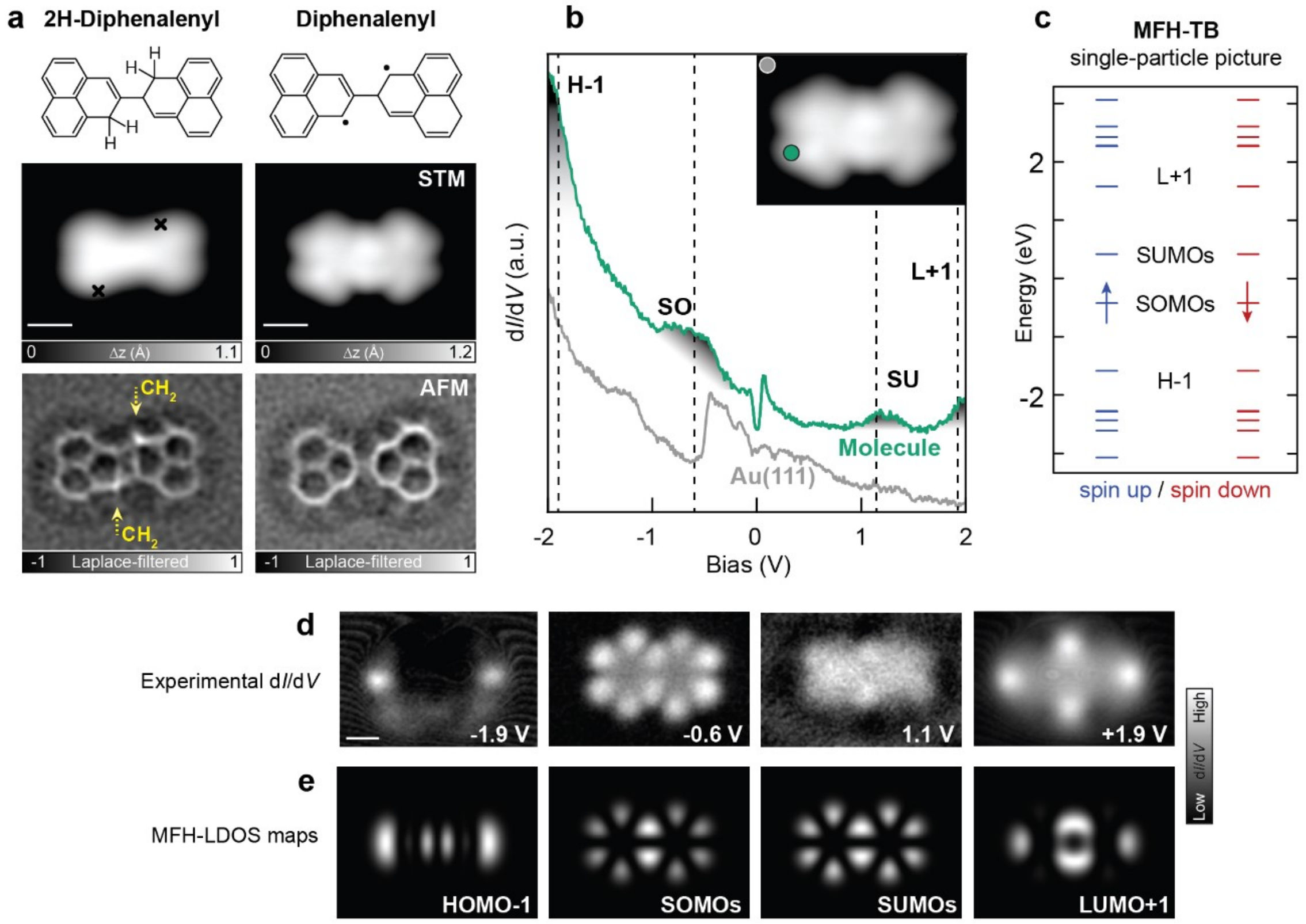

**Figure 1. (a)** Tip-induced activation of diphenalenyl on Au(111). STM (top) and nc-AFM (bottom) images before and after dehydrogenation via voltage pulses (black crosses). The images were taken with CO tip at closed feedback with -100 mV/50 pA (STM) and opened feedback on Au(111) with -5 mV/100 pA, $\Delta z$ = 1.8 Å (AFM). (**b**) dI/dV spectroscopy taken with CO tip on diphenalenyl (green) and Au(111) (grey), revealing the SOMOs and SUMOs at -0.6 V and +1.1 V, respectively, as well as the onsets of the HOMO-1 and LUMO+1 at -1.9 V and +1.9 V. The inset displays the positions, where the spectra were taken. Feedback loop opened at -2 V/350 pA, $V_{rms}$ = 20 mV. Dashed lines mark energies of dI/dV maps displayed in (d). (**c**) MFH Energy diagram of the single-particle states. (**d,e**) Constant current dI/dV maps and MFH calculated projected density of states of the corresponding orbitals. HOMO-1 and LUMO+1 taken at 250 pA and $V_{rms}$ = 30 mV, SOMOs and SUMOs at 200 pA and $V_{rms}$ = 14 mV. Scale bars: 0.5 nm (a, d).

The preparation of diphenalenyl is achieved through a combined solution and on-surface synthesis approach. In the solution phase, the 2*H*-diphenalenyl precursor is synthesized in a sequence of nine steps starting from naphthalene (for details, see the SI). The use of hydro precursors has the advantage that activation of the target open-shell compound can be achieved using atom manipulation with a scanning tunneling microscopy (STM) tip[8] and does not necessarily require the catalytic action of metal substrates. We have recently shown that phenalenyl and triangulene can be achieved through selective activation of the corresponding hydro precursors using controlled voltage pulses from the STM tip[2]. Here, we follow a similar approach to sequentially activate the two hydro-phenalenyl subunits of the precursor. The substrate was prepared by sublimation of NaCl on a clean Au(111) surface held at room temperature, which leads to a sub-monolayer coverage of NaCl organized into 1ML and 2ML islands. The molecular 2*H*-diphenalenyl precursor was deposited by sublimation onto the previously prepared sample, kept at a temperature of about 100 K during the deposition and rapidly transferred into the STM. An overview STM image of the so-obtained sample is reported in Figure S2, where, the molecular precursors can be found in sub-monolayer coverage adsorbed both on the Au(111) surface and NaCl islands. Sequential tip-induced cleaving of the hydrogen atoms from the $sp^3$ carbon atoms[2,8] yields the target diphenalenyl diradical, as proven by constant-height nc-AFM measurements of the precursor and the target compound (Figure 1a, bottom) both adsorbed on Au(111). The precursor molecules adsorbed on 1ML NaCl were similarly manipulated into diphenalenyl diradical as shown in Figure S2 (b,c). The change in the electronic structure can also be observed in the STM images (Figure 1a), showing distinct lobes and nodal planes for the activated molecule. Constant-height d*I*/d*V* spectroscopy of the activated molecule adsorbed on Au(111) is shown in Figure 1b, revealing the presence of two distinct resonances at –0.6 eV and 1.1 eV, and the onset of two conductance peaks at ±1.9 eV. To do a first assignment of the observed conductance peaks to the respective molecular orbitals, we used a tight-binding (TB) level of theory, taking into account the electron–electron Coulomb repulsion within the mean-field Hubbard (MFH) approximation. The calculated energy diagram reported in Figure 1c features two frontier states, commonly denoted as singly occupied and unoccupied molecular orbitals, SOMOs and SUMOs, respectively[16]. A comparison of the calculated local density of states (LDOS) and the experimental d*I*/d*V* maps of the molecule's electronic resonances allows a clear assignment of the experimentally observed resonances (Figure 1d,e).

In order to probe the magnetic properties of diphenalenyl, d*I*/d*V* spectroscopy at low bias voltages was conducted both on Au(111) and monolayer NaCl. As displayed in Figure 2a, the spectra show steps in the differential conductance at ± 41 mV and ±48 mV for the molecules on Au(111) and NaCl, respectively. Ovchinnikov's rule[17] and Lieb's theorem[18] predict the single spins of the two phenalenyl units to form an $S = 0$ ground state. For two coupled spins with $S = ½$ and Heisenberg coupling $J\boldsymbol{S}_1 \cdot \boldsymbol{S}_2$, the excited state is $S = 1$ with energy $E = J$.

Therefore, the observed steps in differential conductance can be assigned to spin excitations from the singlet ground state to the triplet excited state.

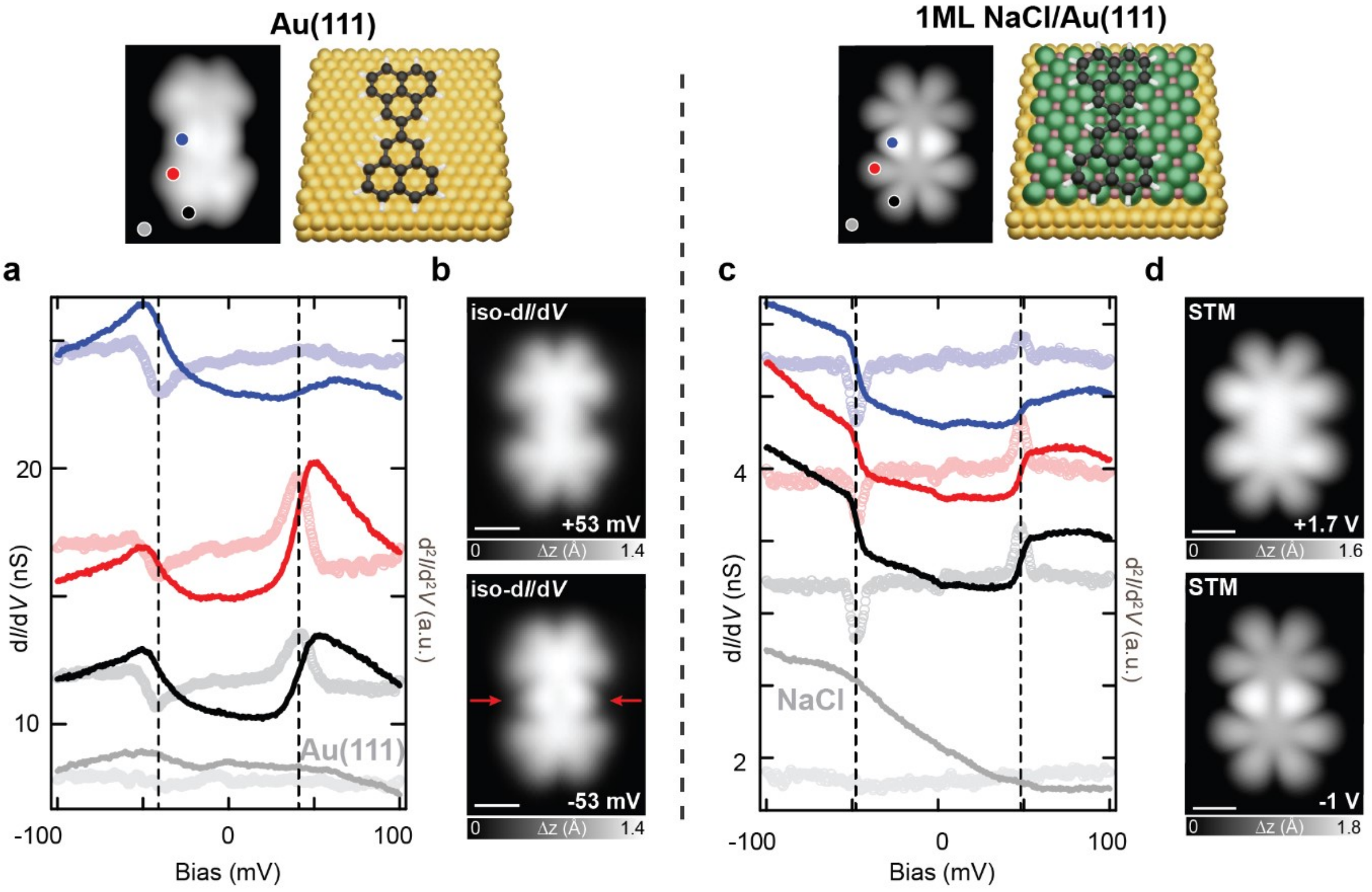

**Figure 2.** Inelastic spectroscopy of the spin excitations of diphenalenyl on Au(111) and 1ML-NaCl/Au(111). (**a**) d*I*/d*V* spectroscopy (solid lines) and second derivative (circles) of diphenalenyl adsorbed on Au(111). Tip positions are marked in the image above and the reference spectrum (bottom) is taken on Au(111). The dotted vertical lines mark the dip/peak positions of the second derivative at ±41 mV. Feedback opened at -100 mV/750 pA; $V_{rms}$ = 2 mV (**b**) Constant d*I*/d*V* images (10 nS, $V_{rms}$ = 3.5 mV) taken at energies above spin excitation. The red arrows in bottom panel highlight the asymmetry for different bias polarities. (**c**) d*I*/d*V* spectroscopy (solid lines) and second derivative (circles) of diphenalenyl on monolayer NaCl. Tip positions are marked in the image above and the reference spectrum (bottom) is taken on NaCl. Dotted vertical lines at dip/peak of second derivative at ±48 mV. Feedback opened at -100 mV/250 pA; $V_{rms}$ = 2.8 mV (**d**) Constant current STM images (20 pA) of diphenalenyl on NaCl/Au(111) taken at the onset energies of the frontier orbitals, displaying the asymmetry of the SOMOs and SUMOs. Scale bars: 0.5 nm (b, d).

A specific feature of our system is the marked asymmetry of the height of the conductance steps in Figure 2a. The d*I*/d*V* spectra taken between the phenalenyl units (blue lines) show a higher step at negative bias polarity, compared to the smaller step at positive bias polarity. This effect can be seen for diphenalenyl on Au(111) as well as on NaCl islands. Figure 2b displays two iso-d*I*/d*V* maps[19], using the d*I*/d*V* signal as feedback, taken at energies just outside of the spin excitation gap. The asymmetry with bias polarity becomes clearly visible. Therefore, the asymmetry of the spectra taken in the center of the dimer is an intrinsic property of the diphenalenyl molecules and independent from the underlying substrate. On the other hand, for the molecule deposited on Au(111), the inelastic steps are broadened and show pronounced triangular overshoots. For the spectra taken at the sides of the molecule these overshoots are significantly larger for positive bias (Figure 2a). This asymmetry is not present when the molecule is adsorbed on NaCl, and thus is non-generic. As discussed below, our calculations are able to account for both asymmetries.

Importantly, there are three main differences between the spectra acquired for diphenalenyl on Au(111) with those taken on NaCl. First, the singlet–triplet excitation energy is significantly higher on NaCl with 48 meV, compared to 41 meV on Au(111). Second, the peak-like features at the excitation steps are not present in the spectra taken on NaCl and, third, the width of the steps is much broader on Au(111) than on NaCl. The latter becomes apparent in the numerical derivation of the d*I*/d*V* signal in Figure 2a.

We now resort to theory to rationalize the main properties of the observed inelastic excitations by addressing i) their different energies and broadening, depending on the presence, or not, of a NaCl decoupling layer, and ii) the bias asymmetry of the excitation line shapes. First, we provide evidence that inelastic excitations observed at around 40 meV are associated with interphenalenyl exchange. For that matter, following our previous works,[4,20,21,] we build a generalized Hubbard model, that also includes long-range Coulomb interactions (see SI for computational methods). Our calculations show that diphenalenyl remains open-shell, possessing an $S = 0$ ground state and an $S = 1$ excited state with energy in the range of the experimental observations. Thus, diphenalenyl hosts two antiferromagnetically coupled unpaired spins.

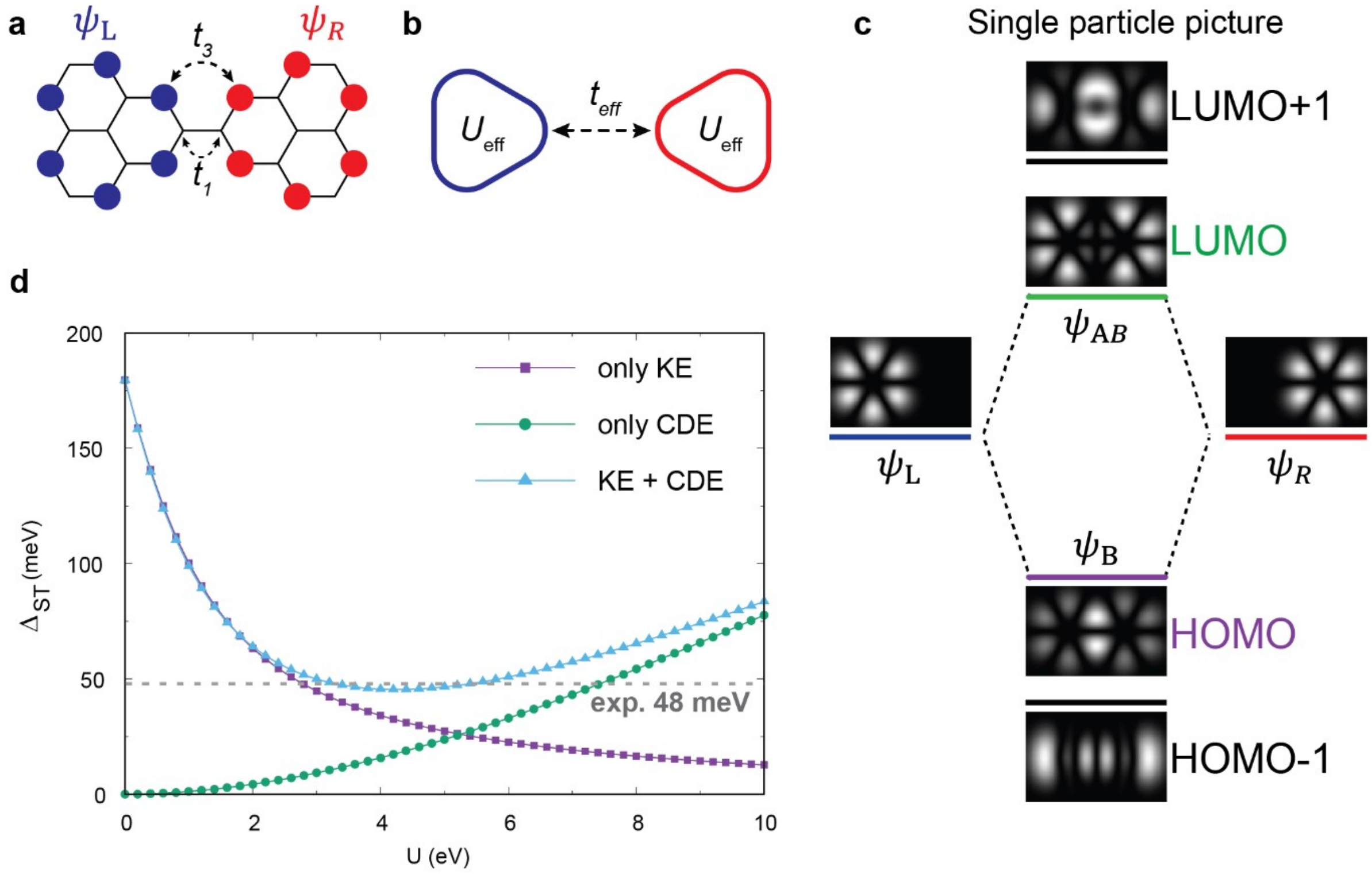

**Figure 3.** Calculations for extended Hubbard model of diphenalenyl (see SI and Ref. [20]for details). (**a**) Zero mode distribution of diphenalenyl. Due to missing weight at the binding site, the intermolecular hybridization between the two phenalenyl units is driven by third-neighbor hopping $t_3$ instead of $t_1$. (**b**) Simplified two-site model of diphenalenyl with effective hopping parameter $t_{\text{eff}}$ and effective Coulomb repulsion $U_{\text{eff}}$. (**c**) Molecular orbitals in the single-particle picture. Hybridization of the SOMOs (blue, red) leads to bonding (HOMO, purple) and anti-bonding (LUMO, green) frontier orbitals. (**d**) Singlet–triplet splitting of diphenalenyl as function of Coulomb repulsion for kinetic exchange only (KE, purple squares) and Coulomb driven exchange (CDE, green circles). Considering both exchanges (light blue triangles) predicts a splitting close to the experimentally observed energies (horizontal dashed line) for a wide range of values for the Coulomb repulsion.

There are two mechanisms for intermolecular exchange in this type of system, as discussed by two of us recently.[20] One is driven by intermolecular hybridization of the zero modes of the phenalenyls, the other by Coulomb-mediated virtual occupation of higher-energy extended molecular orbitals. Based on a model with first-neighbor hopping only, one could expect that intermolecular hybridization vanishes, as the zero-modes have a null weight on the binding sites (as depicted in Figure 3a). Our DFT calculations show that this description is not complete and that hybridization does not vanish (see the SI). This automatically suggests that third-neighbor hopping is non-zero (see figure 3a and also ref.[15]). Therefore, intermolecular hybridization is present, so that the two mechanisms for intermolecular exchange are active and can be accounted for by means of exact diagonalization of the model in a restricted set of multi-electronic states (see the SI). The predictions of this model for the total exchange, and the relative contribution of the two mechanisms are shown in Figure 3d. They show that, for a wide range of values of the Coulomb interactions, the predicted singlet–triplet excitation energy is close to the experimental value. Importantly, the ratio of intermolecular hybridization and intramolecular addition energies show the dimer has a strong diradical character.

We now address the substrate-dependent energy and linewidth of the excitations. To do so, we include in our Hamiltonian the hybridization of the molecular orbitals (MOs) with the conduction electrons in the substrate. Our ab initio calculations show that this hybridization strongly depends on both substrate and MO. (see the SI and refs.[21,22]). The interacting MOs coupled to the conduction electrons in the substrate constitute a multi-orbital Anderson model that we solve in the one-crossing approximation[23] (OCA).

The key quantity that permits to relate the model calculations to the experimental results are the spectral functions of the electrons in the molecule which are directly connected to the d$I$/d$V$ in the tunneling regime[24] and include both the many-body interactions and the influence of the substrate. We compute the spectral functions $A_k(\omega)$ projected MOs within OCA, taking into account the MOs shown in Figure 3c, i.e. the HOMO-1, HOMO, LUMO and LUMO+1 (see the SI and ref.[21] for details). The results are shown in Figure 4. The coupling to the substrate has two major effects on the system: on the one hand, it leads to screening of the Coulomb tensor, generally lowering the Hubbard $U$, and therefore modifying the bare excitation energies. On the other hand, the coupling to the substrate gives rise to finite linewidths of the spectral features and relatedly to Kondo exchange which renormalizes the excitation energies[21,25]. In our calculations, the screening of the Coulomb tensor is taken into account by a screening parameter in our model Hamiltonian (see the SI and ref.[20]). Finite linewidths and Kondo exchange on the other hand are a consequence of solving the Anderson model (by OCA) which includes the single-particle broadening of the MOs due to coupling to the substrate obtained from realistic DFT calculations of the molecule on both surfaces (see the SI and refs.[21,22]).

Our DFT calculations (see the SI) show that in the presence of the NaCl monolayer, the coupling of the molecule to the substrate is very weak. Therefore, renormalization of the excitation energy by Kondo exchange coupling is negligible, and we can account for the experimentally observed excitation energy of $\sim$ 48 meV in our model, if we take $U \sim 5.4$ eV (c.f. Figure 3d). In contrast, for the Au(111) surface, DFT calculations show that hybridization with the molecule is appreciable, and thus leads to a substantial renormalization (see Figure 4f and the SI). At the same time, we expect the Coulomb interaction in the molecule to be smaller for Au(111) than for the NaCl monolayer due to screening by the conduction electrons. We find good agreement with the experimental value for Au(111) of $\sim$ 41 meV for a Coulomb interaction tensor corresponding to a Hubbard-$U$ of 2.5 eV. Therefore, we conclude that the observed energy difference is due to a combination of enhanced substrate-induced renormalization and screening of the interactions when the molecules are deposited on Au(111). As expected from previous work[21], the linewidth of the calculated spin-excitation steps is larger for the molecule on Au(111), in agreement with the experiments. It is important to note, however, that the linewidth is somewhat smaller than the experimental value, which probably reflects the limitations of the OCA method.

Finally, we address the origin of the bias asymmetries, both, the generic one seen for both substrates for spectra taken at the central part of the molecule (blue lines in Figure 2a**,**c), as well as the bias asymmetry observed in the spectra taken at the outer parts of the molecule deposited on Au(111). First, we note that the contribution of the HOMO and the LUMO to the step heights (Figure 4a,d) already results in pronounced and opposite asymmetries for both substrates: the HOMO has a significantly larger step for negative bias than for positive bias, while for the LUMO it is exactly the opposite. This asymmetry is ultimately caused by proximity and height of the ion resonance closest to the step, that is, with the same bias polarity. Additionally, in the case of Au(111) the steps show the characteristic triangular overshoots induced by Kondo exchange with the conduction electrons[21,24–29]. The overshoot is especially pronounced for the positive-bias step in the LUMO spectral function. The reason is the enhancement of the product of the conduction electron density of states $N(E_F)$ and Kondo exchange $J_K$, given by $\Gamma_{LUMO}/E_A$, due to the proximity of the positive ion resonance and thus low electron-addition energy $E_A$.

The bias and position dependence of the d$I$/d$V$ are related to the LDOS that in our many-body picture relates to orbital-resolved spectral function $A_k(\omega)$ as:

$$\rho(\vec{r};\omega) = \sum_k |\psi_k(\vec{r})|^2 A_k(\omega)$$

where the index $k$ labels the MO, and $|\psi_k(\vec{r})|^2$ are the square of the MO wave functions. Figures 4c and 4f show the LDOS computed at three different points over the molecule as indicated by the colors corresponding to the circles of the same color in Figure 4b and 4e, respectively. These show the same bias asymmetries observed in the experiment. First, over the center of the molecule (blue line) the LDOS shows the generic asymmetry common to both substrates where the negative bias step is significantly larger than the positive bias step. The reason is the vanishing of the LUMO wave function at the center of the molecule (c.f. Figure 3c). Therefore, the LDOS is dominated by the HOMO spectral function (purple lines in Figure 4a,d). In contrast, both HOMO and LUMO contribute to the LDOS at the outer parts of the molecule (red and black lines in Figure 4c,f). In the case of the NaCl monolayer, this leads to almost symmetric steps, in agreement with experiment (c.f. Figure 2c). On the other hand, for the gold substrate the dominant height of the positive bias step in the HOMO also leads to a predominance of the positive bias step in the LDOS over the side units with the characteristic Kondo exchange induced overshoot, again in agreement with experiment.

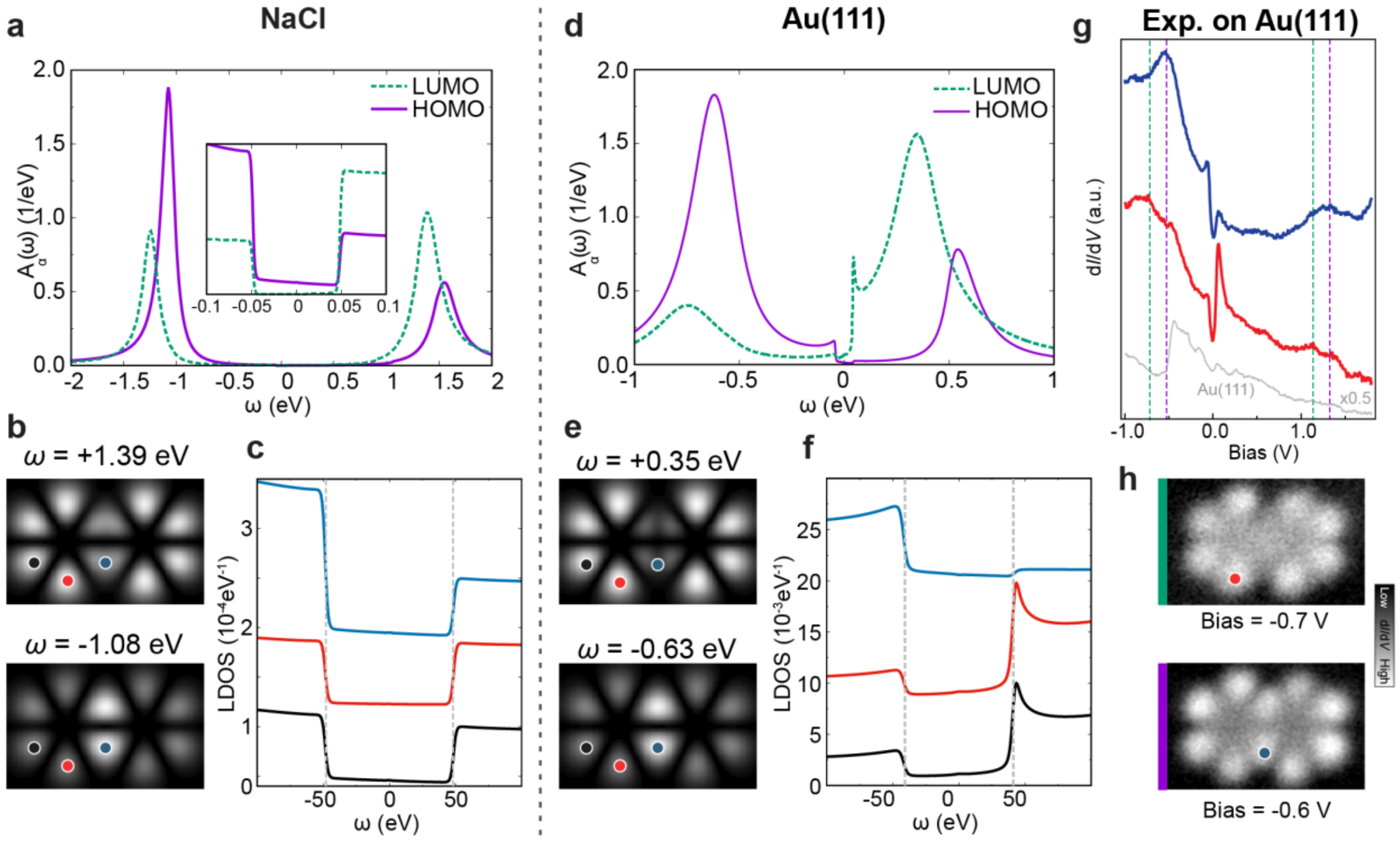

**Figure 4.** (**a,d**) Spectral function of HOMO (purple) and LUMO (green) for Diphenalenyl on ML-NaCl/Au(111) and Au(111), showing significant spectral weight at both bias polarities. The inset in (a) displays a zoom of the inelastic step features. (**b,e**) Constant height local density of states (LDOS) maps at energies of the highest Hubbard peaks of the HOMO and LUMO spectral functions. (**c,f**) Calculated d*I*/d*V* spectra for different positions, marked by circles in (b,e). Vertical dotted lines mark the singlet–triplet excitation energies. (**g**) Measured d*I*/d*V* spectroscopy taken with CO tip on diphenalenyl on Au(111). The vertical dotted lines indicate the positions of the resonances corresponding to the HOMO (purple) and LUMO (green), for both polarities. Position of spectra are marked by circles in (h). Feedback loop opened at -1 V/350 pA, $V_{rms}$ = 14 mV. (**h**) Measured d*I*/d*V* maps at negative bias voltages, matching the LDOS of the LUMO (top) and HOMO (bottom). Maps taken at 200 pA and $V_{rms}$ = 14 mV.

Figure 4b,e shows LDOS maps, $\rho(\vec{r};\omega)$, computed at the energies $\omega$ of the highest Hubbard peak in the HOMO and the LUMO spectral functions for both substrates. These correspond to constant height maps of the d*I*/d*V* at fixed voltage. Clearly, comparison with Figure 3c shows that the LDOS maps at negative energies resemble the density map of the HOMO while those at positive energies resemble the density map of the LUMO. Naturally, the correspondence is not 100% since also the other orbital(s) contribute with some weight.

Further validation of our model, and additional evidence for intermolecular hybridization, comes from the prediction for the molecule on Au(111) of a splitting of the negative ion resonance, that arises from the HOMO–LUMO hybridization splitting, of about 100 meV. In the many-body picture, both MO contribute to the negative ion resonance, at slightly different energies (Figure 4a,b). This prediction is confirmed by STS with a CO-functionalized tip, which, depending on the position of the STM tip over the molecule (shown in the inset), exhibits peaks at different voltages. For the position over one of the lobes at the center of the dimer (blue), where only the HOMO contributes (see Figure 4h, bottom), there is a pronounced peak at around –0.6 V, while at the position away from the center (red), where the LUMO contribution is the strongest (see Figure 4h top), a broader peak around -0.7 V is found. Constant current d*I*/d*V* maps taken at these voltages (Figure 4h, top) further confirm the assignment of these peaks to HOMO and LUMO.

In summary, thorough spectroscopy studies combined with theory portray diphenalenyl as an open-shell molecule, where strong interphenalenyl antiferromagnetic exchange leads to an $S$ = 0 ground state and an $S$ = 1 excited state. Additionally, we have provided strong evidence for the existence of intermolecular hybridization. The peculiar nature of the zero-modes in this class of system makes it possible to unveil the role of third-neighbor hopping, dominant for these states, and very frequently ignored in the modeling of graphene. By comparing the spectra for the same molecule on two different surfaces, we also show the relevant role of coupling to the substrate, that changes not only the lifetimes, but also the energies of the inelastic excitations. These findings need to be taken into account for the design of platforms that exploit phenalenyl and other planar nanographene radicals as molecular building blocks for quantum technologies such as quantum computing, quantum simulation or quantum sensing.

## Associated content

**Supporting Information**

Experimental and computational methods, supporting STM and STS data, additional calculations, and a detailed synthetic description and characterization of chemical compounds reported in this study (PDF).

## Author Contributions

P.R., R.F and M.J conceived the experiments. A.B. synthesized and characterized the precursors in solution. N.K and E.T performed the on-surface synthesis and scanning probe measurements. N.K and E.T performed TB calculations and analyzed the data. D.J, J.F.R and G.G simulated the system using different levels of theory. All authors discussed the results and contributed to the writing of the manuscript.

## Funding Sources

This research was supported by the Swiss National Science Foundation (SNSF; Grant No. CRSII5_205987, 200020_18201, PP00P2_170534 and PP00P2_198900), CarboQuant funded by the Werner Siemens Foundation, the EU Horizon 2020 research and innovation program − Marie Skłodowska-Curie grant no. 813036, and ERC Starting grant (INSPIRAL, grant no. 716139). The work was also financially supported from Grant PID2020-112811GB-I00 funded by MCIN/AEI/10.13039/501100011033 and Grant No. IT1453-22 from the Basque Government. The research was also supported by NCCR MARVEL, a National Centre of Competence in Research, funded by the Swiss National Science Foundation (grant number 205602 and by a grant from the Swiss National Supercomputing Centre (CSCS) under project ID s1142. JFR further acknowledges financial support from FCT (Grant No. PTDC/FIS-MAC/2045/2021), Generalitat Valenciana funding Prometeo2021/017 and MFA/2022/045, and MICIIN-Spain (Grant No. PID2019-109539GB-C41).

## Acknowledgement

We thank Oliver Gröning, Kristjan Eimre and Carlo Antonio Pignedoli for the fruitful scientific discussions. Skillful technical assistance by Lukas Rotach is gratefully acknowledged.

# Supplementary Information

## Exchange interactions and intermolecular hybridization in a spin-1/2 nanographene dimer

N. Krane, E. Turco, A. Bernhardt, D. Jacob, G. Gandus, D. Passerone, M. Luisier, M. Juríček, R. Fasel, J. Fernández-Rossier, and P. Ruffieux

### Contents

*The raw NMR data is available free of charge on a public repository Zenodo under the link https://zenodo.org/record/8128962 (DOI: 10.5281/zenodo.8128962).*

# 1. Synthetic procedures

## Chemicals

All chemicals and solvents were used without further purification if not specified otherwise. Chemicals and solvents were purchased from abcr GmbH, Acros Organics, Alfa Aesar, Merck/Sigma-Aldrich, and Flourochem. Dry solvents were purchased from Acros Organics and used without further treatment if not described otherwise.

## Remarks for reaction procedures and analytical methods

For reactions that were carried out under inert conditions, nitrogen was used as an inert gas. Glassware was oven-dried at 150 °C, cooled down in vacuum (oil pump, ~ $10^{-3}$ mbar), and flushed with nitrogen. After adding solid reagents and a magnetic stirring bar under ambient conditions, the closed vessels were evacuated and flushed with nitrogen three times. Dry solvents and liquid reagents were added afterwards under a flow of nitrogen. If an air- or moisture-sensitive solid was used, it was transferred into the reaction vessel in a nitrogen-filled glovebox. Solvents and liquid reagents were deoxygenated by bubbling with nitrogen while sonication in an ultrasonic bath for 15–30 min (method A) or by freeze-pump-thaw technique in three cycles under a nitrogen atmosphere (method B). Reactions that were carried out at room temperature were running between 20 and 25 °C depending on fluctuating external factors as the season and ventilation. Custom-made glassware was used for the column chromatography under inert conditions (Figure S1a) as well as for photochemical brominations (Figure S1b). Reactions that were carried out in sealed tubes were heated in a custom-made aluminum heating block (Figure S5).

**Figure S1**

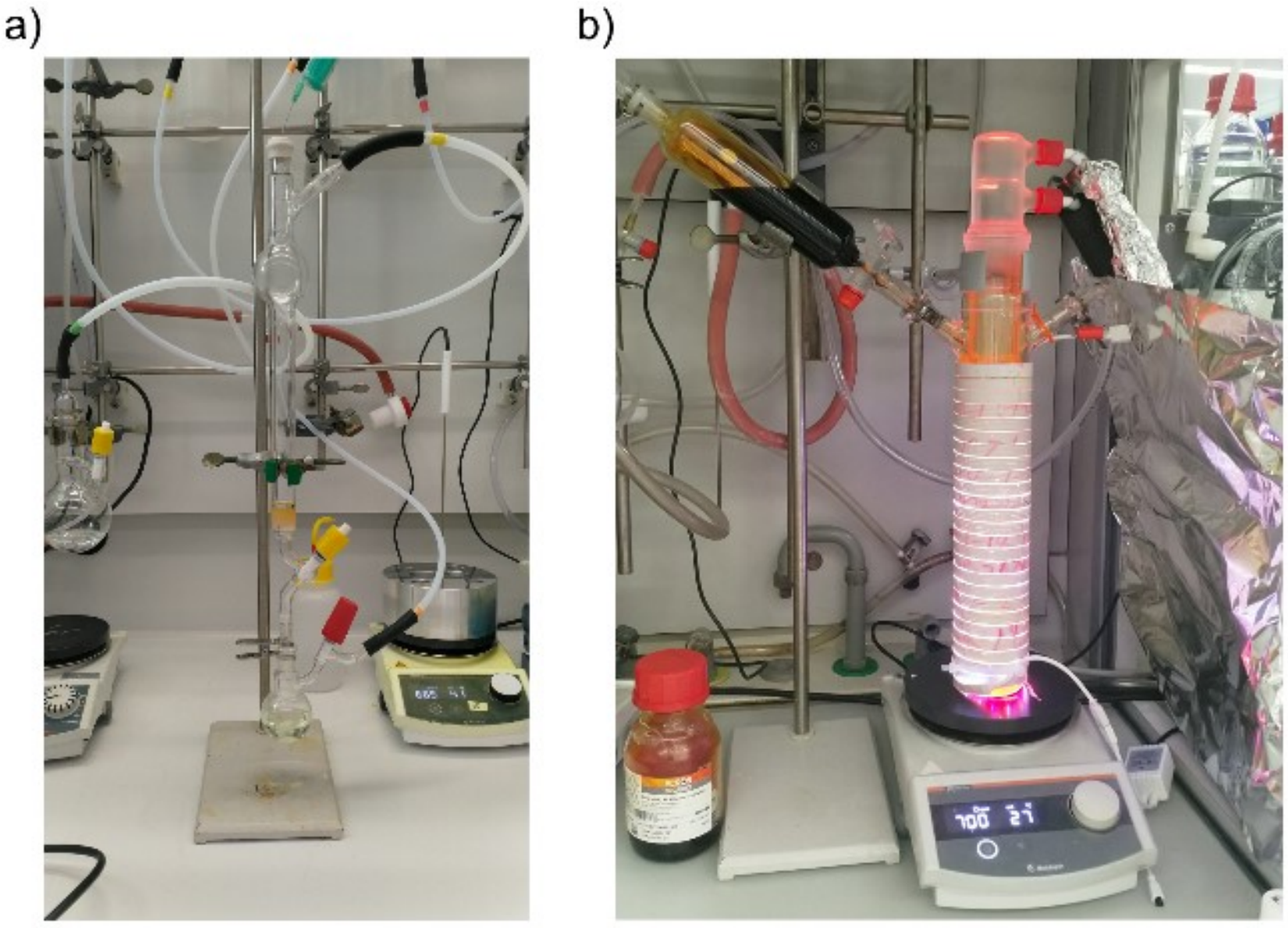

Figure S5. Custom-made glassware for (a) column chromatography under inert conditions and aluminum heating block. (b) Custom-made reaction vessel for photobrominations.

For inert chromatography, the column was equipped with a connector for gas supply on its top, the bottom was connected to a Schlenk flask to evacuate the system. Before use, the column was washed with deoxygenated solvent. The eluent and the crude mixtures were inserted through the septum.

The photoreactor consists of a glass tube that is constructed similarly to a cold trap and is circulated with coolant (*i*-PrOH cooled with a chiller) during the reaction. This tube was inserted into the reaction vessel and immersed into the reaction solution. The reaction vessel was wrapped with a commercial flexible LED stripe for the photoirradiation. Two more joints at the upper rim of the reaction vessel allow to connect a dropping funnel with bromine and a gas washing bottle for the neutralization of the bromine gas.

Thin-layer chromatography (TLC) was performed using analytical silica gel aluminum plates 60 $F_{254}$ by Merck. Flash column chromatography was performed by standard technique using silica gel 60 mesh or Florisil®.

Proton and decoupled carbon NMR spectra were recorded with a Bruker FT-NMR Avance III 400 and Avance 600 spectrometers at 298 K unless indicated otherwise. The frequency and the solvent are given separately for each substance. Chemical shifts are given in units of the $\delta$-scale in ppm. Shifts for $^{1}H$ and $^{13}C$ NMR spectra are given relative to the residual proton and carbon signal of the indicated solvent, respectively: $CDCl_3$ (7.26 and 77.16), $CH_2Cl_2$ (5.32 and 53.84), DMSO-$d_6$ (2.50 and 39.52), acetone-$d_6$ (2.05 and 29.84), THF-$d_8$ (1.72 and 25.31), toluene-$d_8$ (2.08 and 20.43) and benzene-$d_6$ (7.16 and 128.06).[1] Coupling constants are given in Hertz (Hz). Processing and interpretation was performed using MestReNova 14.1.

High-resolution mass spectra were measured as EI (ThermoFischer Scientifics DFS), ESI or APCI (ThermoFischer Scientifics QExactive MS).

## Solution synthesis of 2*H*-diphenalenyl

The target molecule 2*H*-diphenalenyl (**10**) was prepared from naphthalene in a sequence of nine synthetic steps, involving the preparation of functionalized phenalenone derivatives. The dimeric structure was constructed via a Suzuki cross-coupling. The subsequent reduction yielded the target compound 2*H*-diphenalenyl that was subjected to experiments on a Au(111) surface (Scheme 1).

## Scheme 1

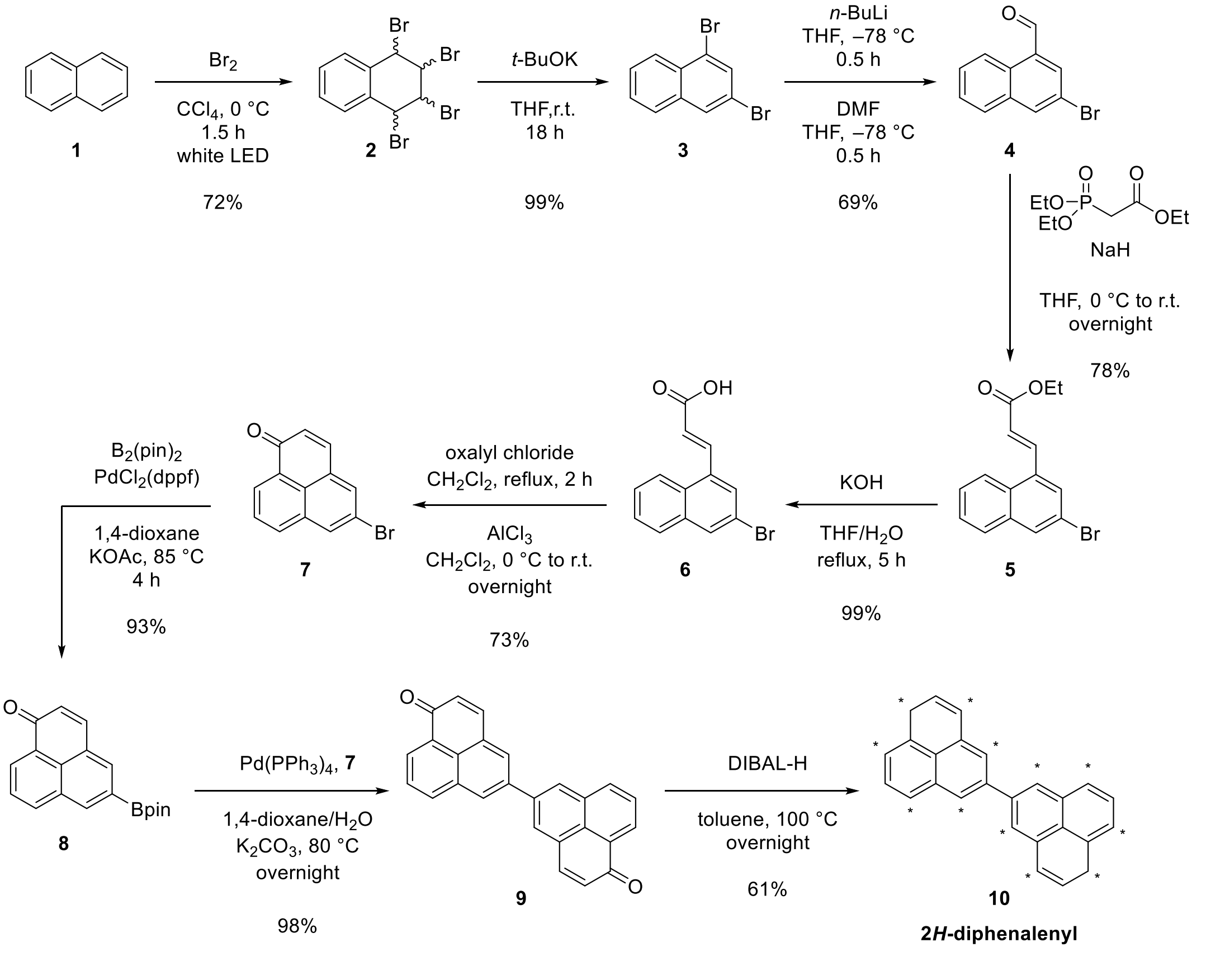

Scheme 1. Synthetic route to 2*H*-diphenalenyl (**10**) starting from naphthalene (**1**).[2–7]

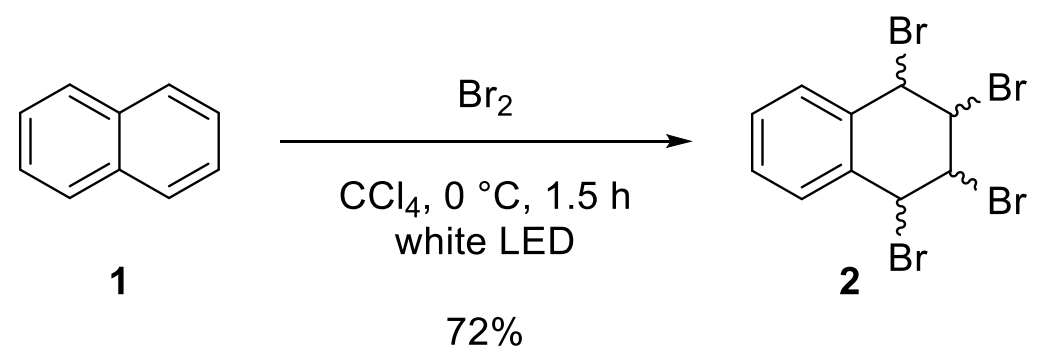

**1,2,3,4-Tetrabromo-1,2,3,4-tetrahydronaphthalene (2).** The reaction was carried out using a modified literature procedure[2] in a custom-made photoreactor, where a tube with cooling medium (isopropanol cooled by a chiller to –2 °C) was immersed into the reaction vessel equipped with a stirring bar. A commercial LED stripe emitting white light was fitted around the vessel. A solution of naphthalene (**1**; 9.00 g, 70.2 mmol) in carbon tetrachloride (140 mL) was cooled to 0 °C and irradiated. A solution of bromine (14.4 mL, 281 mmol) in carbon tetrachloride (66 mL) was added dropwise. After the addition of bromine was complete, the irradiation was continued at 0 °C. The reaction was monitored by TLC ($SiO_2$, cyclohexane). After 1.5 h, the starting material was consumed and some of the target compound was precipitated. The reaction mixture was quenched with a solution of $NaHSO_3$ (36.5 g, 351 mmol) dissolved in water (150 mL). After phase separation, the organic layer was washed with water followed by saturated aqueous $NaHCO_3$ before carbon tetrachloride was removed under reduced pressure and recovered. The previously obtained aqueous layer after quenching the reaction was again extracted with dichloromethane. The new organic layer was also washed with water, followed by saturated aqueous $NaHCO_3$ and then brine. The combined organic layers were combined with the remaining solids and the solvent was removed under reduced pressure. The crude product was recrystallized from *n*-hexane and dichloromethane in a refrigerator. The desired compound was obtained as a colorless crystalline solid (22.3 g, 49.8 mmol, 72%). The obtained NMR spectra are in accord with the previously reported data.[8] **$^1$H NMR (400 MHz, $CDCl_3$, ppm):** $\delta$ 7.61 (dd, *J* = 5.7, 3.4 Hz, 2H), 7.42 (dd, *J* = 5.8, 3.3 Hz, 2H), 5.77–5.67 (m, 2H), 5.06–4.95 (m, 2H). **$^{13}$C NMR (101 MHz, $CDCl_3$, ppm):** $\delta$ 133.0, 130.2, 129.8, 54.2, 50.3.

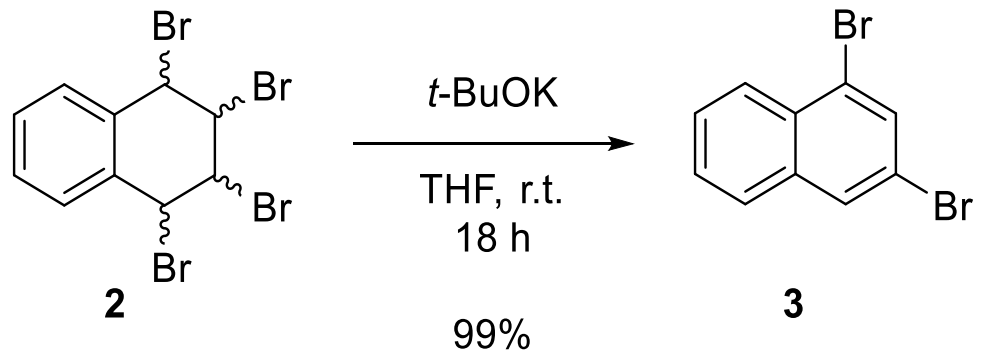

**1,3-Dibromonaphthalene (3).**[2] The reaction was carried out under inert conditions. To a solution of 1,2,3,4-tetrabromo-1,2,3,4-tetrahydronaphthalene (**2**; 20.4 g, 45.6 mmol) in THF (200 mL), a solution of *t*-BuOK (11.8 g, 105 mmol) in THF (110 mL) was added dropwise at room temperature. After the addition was completed, the reaction mixture was stirred overnight before it was quenched with brine and extracted with methyl *tert*-butyl ether. The combined organic layers were washed with water, followed by brine and dried over $MgSO_4$. After evaporation of the solvent under reduced pressure, the crude product was purified over a short plug ($SiO_2$, cyclohexane). The desired compound was obtained as a colorless solid in almost quantitative yield (12.9 g, 45.0 mmol, 99%). The obtained NMR spectra are in accord with the previously reported data.[8] **$^1$H NMR (400 MHz, $CDCl_3$, ppm):** $\delta$ 8.19 (d, *J* = 8.3 Hz, 1H), 7.97 (d, *J* = 2.3 Hz, 1H), 7.89 (d, *J* = 1.6 Hz, 1H), 7.74 (dd, *J* = 8.1, 1.7 Hz, 1H), 7.60 (ddd, *J*

= 8.6, 6.9, 1.4 Hz, 1H), 7.55 (ddd, *J* = 8.1, 6.7, 1.2 Hz, 1H). **$^{13}$C NMR (101 MHz, CDCl$_3$, ppm):** *δ* 135.3, 132.7, 130.8, 130.1, 127.9, 127.8, 127.6, 127.4, 123.7, 119.0.

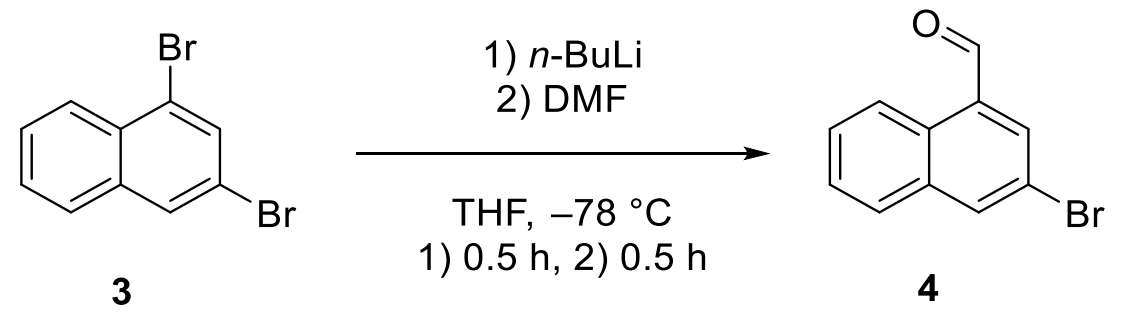

**3-Bromo-1-naphthaldehyde (4).** The reaction was carried out under inert conditions. To a cooled (–78 °C) solution of 1,3-dibromonaphthalene (**3**; 2.12 g, 7.42 mmol) in dry THF (40 mL), cooled *n*-BuLi (2.82 mL, 7.05 mmol, 2.5 M in *n*-hexane) was added dropwise while the temperature of the reaction mixture was kept below –70 °C. The mixture was stirred for 0.5 h before DMF (861 µL, 11.1 mmol) was added. After the addition, the reaction mixture was stirred for 1 h at the same temperature before it was quenched with brine. The layers were separated and the aqueous layer was extracted with methyl *tert*-butyl ether. The combined organic layers were dried over $MgSO_4$ and the solvent was removed under reduced pressure. The crude product was purified by flash column chromatography ($SiO_2$, cyclohexane/EtOAc 5:1 v/v) to afford the desired compound as a colorless solid (1.21 g, 5.13 mmol, 69%). The yield varied between 32 and 69%, the best result was obtained with freshly distilled DMF and a constant temperature of less than –70 °C. The obtained NMR spectra are in accord with the previously reported data.[3] **$^1$H NMR (400 MHz, CDCl$_3$, ppm):** *δ* 10.36 (s, 1H), 9.16 (dd, *J* = 8.6, 1.2 Hz, 1H), 8.26 (d, *J* = 2.1 Hz, 1H), 8.06 (d, *J* = 2.0 Hz, 1H), 7.84 (d, *J* = 8.1 Hz, 1H), 7.71 (ddd, *J* = 8.5, 6.9, 1.4 Hz, 1H), 7.62 (ddd, *J* = 8.2, 7.0, 1.2 Hz, 1H). **$^{13}$C NMR (101 MHz, CDCl$_3$, ppm):** *δ* 192.1, 138.9, 137.0, 135.3, 133.0, 129.5, 129.2, 128.1, 127.8, 125.1, 118.7.

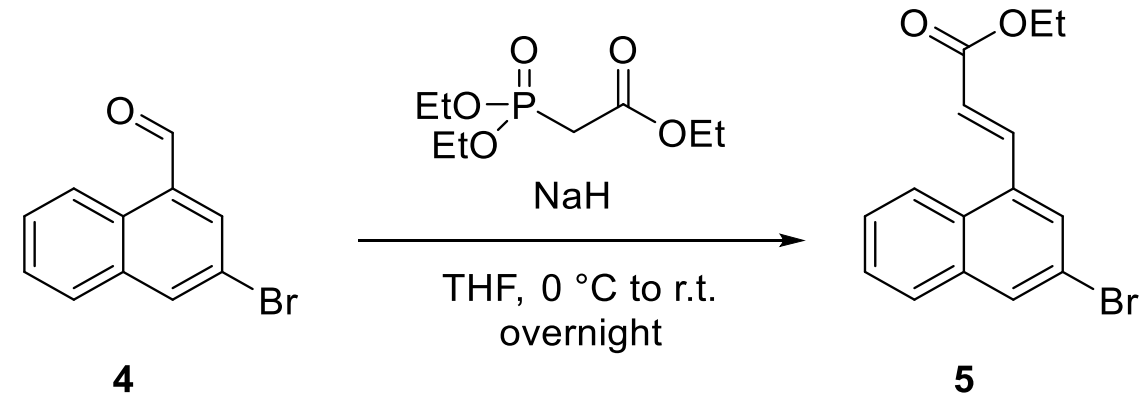

**Ethyl (*E*)-3-(3-bromonaphthalen-1-yl)acrylate (5).** The reaction was carried out under inert conditions using a modified literature procedure.[4] A suspension of NaH (562 mg, 23.4 mmol) in THF (80 mL) was cooled in an ice bath before triethyl phosphonoacetate (3.87 mL, 19.5 mmol) was added dropwise. The mixture was stirred for 0.5 h at the same temperature before a solution of 3-bromo-1-naphthaldehyde (**5**; 4.09 g, 13.0 mmol) in dry THF (40 mL) was added. The reaction mixture was allowed to warm to room temperature and was stirred overnight. Subsequently, the reaction mixture was quenched with brine and the layers were separated. The aqueous layer was extracted with methyl *tert*-butyl ether and the combined organic layers were dried over $MgSO_4$. After removal of the solvent under reduced pressure, the crude product was purified by flash column chromatography ($SiO_2$, cyclohexane/EtOAc 5:1 v/v) to afford the desired compound as a colorless solid (3.09 g, 10.1 mmol, 78%). The obtained NMR spectra are in accord with the previously reported data.[3] **$^1$H NMR (400 MHz,**

**$CDCl_3$, ppm):** $\delta$ 8.40 (d, *J* = 15.7 Hz, 1H), 8.12 (d, *J* = 8.1 Hz, 1H), 8.01 (d, *J* = 1.9 Hz, 1H), 7.79 (d, *J* = 1.9 Hz, 1H), 7.76 (d, *J* = 7.9 Hz, 1H), 7.57 (ddd, *J* = 8.1, 7.0, 1.6 Hz, 1H), 7.53 (ddd, *J* = 7.7, 6.9, 1.4 Hz, 1H), 6.51 (d, *J* = 15.7 Hz, 1H), 4.32 (q, *J* = 7.1 Hz, 2H), 1.38 (t, *J* = 7.1 Hz, 3H). **$^{13}C$ NMR (101 MHz, $CDCl_3$, ppm):** $\delta$ 166.6, 140.2, 134.9, 134.0, 132.1, 130.0, 128.0, 127.9, 127.4, 127.3, 123.7, 122.5, 119.5, 60.9, 14.5.

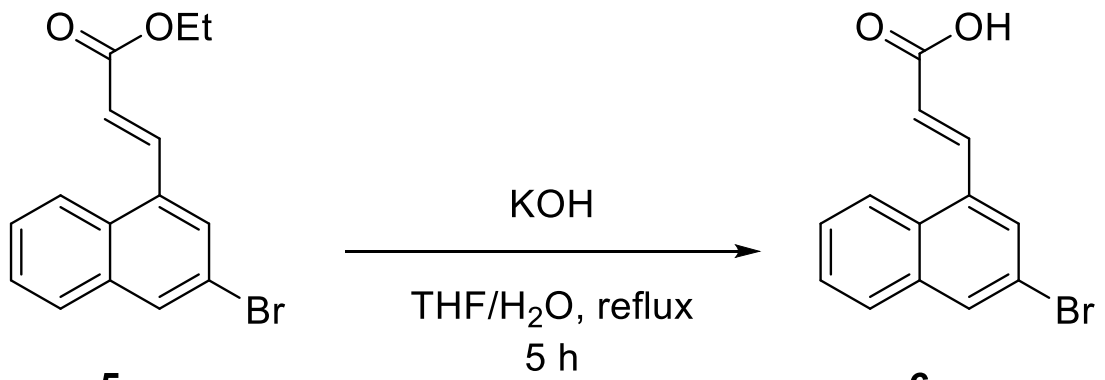

5 → 6

75%

**(*E*)-3-(3-Bromonaphthalen-1-yl)acrylic acid (6).** Ethyl (*E*)-3-(3-bromonaphthalen-1-yl)acrylate (**5**; 2.39 g, 7.80 mmol) was dissolved in THF (63 mL), a solution of KOH (875 mg) in water (7 mL) was added and the reaction mixture was refluxed for 4 h. After cooling to room temperature, most of the solvent was evaporated under reduced pressure. The residing liquid was acidified with concentrated HCl. The formed precipitate was filtered, washed several times with water and then dissolved in dichloromethane. Evaporation of the solvent under reduced pressure yielded the desired compound as a white solid (1.61 g, 5.81 mmol, 75%). The obtained NMR spectra are in accord with the previously reported data.[3] **$^1H$ NMR (400 MHz, DMSO-$d_6$, ppm):** $\delta$ 12.66 (br s, 1H), 8.30 (d, *J* = 15.6 Hz, 1H), 8.29 (s, 1H), 8.18 (d, *J* = 7.4 Hz, 1H), 8.04 (s, 1H), 7.98 (d, *J* = 7.6 Hz, 1H), 7.70–7.59 (m, 2H), 6.68 (d, *J* = 15.7 Hz, 1H). **$^{13}C$ NMR (101 MHz, DMSO-$d_6$, ppm):** $\delta$ 167.1, 138.6, 134.5, 133. 6, 131.7, 129.3, 128.0, 127.61, 127.57, 127.5, 123.7, 123.4, 118.9.

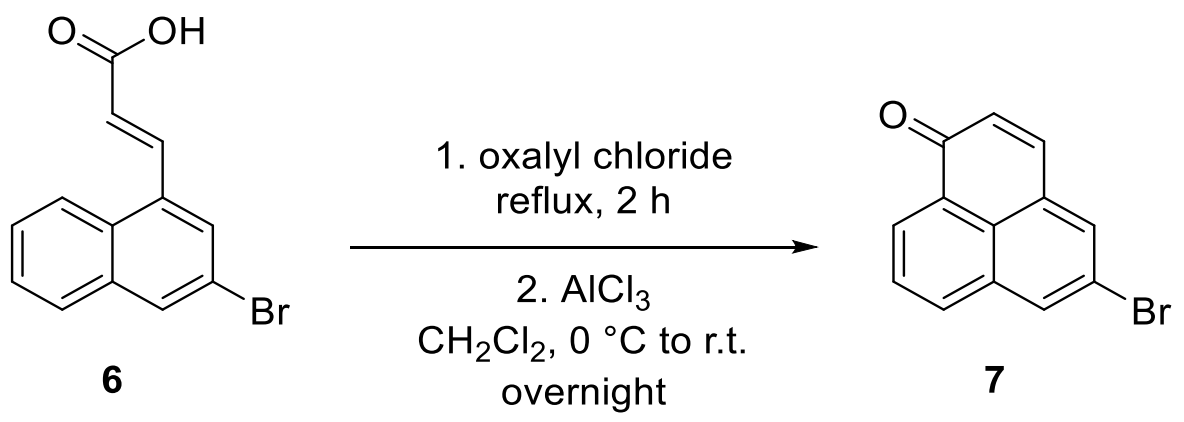

6 → 7

73%

**5-Bromo-1*H*-phenalen-1-one (7).** The reaction was carried out under inert conditions using a modified literature procedure.[3] (*E*)-3-(3-Bromonaphthalen-1-yl)acrylic acid (**6**; 1.35 g, 4.87 mmol) was dissolved in oxalyl chloride (12 mL) and the mixture was refluxed for 2 h. After cooling to room temperature, excessive oxalyl chloride was removed under reduced pressure and the remaining yellow solid was dissolved in dichloromethane (35 mL). The solution was cooled in an ice bath before solid $AlCl_3$ (1.95 g, 14.6 mmol) was added. The ice bath was removed and the reaction mixture was stirred overnight at room temperature. Subsequently, the reaction mixture was poured into water, the layers were separated and the aqueous layer was extracted with dichloromethane. The combined organic layers were washed with aqueous $NaHCO_3$ and dried over $MgSO_4$ before the solvent was removed under reduced pressure. The crude product was purified by flash column chromatography ($SiO_2$, $CH_2Cl_2$) to afford the desired compound as a yellow solid (917 mg, 3.54 mmol, 73%). The obtained NMR spectra are in accord with the previously reported data.[3] **$^1$H NMR (400 MHz, $CDCl_3$, ppm):** $\delta$ 8.60 (dd, $J$ = 7.3, 1.2 Hz, 1H), 8.18 (d, $J$ = 1.8 Hz, 1H), 8.11 (dd, $J$ = 8.1, 1.2 Hz, 1H), 7.83 (d, $J$ = 1.9 Hz, 1H), 7.79 (dd, $J$ = 8.0, 7.4 Hz, 1H), 7.67 (d, $J$ = 9.8 Hz, 1H), 6.75 (d, $J$ = 9.8 Hz, 1H). **$^{13}$C NMR (101 MHz, $CDCl_3$, ppm):** $\delta$ 185.3, 140.6, 134.0, 133.8, 133.4, 133.3, 130.6, 130.5, 129.8, 129.6, 128.3, 126.3, 120.6.

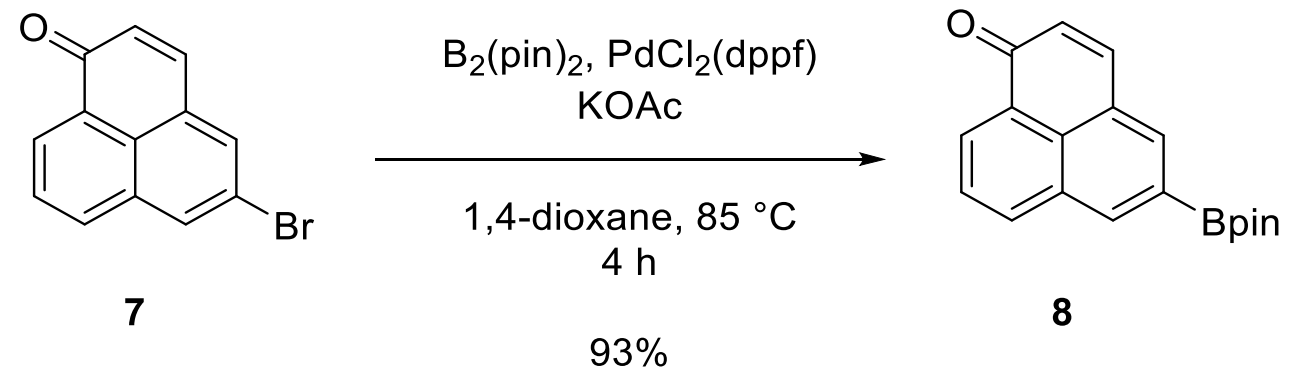

7 8

**5-(4,4,5,5-Tetramethyl-1,3,2-dioxaborolan-2-yl)-1*H*-phenalen-1-one (8).** The compound was prepared using a modified literature procedure.[5] The reaction was carried out under inert conditions using deoxygenated solvents (method A). 5-Bromo-1*H*-phenalen-1-one (**7**; 80 mg, 0.31 mmol), $B_2pin_2$ (78 mg, 0.31 mmol), potassium acetate (60 mg, 0.61 mmol) and $PdCl_2$(dppf) (11 mg, 0.015 mmol) were suspended in 1,4-dioxane (6 mL) and the mixture was heated for 3 h at 80 °C. After cooling to room temperature, the reaction mixture was diluted with methyl *tert*-butyl ether and washed with water followed by brine. The phases were separated and the organic layer was dried over $MgSO_4$. After evaporation of the solvents under reduced pressure, the crude product was passed through a pad of silica with methyl *tert*-butyl ether as an eluent. The desired compound was obtained as a yellow-brown solid (89 mg, 0.29 mmol, 93%). **$^1$H NMR (600 MHz, $CDCl_3$, ppm):** $\delta$ 8.66 (dd, $J$ = 7.3, 1.2 Hz, 1H), 8.53 (s, 1H), 8.25 (d, $J$ = 8.0 Hz, 1H), 8.13 (s, 1H), 7.79 (dd, $J$ = 7.6, 7.6 Hz, 1H), 7.79 (d, $J$ = 9.7 Hz, 1H), 6.74 (d, $J$ = 9.7 Hz, 1H), 1.42 (s, 12H). **$^{13}$C NMR (151 MHz, $CDCl_3$, ppm):** $\delta$ 185.8, 142.3, 140.2, 136.5, 135.8, 131.8, 131.5, 129.7, 129.4, 129.2, 127.3, 127.2, 84.6, 25.1 (the resonance of $C_q$(B) was not detected, presumably due to long relaxation time). **HRMS (EI) *m/z*:** [*M*]$^+$ Calcd for $C_{19}H_{19}O_3B$ 306.1422; Found 306.1426.

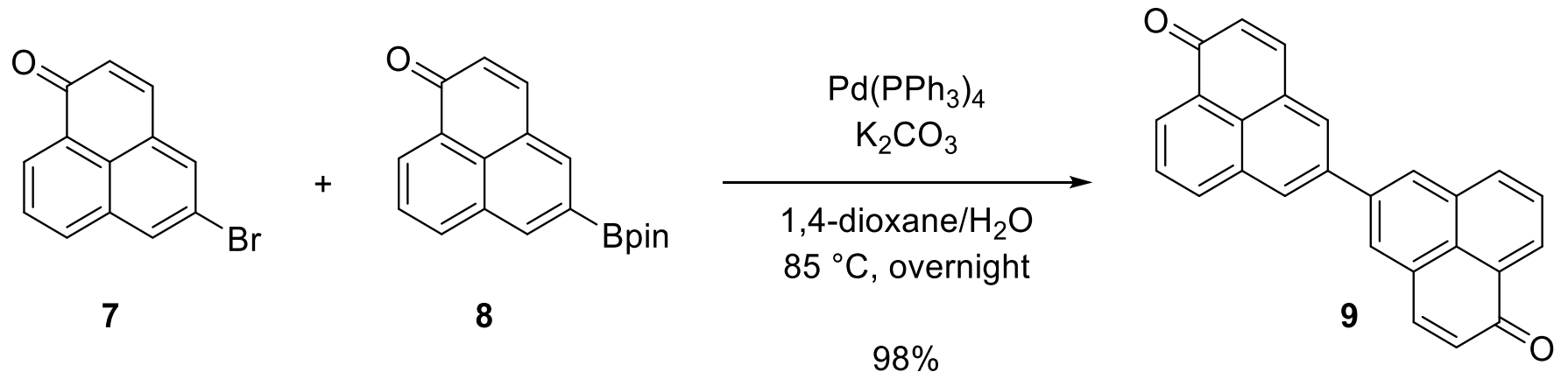

7 8 9

**6*H*,6'*H*-[2,2'-Biphenalene]-6,6'-dione (9).** The reaction was carried out under inert conditions with deoxygenated solvents (method A) using a modified literature procedure.[6] 5-Bromo-1*H*-phenalen-1-one (**7**; 20 mg, 80 µmol), 5-(4,4,5,5-tetramethyl-1,3,2-dioxaborolan-2-yl)-1*H*-phenalen-1-one (**8**; 24 mg, 80 µmol), potassium carbonate (21.3 mg, 150 µmol) and $Pd(PPh_3)_4$ (0.89 mg, 10 mol%) were suspended in 1,4-dioxane (1.6 mL) and water (0.4 mL), and the reaction mixture was heated overnight at 85 °C. After cooling to room temperature, the reaction mixture was diluted with water. The formed yellow precipitate was filtered off and washed with water, acetone and $CH_2Cl_2$ before it was dried under reduced pressure. The desired compound was obtained as a yellow solid (34.2 mg, 75.0 µmol, 98%). **$^1$H NMR (600 MHz, $CDCl_3$, ppm):** $\delta$ 8.67 (dd, $J$ = 7.3, 1.2 Hz, 2H), 8.38 (d, $J$ = 1.7 Hz, 2H), 8.34 (ddd, $J$ = 8.0, 1.2, 0.6 Hz, 2H), 8.15 (d, $J$ = 1.7 Hz, 2H), 7.89 (d, $J$ = 9.7 Hz, 2H), 7.88 (dd, $J$ = 8.0, 7.3 Hz, 2H), 6.83 (d, $J$ = 9.7 Hz, 2H). **$^{13}$C NMR (151 MHz, $CDCl_3$, ppm):** $\delta$ 185.7, 141.6, 138.5, 135.3, 132.8, 130.8, 130.6, 130.2, 129.9, 129.6, 129.0, 128.1, 127.3. **HRMS (EI) *m/z*:** [*M*]$^+$ Calcd for $C_{26}H_{14}O_2$ 358.0988; Found 358.0989.

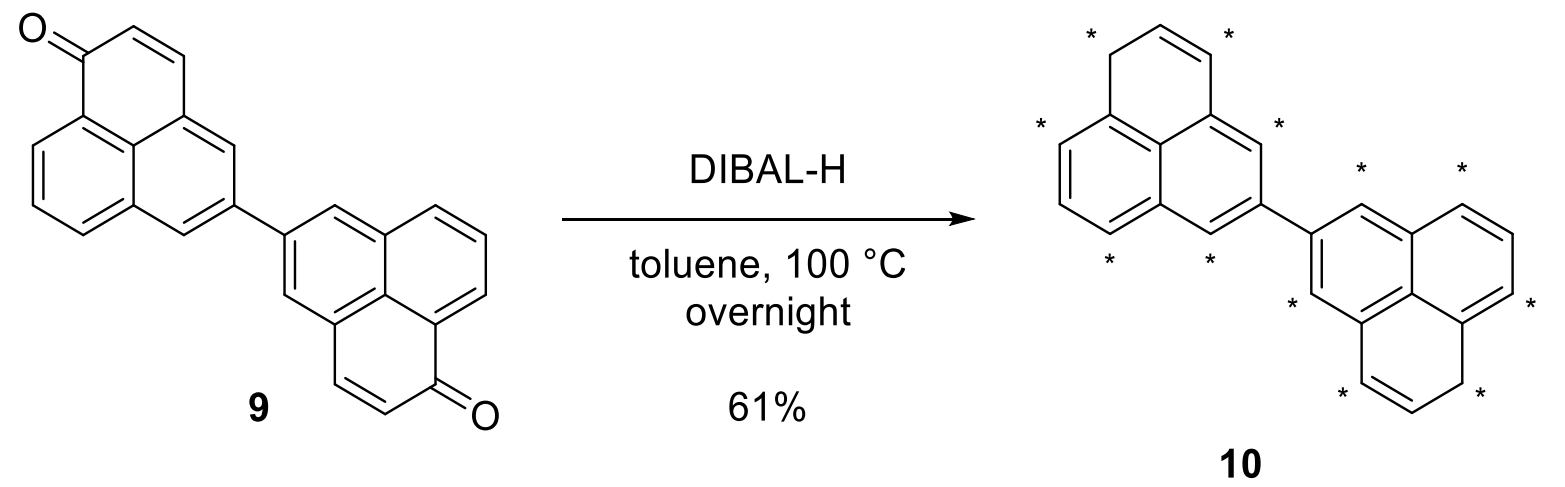

**10**
**2*H*-diphenalenyl**

**6*H*,6'*H*-2,2'-Biphenalene (10, 2*H*-diphenalenyl).** The compound was prepared using a modified literature procedure.[7] The reaction was carried out under inert conditions with deoxygenated solvents (method B). 6*H*,6'*H*-[2,2'-Biphenalene]-6,6'-dione (**9**; 50 mg, 0.56 mmol) was suspended in toluene (5 mL). A solution of DIBAL-H (0.56 mL, 0.56 mmol, 1.0 M in toluene) was added dropwise at room temperature and the reaction mixture was heated at 100 °C overnight. After cooling to room temperature, the reaction mixture was passed through a pad of Florisil® with toluene as an eluent under the exclusion of air. After removal of the solvent in vacuum, the desired compound was obtained as a light-yellow solid (28 mg, 0.14 mmol, 61%). **$^1$H NMR (400 MHz, $CD_2Cl_2$):** Compound **10** is obtained as a mixture of regioisomers, which differ by positions of the methylene groups that can occupy any $\alpha$-position (marked with asterisks) of the phenalenyl subunit. This mixture of regioisomers gives a complex proton NMR spectrum with characteristic singlets for the methylene groups. The ratio between the integrated signal intensity of all methylene groups and all aromatic protons is 2:7, matching the expected value. **HRMS (EI) *m/z*:** [*M*]$^+$ Calcd for $C_{26}H_{18}$ 330.1403; Found 330.1390.

## 2. Additional STM/STS data

### Figure S2

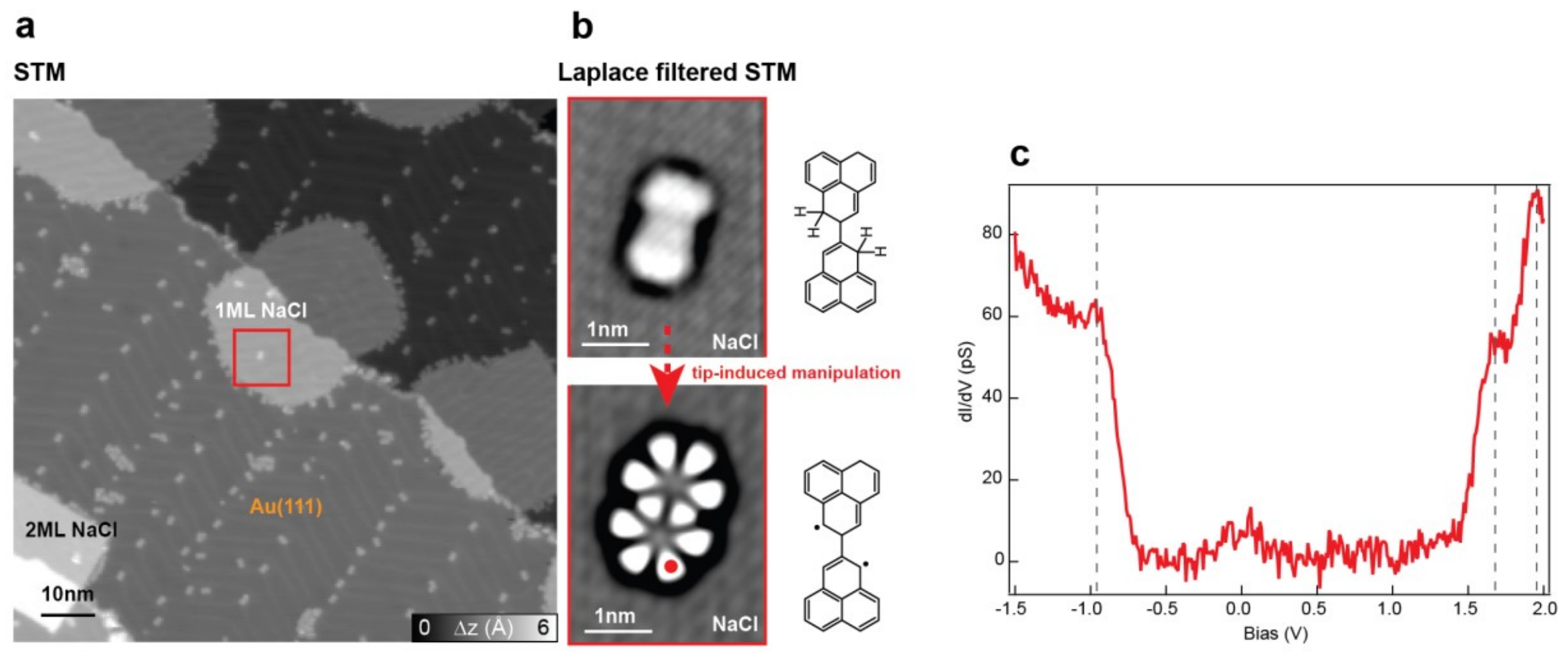

Figure S2. (**a**) STM image of diphenalenyl and NaCl on Au(111). The NaCl islands occur as rectangular shaped double layers (bottom left) and more roundish monolayer (ML) islands (center). The red square marks a diphenalenyl on ML-NaCl/Au(111). Image taken at *V* = 1.5 V and *I* = 10 pA. (**b**) Laplace-filtered STM images of passivated and activated diphenalenyl and underlying atomic structure of the NaCl island. Images taken at 1.5 V/20 pA (top) and -1.5 V/20 pA (bottom). (**c**) d*I*/d*V* spectroscopy taken at position marked by circle in (b). The vertical dotted lines mark the energies of the Hubbard peaks of the HOMO (-0.95 V, +1.95 V) and LUMO (+1.7 V).

### Figure S3

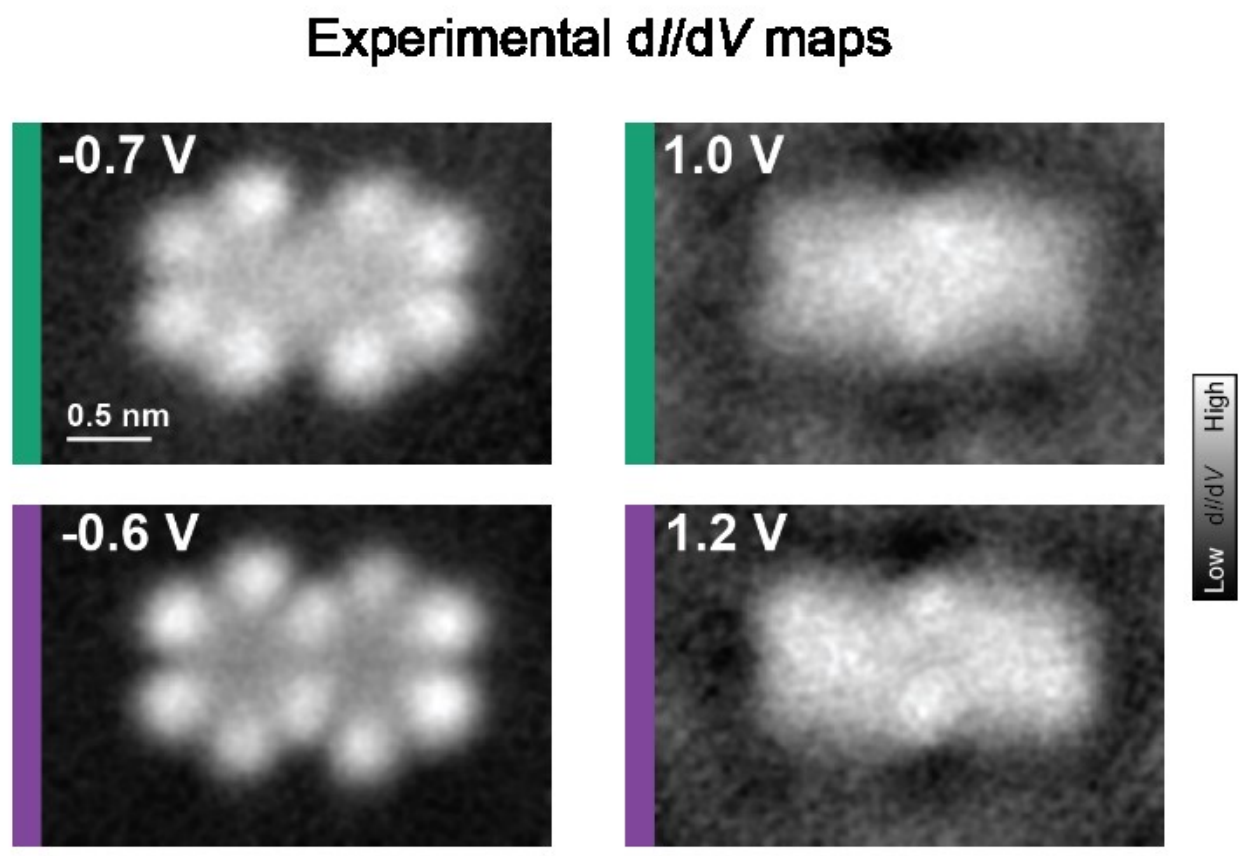

Figure S3. Constant current d*I*/d*V* maps of diphenalenyl on Au(111) at negative (left) and positive (right) bias polarity, resolving the many-body state due to entanglement of LUMO (green, top) and HOMO (purple, bottom). Maps taken at *I* = 200pA with a modulation voltage $V_{rms}$ = 14mV.

## 3. Experimental methods

All experiments were conducted on a Au(111) surface after repeated sputtering and annealing cycles until an atomically clean surface was achieved. NaCl was then deposited from a Knudsen cell evaporator at 730 °C, while the sample temperature was kept around 270 K to promote the growth of monolayer NaCl. The 2H-Diphenalenyl precursor (for synthesis see SI section 1) was deposited from a Knudsen cell evaporator onto the sample at $T_{\text{sample}}$ = 110 K. The two additional hydrogen atoms on the 2H-diphenalenyl passivate the precursor for better handling before the deposition.

On the surface, sequential tip-induced cleaving of the hydrogen atoms from the $sp^3$ carbon atoms[1,2] yields the target diphenalenyl diradical, as was proven by constant height AFM measurements shown in Fig. 1a. In order to activate a 2H-diphenalenyl precursor on monolayer NaCl, the tip was placed above the precursor at 1.5 V bias voltage and a current set point of 20 pA, then retracted by 5-6 Å and for a few seconds a bias voltage of 4 V was applied[2]. The successful cleavage of the hydrogen atoms and change in the electronic structure can be observed in the STM images, showing distinct lobes and nodal planes for the activated molecule. A precursor adsorbed on the Au(111) surface can be activated by tunneling currents at bias voltages around -2V[1].

## 4. Tight-binding (TB) and mean-field Hubbard (MFH) calculations

TB-MFH calculations were performed by numerically solving the mean-field Hubbard Hamiltonian with third-nearest-neighbor hopping

$$\hat{H}_{MFH} = \sum_{j} \sum_{\langle\alpha,\beta\rangle_j,\sigma} -t_j c^{\dagger}_{\alpha,\sigma} c_{\beta,\sigma} + U \sum_{\alpha,\sigma} \langle n_{\alpha,\sigma} \rangle n_{\alpha,\overline{\sigma}} - U \sum_{\alpha} \langle n_{\alpha,\uparrow} \rangle \langle n_{\alpha,\downarrow} \rangle. \quad \text{(SE1)}$$

Here, $c^{\dagger}_{\alpha,\sigma}$ and $c_{\beta,\sigma}$ denote the spin selective ($\sigma \in \{\uparrow,\downarrow\}$ with $\overline{\sigma} \in \{\downarrow,\uparrow\}$) creation and annihilation operator at sites $\alpha$ and $\beta$, $\langle\alpha,\beta\rangle_j$ ($j = \{1,2,3\}$) denotes the nearest-neighbor, second-nearest-neighbor and third-nearest-neighbor sites for $j$= 1, 2 and 3, respectively, $t_j$ denotes the corresponding hopping parameters (with $t_1$= 2.7 eV, $t_2$= 0.1 eV and $t_3$= 0.27 eV for nearest-neighbor, second-nearest-neighbor and third-nearest-neighbor hopping[1]), $U$ denotes the on-site Coulomb repulsion, $n_{\alpha,\sigma}$ denotes the number operator, and $\langle n_{\alpha,\sigma} \rangle$ denotes the mean occupation number at site $\alpha$. Orbital electron densities, $\rho$, of the $n^{\text{th}}$-eigenstate with energy $E_n$ have been simulated from the corresponding state vector $a_{n,i,\sigma}$ by

$$\rho_{n,\sigma}(\vec{r}) = \left| \sum_{i} a_{n,i,\sigma} \phi_{2p_z}(\vec{r} - \vec{r}_i) \right|^2, \quad \text{(SE2)}$$

where $i$ denotes the atomic site index and $\phi_{2p_z}$ denotes the Slater $2p_z$ orbital for carbon.
All the TB-MFH calculations presented in the manuscript were done in the third-nearest-neighbor approximation and using an on-site Coulomb term $U$ = 3.5 eV.
The TB-MFH software library[2] was developed within the Python programming language. The code is open-source, available at https://github.com/eimrek/tb-mean-field-hubbard.

## 5. DFT results of diphenalenyl on Au(111) and NaCl

The initial setup of the molecular structures was carried out using the Atomic Simulation Environment (ASE) package. [1] The electronic structure calculations in this study were performed using the Generalized Projector Augmented Wave (GPAW) software package, [2] employing the Perdew-Burke-Ernzerhof (PBE) exchange-correlation functional and the double-ζ polarized (dzp) basis set. To ensure the structural accuracy, the geometries were optimized until the maximum force on any atom was below the threshold of 0.05 Hartree Bohr$^{-1}$. Single-point energy calculations were carried out after the geometry optimization, followed by non-equilibrium Green's function (NEGF) calculations to simulate the electronic properties of the open system coupled to the semi-infinite Au(111) substrate. Open boundary conditions (OBCs) were incorporated at the bottom of the simulation cell through a self-energy matrix, coupling the part of the Au(111) substrate included in the simulation cell to the semi-infinite bulk Au(111) substrate. The OBCs were computed using the method described in Ref. [3] The bulk Au(111) substrate was modeled by a three-layer thick Au(111) slab and sampled with a 3×1×1 k-mesh along the transport direction.

The diphenalenyl molecule was investigated when directly absorbed on top of Au(111) and when separated by a monolayer of NaCl, as shown in Figs. S4 and S5, respectively. The molecule absorbed roughly 2.67 Angstrom above the substrate when in contact with Au(111), while the absorption distance increased to approximately 3.31 Angstrom when a layer of NaCl was placed in-between, indicating weaker hybridization of the molecule with the substrate. The projected density of states (PDOS) for the four frontier MOs, ranging from HOMO-1 to LUMO+1, for Au(111) and NaCl are reported in Fig. S6, left and right panels respectively. The MOs hybridize with the bulk states of the substrate, resulting in a finite lifetime for each MO, and the strength of hybridization determines the ease with which the two systems can exchange electrons via the corresponding MO. The PDOS shows that the delta peaks at the position of the MOs eigenenergies broaden into Lorentzians L(x)=a/π Γ/2 1/((x-μ)2-(Γ/2)2), where the value of Γ reflects the strength of hybridization. The values of Γ for the four frontier MOs estimated from fitting the PDOS computed using NEGF with the function L(x) can be found in the legend of Fig. S6, with the HOMO and LUMO showing hybridization strength more than one order of magnitude stronger when in direct contact with Au(111). The imaginary part of the hybridization functions [4], which describe the coupling strength dispersion as a function of energy, are shown in Fig. S7 (left and right panels) for the same MOs, normalized to the integral of the corresponding PDOS. The hybridization functions exhibit a nearly constant behavior around the MO eigenenergies, resulting in a well-defined and symmetric Lorentzian shape of the corresponding PDOS. The legend in Fig. S7, left and right panels, provides the values of the hybridization functions at the energy of the corresponding MO eigenenergies, which give a complementary estimate for Γ. These values agree closely with the values of Γ obtained from the PDOS, supporting the conclusion that the MOs hybridize more strongly with Au(111) than with NaCl.

NOTE: The hybridization function, which describes the coupling between the molecule and the substrate, is computed using the formula $\Delta(z) = (G^0_M)^{-1}(z) - G_M^{-1}(z)$, where $G_0^{-1}$ represents the bare Green's function of the molecule without any interaction with the substrate. This is

given by $(G^0_M)^{-1}(z) = zI_M - H_M$, where $H_M$ is the Hamiltonian block of the molecule in the Hamiltonian matrix $H_C$ of the simulation cell, $I_M$ is an identity matrix, and $z=E+i0^+$ is a complex energy with a small positive imaginary shift. The retarded Green's function $G_M$ projected onto the molecular subspace is given by [4] $G_M(z) = S_{MC}G_C(z)S_{CM}$, where $S_{MC}$ is the overlap matrix between the molecule orbitals with all orbitals, and $G_C$ is the Green's function of the simulation cell, which can be expressed as $G_C(z) = zS_C-H_C-\Sigma_L(z)$, taking into account the OBCs through the self-energy $\Sigma_L$. The PDOS can be computed from $G_M$ as $PDOS_i(E)=-1/\pi \, Im[G_M(z)]_i$, where the subscript i denotes the i-th diagonal component of the Green's function.

**Figure S4**

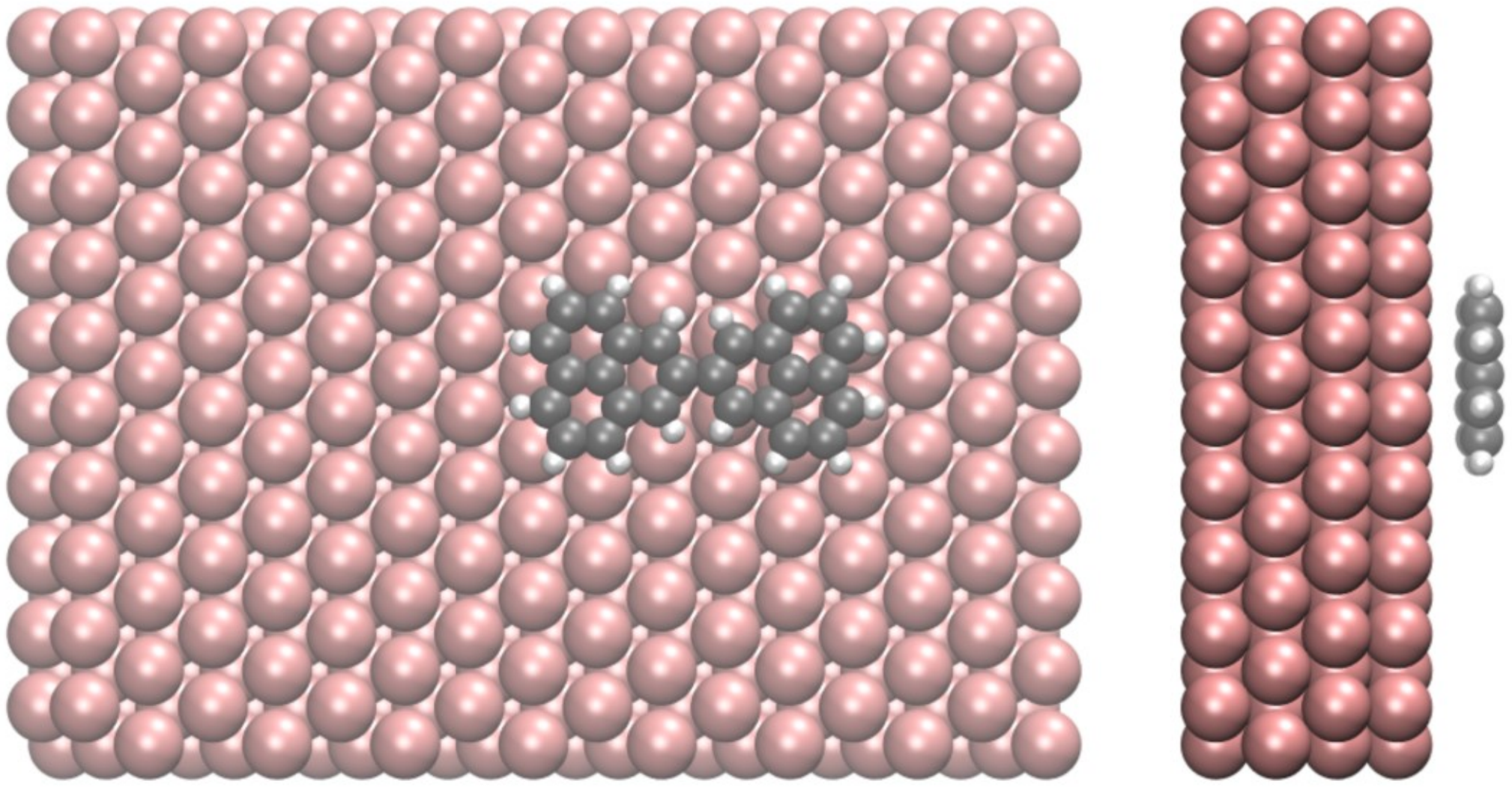

Figure S4. Atomic structure of diphenalenyl molecule adsorbed on Au(111) surface. The pink spheres represent the Au atoms, the grey spheres represent the C atoms, and the white spheres the H atoms. The molecule absorbs approximately 2.67 Angstrom above the substrate.

**Figure S5**

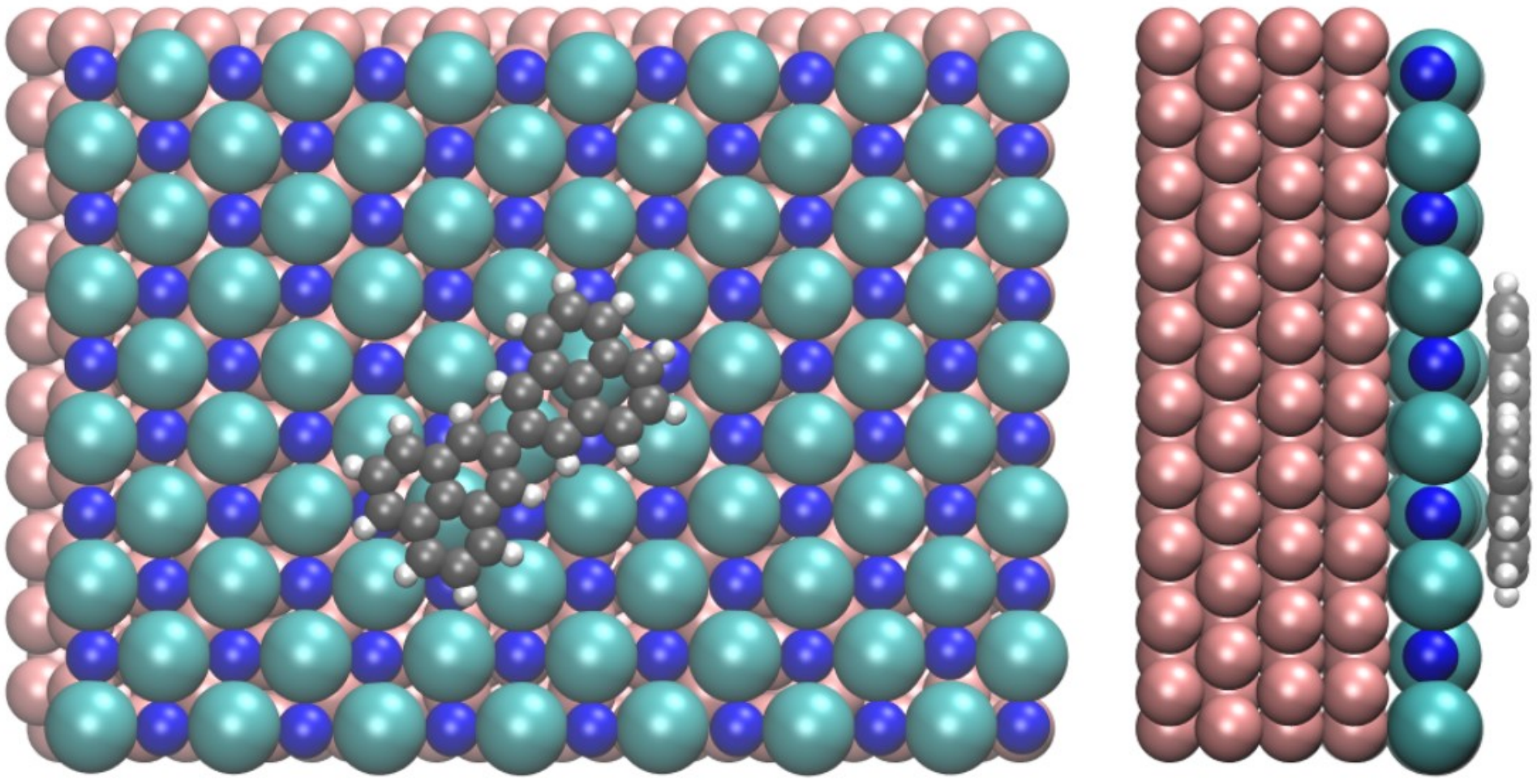

Figure S5. Atomic structure of diphenalenyl molecule adsorbed on 1ML NaCl/Au(111) surface. The pink spheres represent the Au atoms, the green spheres the Cl atoms, the blue spheres the Na atoms, the grey spheres the C atoms, and the white spheres the H atoms. The molecule absorbs approximately 3.31 Angstrom above the NaCl layer.

**Figure S6**

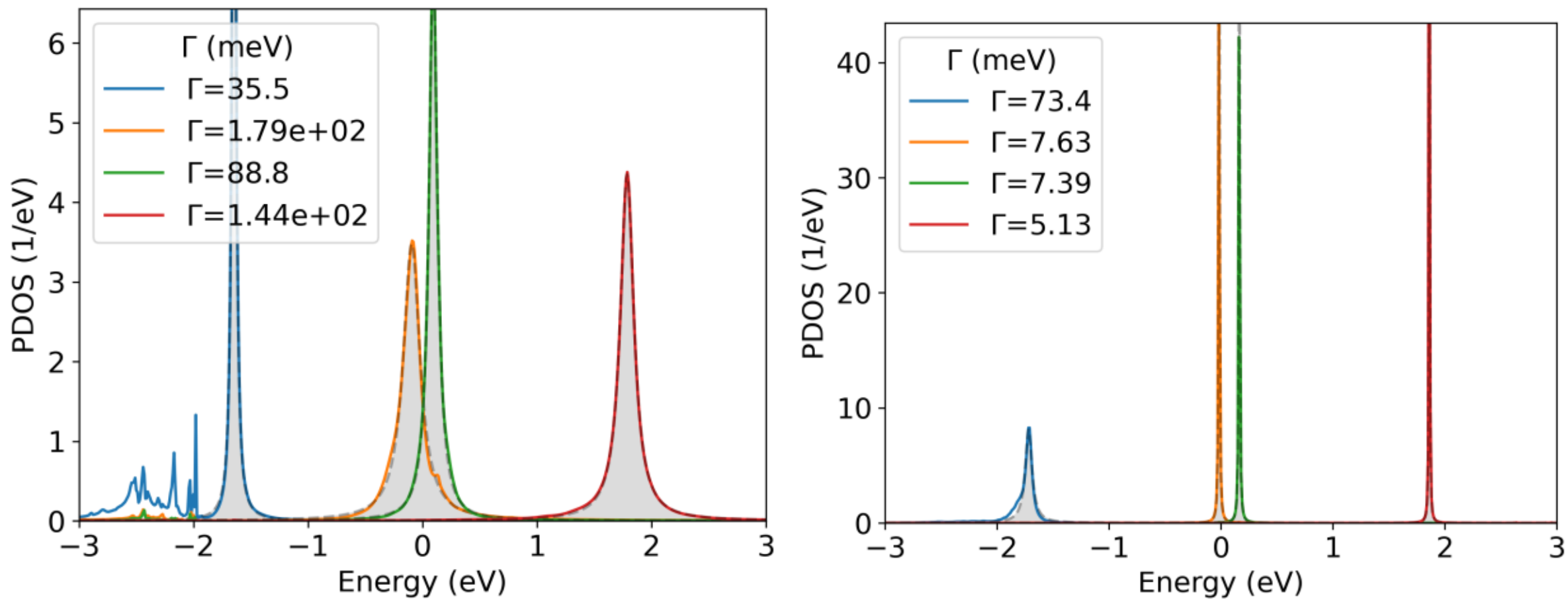

Figure S6. The figure displays the Projected Density of States (PDOS) of the diphenalenyl molecule adsorbed on Au(111) and NaCl. The coloured lines show the PDOS computed from NEGF, while the grey shade represents the PDOS fitted to the formula reported in the text. The Γ values in the legend are estimated from the fitting. The PDOS of the molecule on Au(111) exhibits broader peaks compared to NaCl, especially for the HOMO and LUMO, indicating a more significant interaction between the molecule and the Au(111) substrate.

**Figure S7**

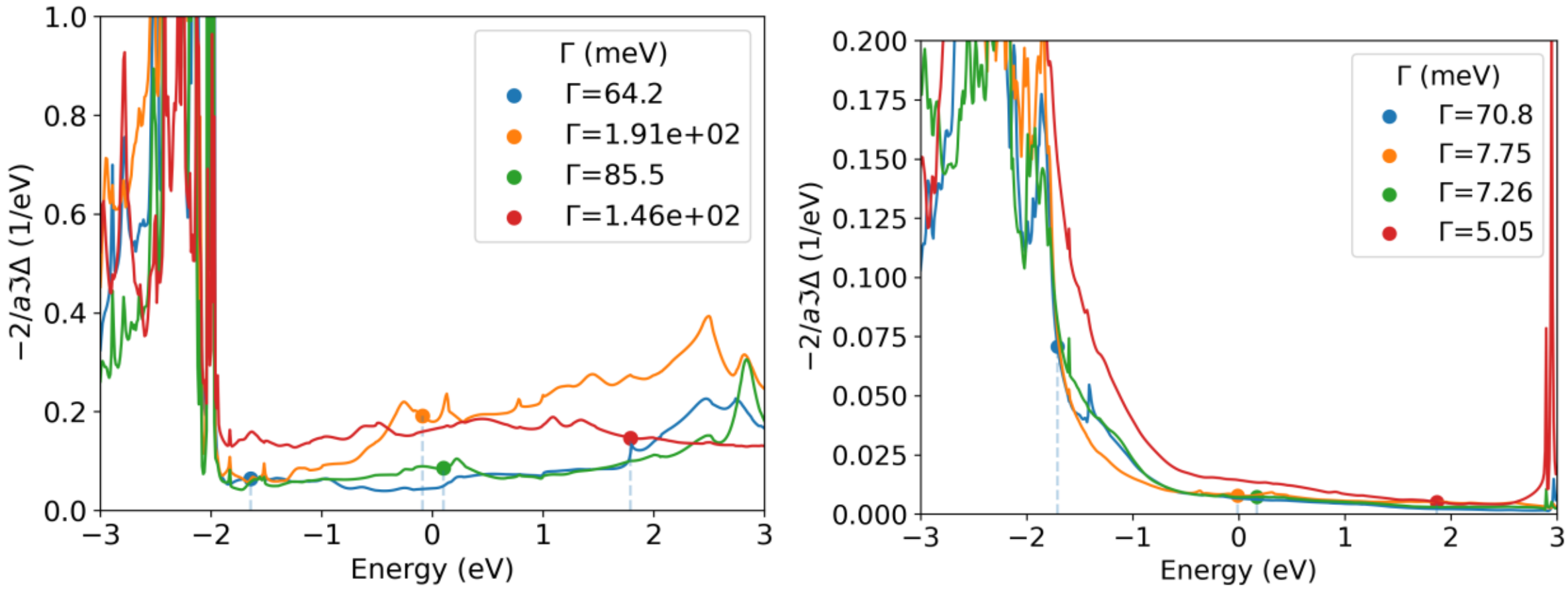

Figure S7. The figure illustrates the computed hybridization functions of the diphenalenyl molecule adsorbed on Au(111) and NaCl, using NEGF. The observed hybridization functions verify the trend of a stronger coupling between the diphenalenyl molecule and Au(111) compared to NaCl, as demonstrated by the values of Γ at the positions of the MOs eigenenergies. Furthermore, these values are in agreement with those obtained from the fitting of the PDOS to Lorentzians.

The non-vanishing intermolecular coupling is verified using the local orbital (LO) approach [5]. It involves constructing a set of localized wavefunctions tied to specific atoms in the system to describe the KS states in a particular energy window of interest. In this study, the LO technique is employed to construct a minimal model of pz-like orbitals, with one orbital for each Carbon atom of the adsorbed molecule, to capture the relevant physics around the Fermi level, including the energy and electronic distribution of the DFT frontier molecular orbitals. The LO approach is chosen due to its proven accuracy in yielding minimal models for organic compounds [5]. The Hamiltonian of the minimal model is obtained by downfolding the electronic structure of the open system, i.e., including the coupling to the semi-infinite Au(111) substrate, onto the pz local orbitals (pz-LOs). This is achieved practically by incorporating the static component of the pz-LOs' hybridization function, evaluated at the Fermi level, into the Hamiltonian of the pz-LOs (see also Eq. (11) of [6] for more details). The resulting on-site energy and hopping values between the pz-LOs models for the diphenalenyl molecule on Au(111) and NaCl are reported in Fig. S8, left and right panel respectively, with a cut-off value of 0.15 eV set for the hoppings. It can be observed that the Hamiltonian matrix elements are very similar in the two models, and the hopping values between 1$^{st}$ and 2$^{nd}$ neighbors are comparable, with approximate magnitudes of $|t_1|$ ~ 2.8 eV and $|t_2|$ ~ 0.2 eV or $|t'_2|$ ~ 0.27 eV (see Fig. S8), respectively. The computed intermolecular exchange is estimated to be |t3| ~ 0.25 eV. These values are subsequently used to parameterize the tight-binding model. $|t_3|$ also corroborates the hypothesis of a non-negligible intermolecular coupling that mixes the zero modes of the two phenalenyl units.

**Figure S8**

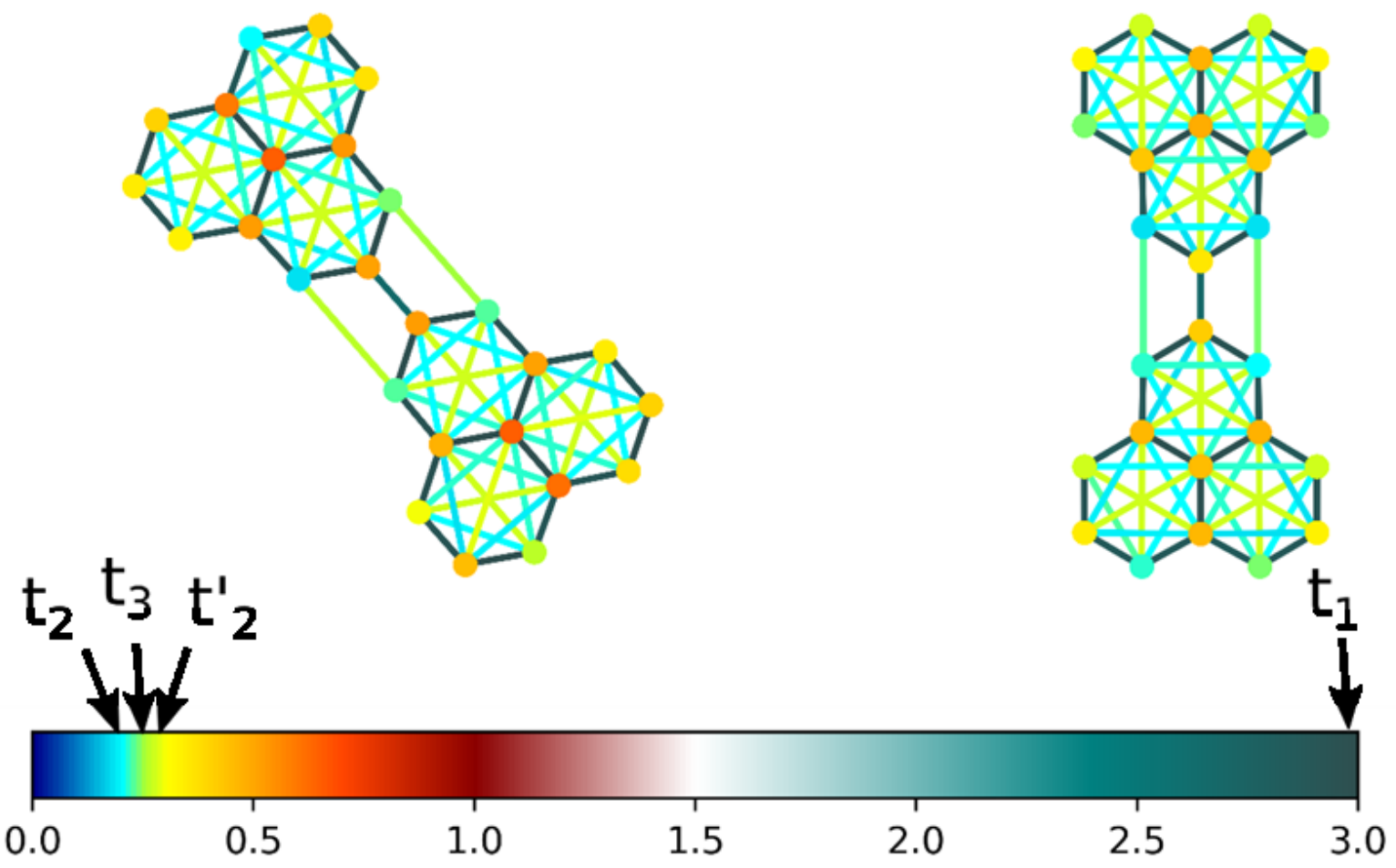

Figure S8. The figure depicts the on-site energy and hopping values between the pz-LOs of the minimal models for the diphenalenyl molecule adsorbed on Au(111) and NaCl, respectively. The calculated hopping value of $|t_3|$ ~ 0.25 eV supports the hypothesis of significant hybridization between the two phenalenyl units. Additionally, the hopping values for the first and second nearest neighbors are estimated to be $|t_1|$ ~ 2.8 eV and $|t_2|$ ~ 0.2 eV or $|t'_2|$ ~ 0.27 eV, respectively, with $|t_2|$ representing the hopping with the second farthest carbon in the phenyl ring, while $|t'_2|$ ~ 0.27 eV represents the hopping with the opposite carbon in the ring.

## 6. Many-body calculations for Phenalenyl dimer model

### i. Extended Hubbard model for Phenalenyl dimer

Similar to previous work by two of us [1] we consider an extended Hubbard model, including first and third nearest-neighbor hopping, as well as parts of the long-range (LR) Coulomb interaction in addition to the local (on-site) Hubbard interaction:

$$\hat{H} = \hat{H}_0 + \hat{W} = \sum_{i,j,\sigma} t_{ij} \left( c^{\dagger}_{i\sigma} c_{j\sigma} + \text{h.c.} \right) + \frac{1}{2} \sum_{\substack{i,j,k,l \\ \sigma,\sigma'}} W_{ijkl} \, c^{\dagger}_{i\sigma} c^{\dagger}_{j\sigma'} c_{l\sigma'} c_{k\sigma} \quad \text{(S1)}$$

where $t_{ij}$ is the hopping between carbon sites $i$ and $j$ and $W_{ijkl}$ is the Coulomb matrix in the site basis. We assume nearest-neighbor hopping $t = -2.7$eV and third-neighbor hopping $t_3 = 0.1t$. As before [1] we take into account the Hubbard on-site interaction $U \equiv W_{iiii}$, the Coulomb repulsion $W_{ijij}$ and Coulomb exchange $W_{ijji}$ between electrons on different sites $i$ and $j$. Additionally, here we also take into account the pair-hopping $W_{iijj}$ and the density assisted hopping $W_{ijkj}$. Following Ref. [1] we compute the Coulomb matrix elements $W_{ijkl}$ in the site basis $\{|i\rangle\}$ with the Gaussian09 quantum chemistry code as follows: first, the matrix elements of the bare Coulomb interaction $\hat{v}_c = 1/r$ are computed in Gaussian09 for the $p_z$ orbitals of the carbon sites in the phenalenyl dimer in the LANL2MB minimal basis set, $V_{ijkl} = \langle i,j| \, \hat{v}_c \, |k,l\rangle$. Second, screening effects by the other orbitals in the molecule and by the substrate are simply taken into account by a dielectric constant: $W_{ijkl} = V_{ijkl}/\epsilon$. The dielectric constant $\epsilon$ is related to the on-site Hubbard interaction by $U = W_{iiii} = V_{iiii}/\epsilon$. The parameter $U$ (or equivalently $\epsilon$) is adjusted in the model such that the OCA calculation (see below) yields the (renormalized) spin excitation energy observed in experiment. Thus for the molecule on the NaCl ML we use $U = 5.4$eV, while for the molecule directly on Au(111) we use $U = 2.5$eV.

### ii. Complete Active Space (CAS)

Diagonalization of the single-particle part $\hat{H}_0$ of (S1) yields the molecular orbitals $|\psi_k\rangle$:

$$\hat{H}_0 |\psi_k\rangle = \epsilon_k |\psi_k\rangle \quad \text{(S2)}$$

which can be expanded in the site basis $\{|i\rangle\}$ as $|\psi_k\rangle = \sum_j \psi_k(j)|j\rangle$. In the absence of third-neighbor hopping ($t_3 = 0$), the diagonalization yields two zero modes ($\epsilon_k = 0$) which may be localized on each of the phenalenyl units (see Fig. 3a in main text). Switching on $t_3$ the zero modes hybridize and split in energy forming HOMO and LUMO (see Fig. 3c in main text). The interactions between the zero modes via $t_3$ and via Coulomb interactions with the HOMO-1 and LUMO+1 give rise to kinetic and Coulomb-driven exchange interactions, respectively. In the case of the phenalenyl dimer these are both antiferromagnetic in nature [1]. Thus here for the many-body calculations we take the four orbitals HOMO-1, HOMO, LUMO and LUMO+1, shown in Fig. 3c in the main text, as the complete active space (CAS). The extended Hubbard Hamiltonian (S1) projected onto the CAS can then be written as

$$\hat{H}_{\text{C}} = \sum_{k \in \text{C}} \epsilon_k \, \hat{N}_k + \frac{1}{2} \sum_{\substack{k,k' \\ q,q' \\ \sigma,\sigma'}} \mathcal{W}_{kk'qq'} \, C^{\dagger}_{k\sigma} C^{\dagger}_{k'\sigma'} C_{q'\sigma'} C_{q\sigma} \quad \text{(S3)}$$

where $C^\dagger_{k\sigma}$ ($C_{k\sigma}$) creates (destroys) one electron of spin $\sigma$ in MO $\psi_k$, $N_k = \sum_\sigma C^\dagger_{k\sigma} C_{k\sigma}$ measures the total occupation of MO $\psi_k$, and the Coulomb interaction tensor between orbitals in the CAS is given by $\mathcal{W}_{k_1k_2k_3k_4} \equiv \langle \psi_{k_1}, \psi_{k_2} | \hat{W} | \psi_{k_3}, \psi_{k_4} \rangle$. Exact diagonalization of this model for different values of $U$ (corresponding to a dielectric constant $\epsilon = V_{iiii}/U$, see above) then yields the blue curve in Fig. 3d in the main text. Kinetic exchange can be switched off by switching off $t_3$ (green curve), while Coulomb-driven exchange can switched off by restriction to HOMO-LUMO model (purple curve), see Ref. [1].

### iii. Anderson impurity model

The CAS described by the Hamiltonian (S3) and coupled to the conduction electrons in the substrate defines an Anderson impurity model (AIM):

$$\hat{H}_{\mathrm{AIM}} = \hat{H}'_{\mathrm{C}} + \hat{H}_{\mathrm{B}} + \hat{V}_{\mathrm{hyb}} \tag{S4}$$

where $\hat{H}'_{\mathrm{C}} = \hat{H}_{\mathrm{C}} + v\,\hat{N}_{\mathrm{C}}$, $\hat{H}_{\mathrm{B}}$ is the Hamiltonian of the conduction electron bath in the metallic substrate and the hybridization term $\hat{V}_{\mathrm{hyb}}$ describes the coupling between bath and the impurity space. The coupling to the bath leads to charge fluctuations on the impurity and defines a chemical potential which we set to zero. The single-particle potential $v$ is set such that the coupled molecule is approximately charge neutral, i.e., $N_{\mathrm{C}} \approx 4$. This is achieved by taking the mean-field approximation of the interaction part $\hat{W}$ for $N_{\mathrm{C}} = 4$, averaging over the four MOs, i.e. $v = -\langle \hat{W} \rangle_{\mathrm{MFA}}$, and fine-tuning until the good agreement with the experimental spectra is reached. This yields $v = -3.65$eV in the case of NaCl and $v = -1.91$ eV in the case of Au.

Integrating out the bath degrees of freedom we obtain the so-called hybridization function for each MO included in the CAS $\Delta_k(\omega) = \sum_b |V_{k,b}|^2/(\omega^+ - \epsilon_b)$. Its negative imaginary part $\Gamma_k(\omega)$ yields the single-particle broadening of the impurity orbitals due to the coupling to the bath. Here we will assume the wide-band limit, i.e. constant broadening $\Gamma_k$ and vanishing real part of the hybridization function. For the Au case the $\Gamma_k$ are obtained from *ab initio* DFT calculations described above. This yields $\Gamma_{\mathrm{homo-1}} \sim 54\,\mathrm{meV}$, $\Gamma_{\mathrm{homo}} \sim 81\,\mathrm{meV}$, $\Gamma_{\mathrm{lumo}} \sim 35\,\mathrm{meV}$ and $\Gamma_{\mathrm{lumo+1}} \sim 49\,\mathrm{meV}$. Expectedly, for the NaCl monolayer the hybridization functions obtained from DFT are smaller by roughly an order of magnitude (~5-10meV). Unfortunately, this leads to numerical problems in the OCA calculations. We thus assume somewhat larger value, $\Gamma_k \sim 15\,\mathrm{meV}$, and the same for all orbitals. Thus for the NaCl case the OCA spectra are somewhat more broadened than is to be expected. But the renormalization of the spin excitation energy by Kondo exchange is still negligible.

### iv. One-crossing approximation

In order to solve the AIM, we use the one-crossing approximation (OCA) which expands the propagators corresponding to the eigenstates of the *isolated* impurity (i.e. the CAS) in the coupling to the bath [2]. The first step, is an exact diagonalization of the impurity Hamiltonian, i.e. $\hat{H}'_{\mathrm{C}} |\Psi_m\rangle = E_m |\Psi_m\rangle$. The eigenstates are simultaneous eigenstates of the total number of electrons $N$ and the total spin $S^2$. Here we consider the system at half-filling ($N_{\mathrm{C}} = 4$). Coupling to the substrate leads to fluctuations of electrons in the impurity, i.e. to excitations to the charged sectors with $N_{\mathrm{C}} \pm 1$ electrons. OCA consists in a diagrammatic expansion of the propagators $G_m(\omega)$ associated with the many-body eigenstates $|\Psi_m\rangle$ of the isolated impurity Hamiltonian $\hat{H}'_{\mathrm{C}}$ in terms of the hybridization functions $\Delta_k(\omega)$, summing only the first and second order diagrams (only those where conduction electron lines cross at most twice) but to infinite order (i.e. self-consistently). The spectral functions $A_k(\omega)$ projected on individual molecular orbitals $|\psi_k\rangle$ are obtained from convolutions of the propagators $G_m(\omega)$ with the hybridization functions $\Delta_k(\omega)$. Further details of the OCA method and its application to the description of spin excitations in adatoms and nanographenes can be found in previous works [3,4,5].

## 8. Copies of NMR and HRMS spectra

**1,2,3,4-Tetrabromo-1,2,3,4-tetrahydronaphthalene (2).** $^{1}H$ NMR / 400 MHz / $CDCl_3$ / 25 °C

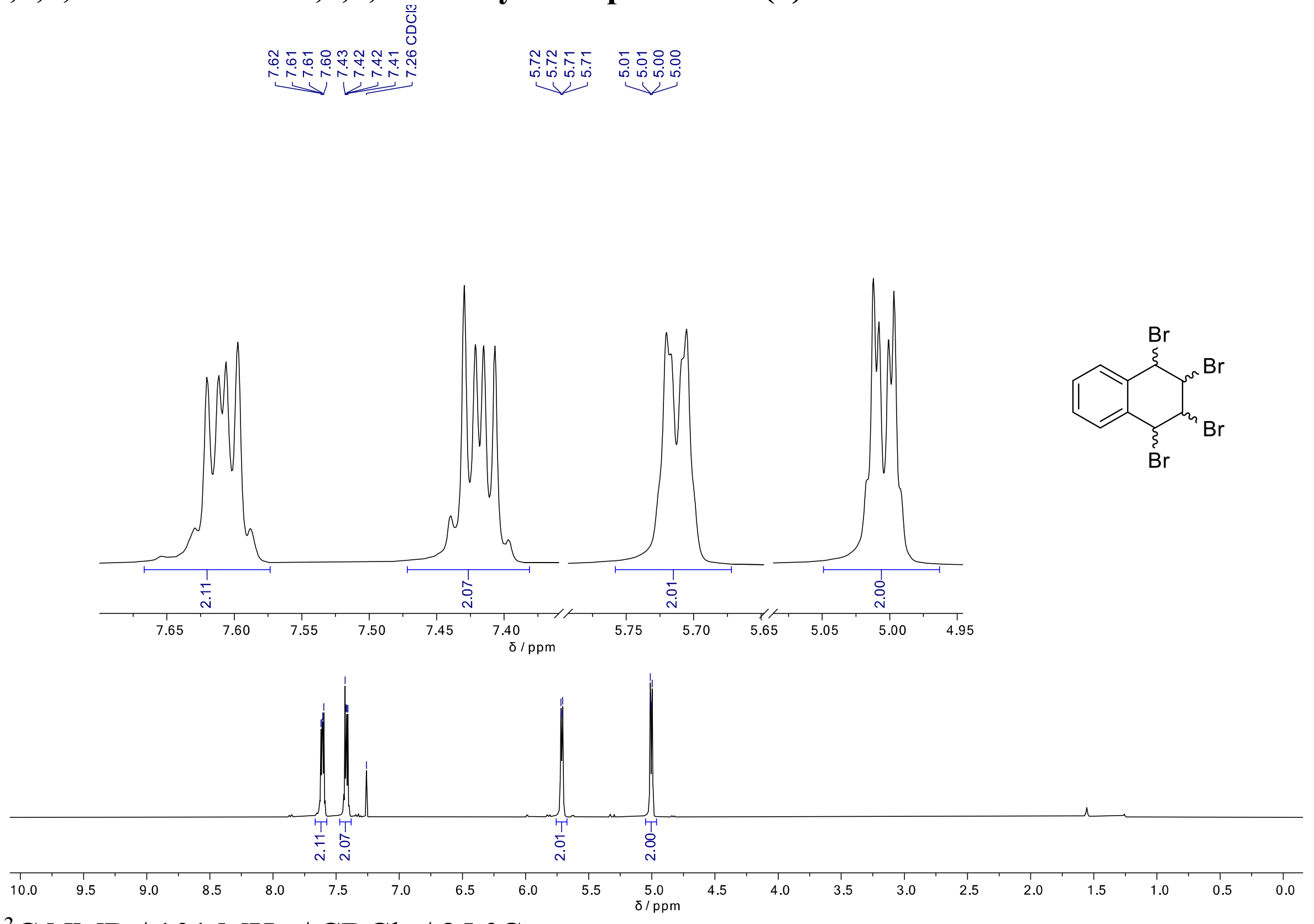

$^{13}C$ NMR / 101 MHz / $CDCl_3$ / 25 °C

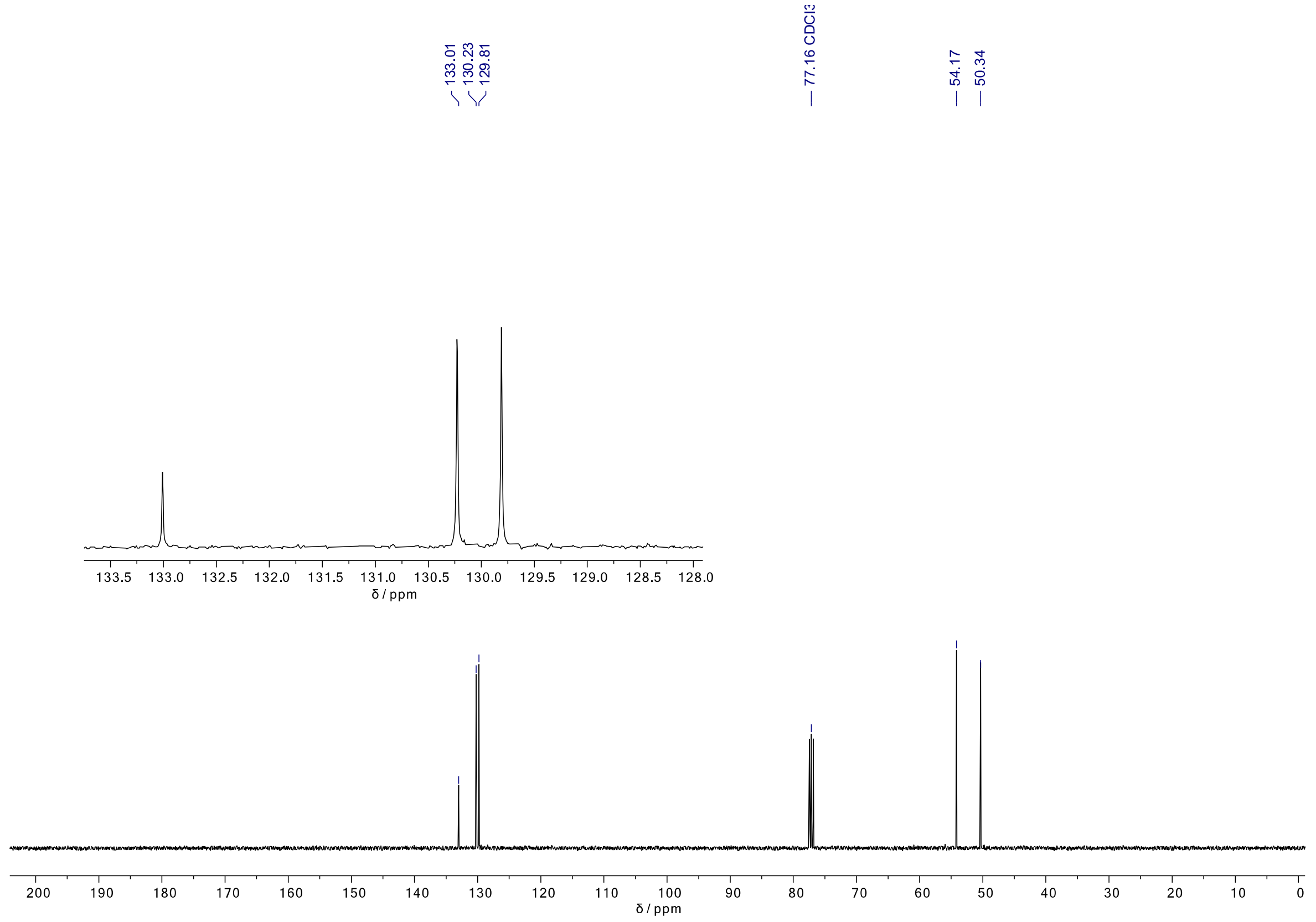

**1,3-Dibromonaphthalene (3).** $^{1}$H NMR / 400 MHz / $CDCl_3$ / 25 °C

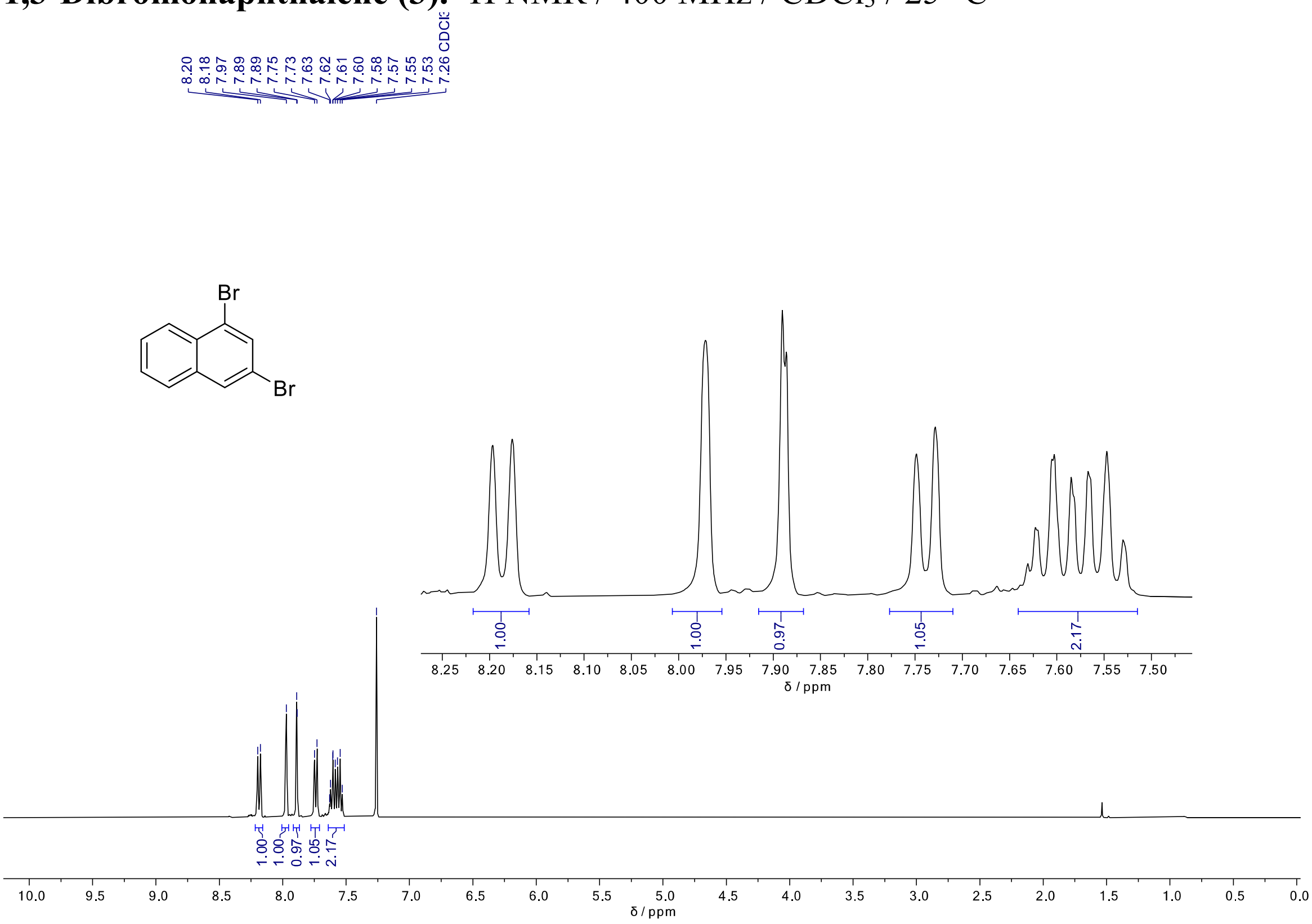

$^{13}$C NMR / 101 MHz / $CDCl_3$ / 25 °C

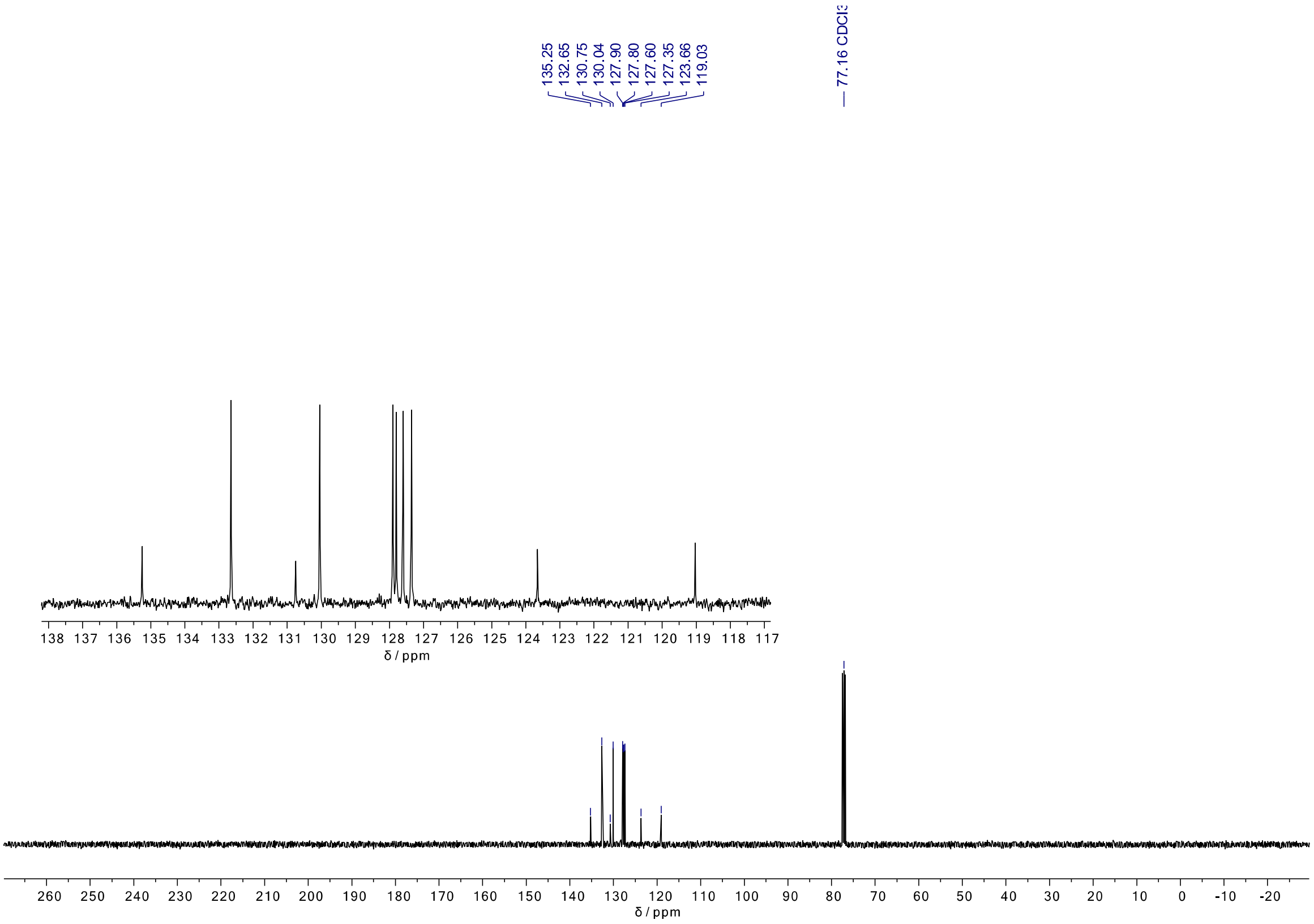

**3-Bromo-1-naphthaldehyde (4).** $^{1}$H NMR / 400 MHz / $CDCl_3$ / 25 °C

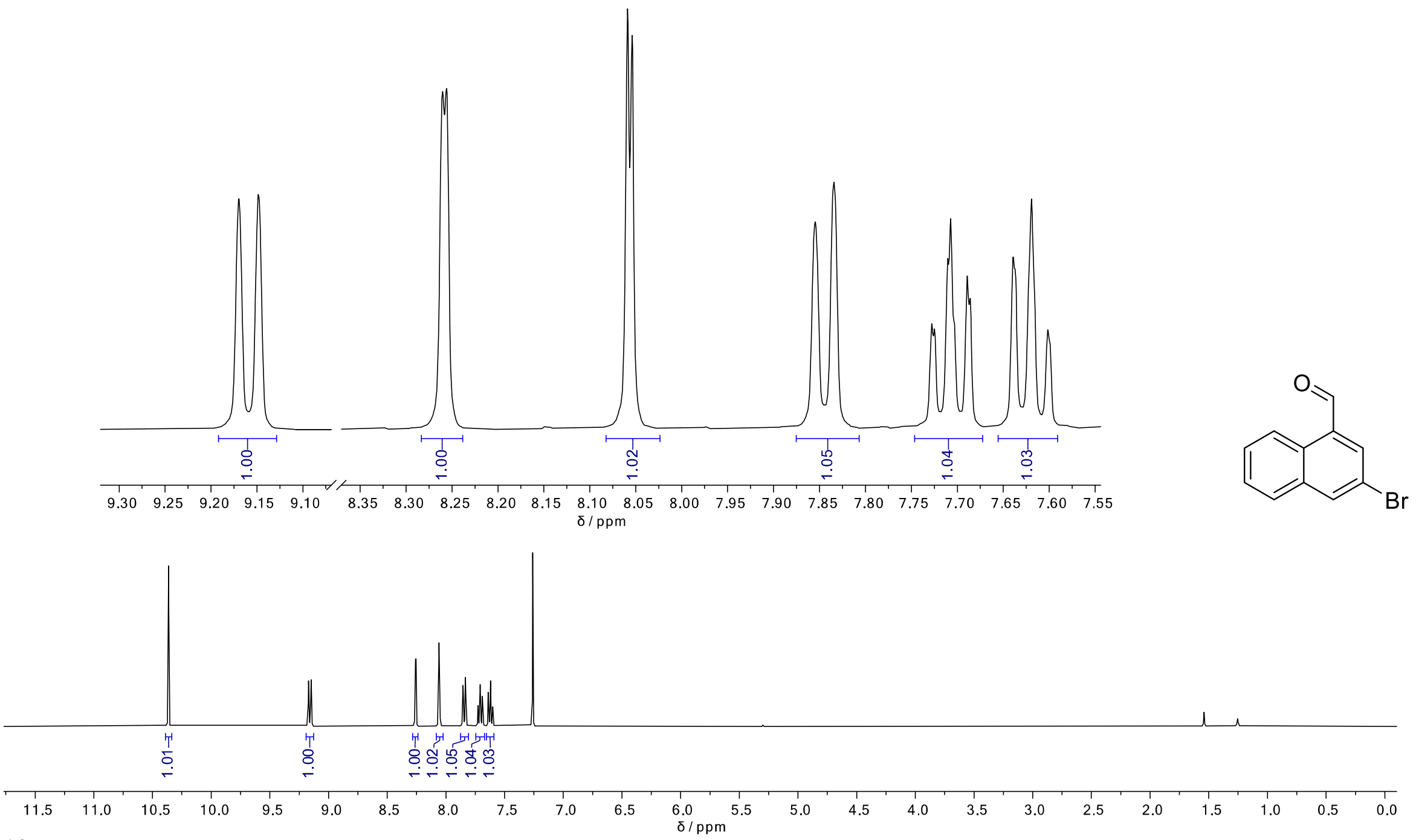

$^{13}$C NMR / 101 MHz / $CDCl_3$ / 25 °C

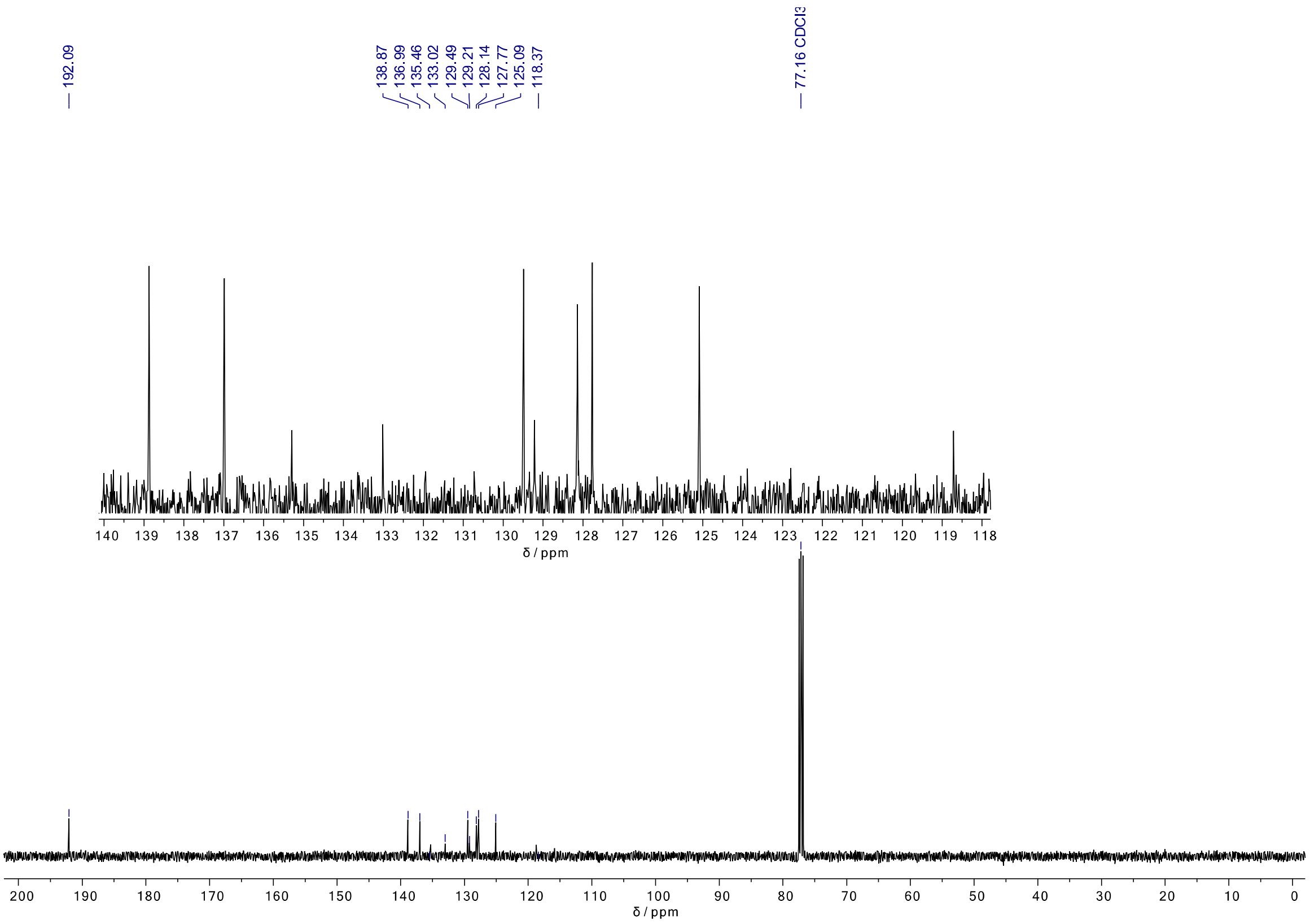

**Ethyl (*E*)-3-(3-bromonaphthalen-1-yl)acrylate (5).** $^{1}$H NMR / 400 MHz / $CDCl_3$ / 25 °C

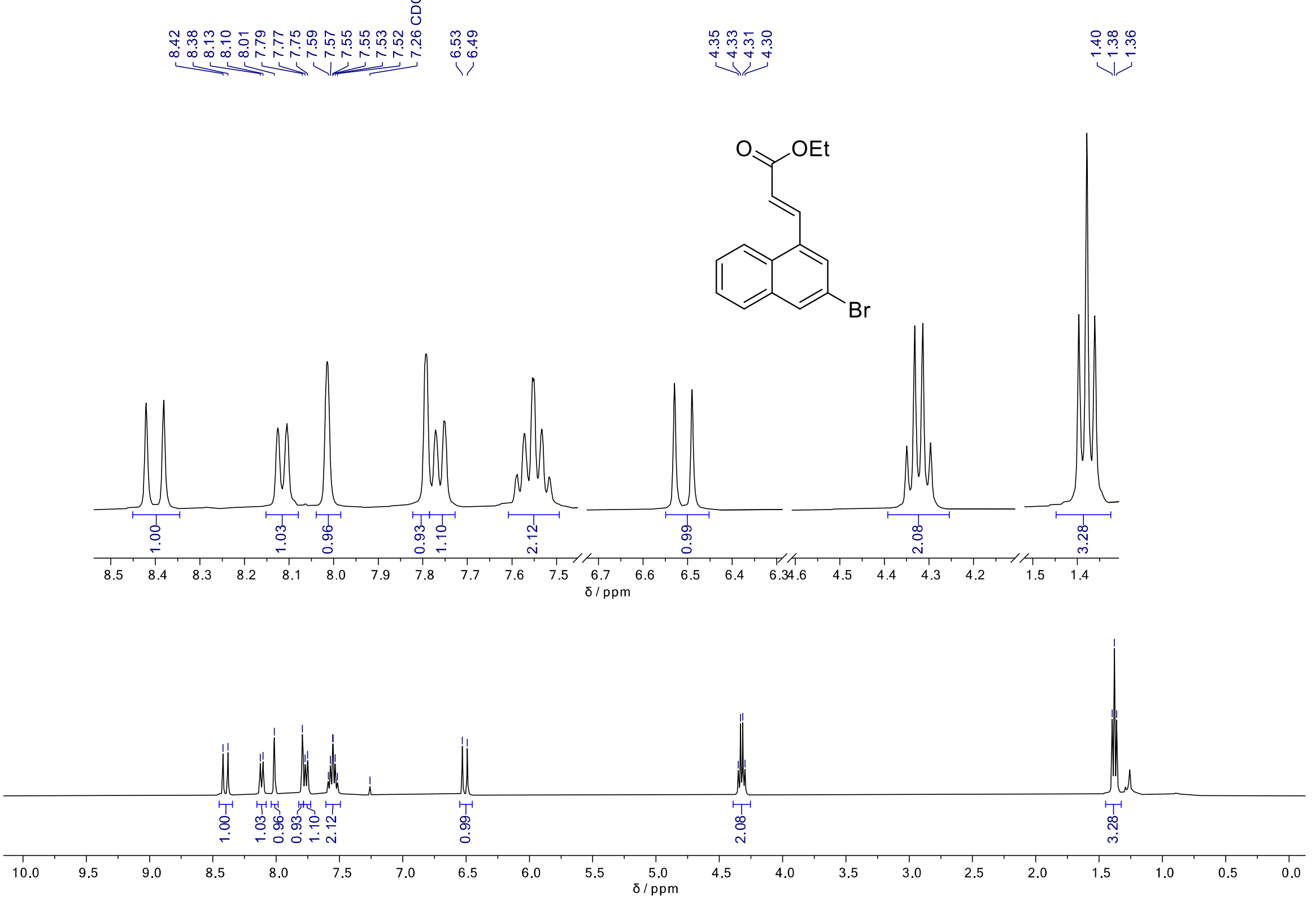

$^{13}$C NMR / 101 MHz / $CDCl_3$ / 25 °C

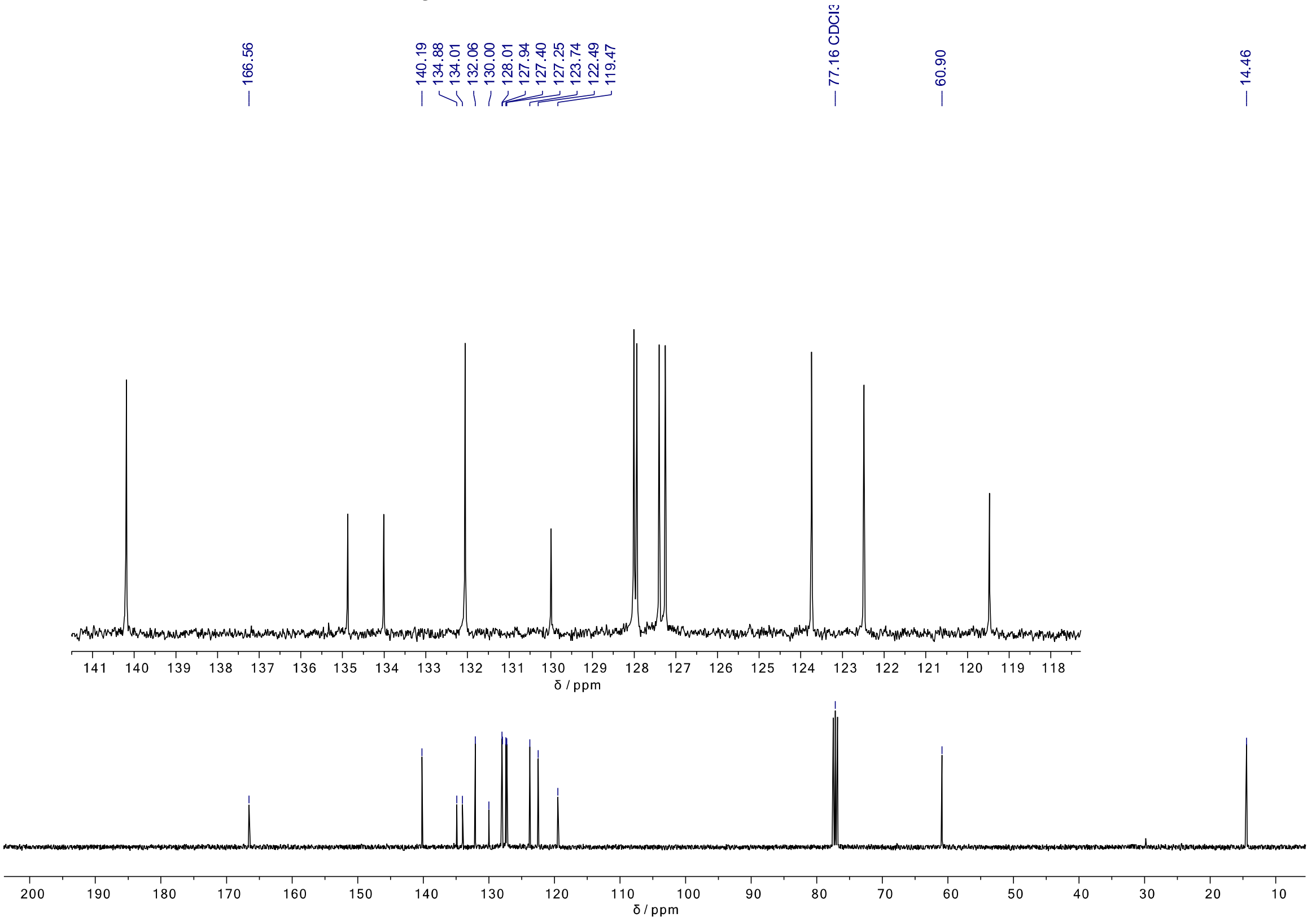

**(*E*)-3-(3-Bromonaphthalen-1-yl)acrylic acid (6).** $^{1}$H NMR / 400 MHz / DMSO-$d_6$ / 25 °C

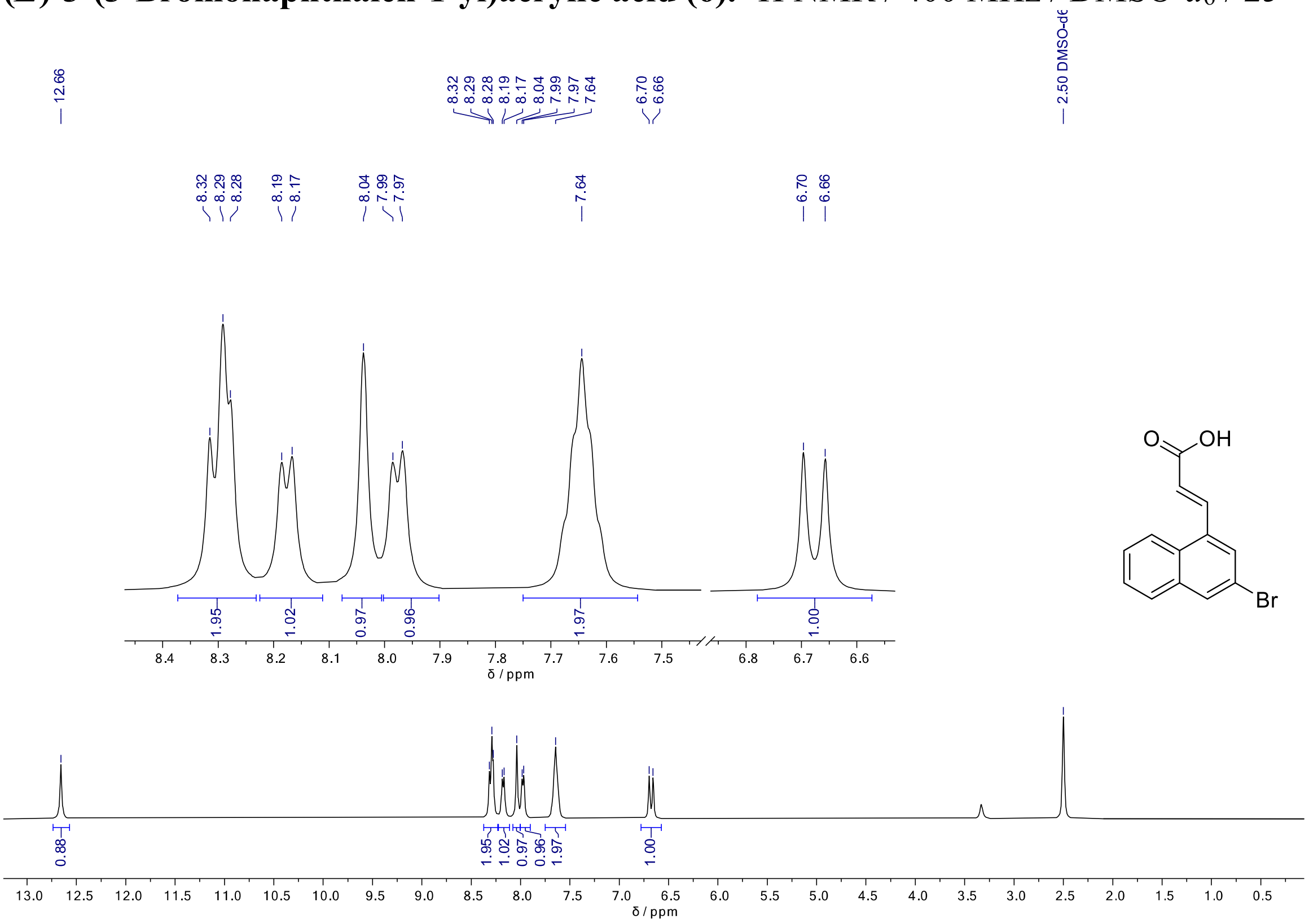

$^{13}$C NMR / 101 MHz / DMSO-$d_6$ / 25 °C

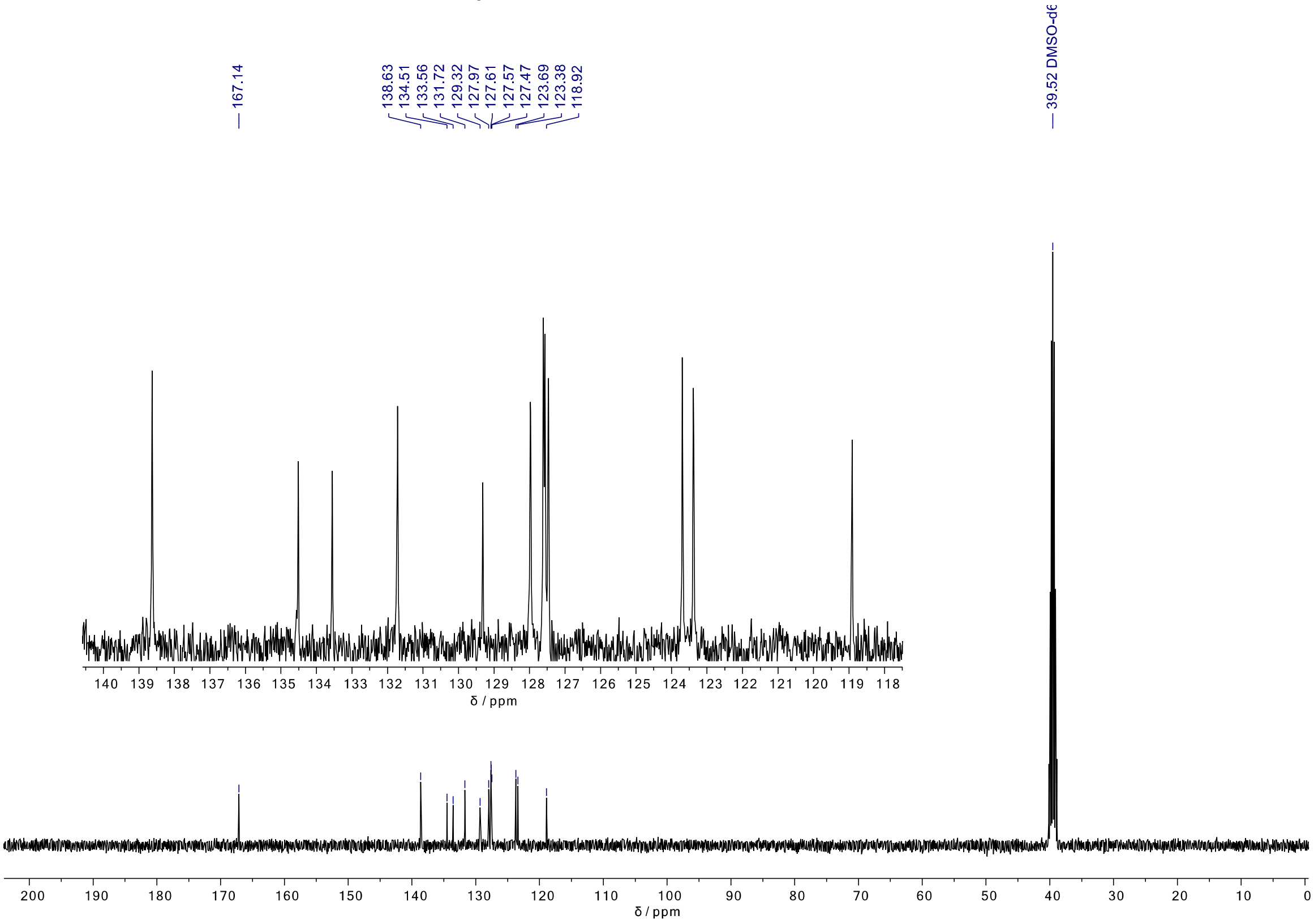

**5-Bromo-1*H*-phenalen-1-one (7).** $^{1}$H NMR / 600 MHz / $CDCl_3$ / 25 °C

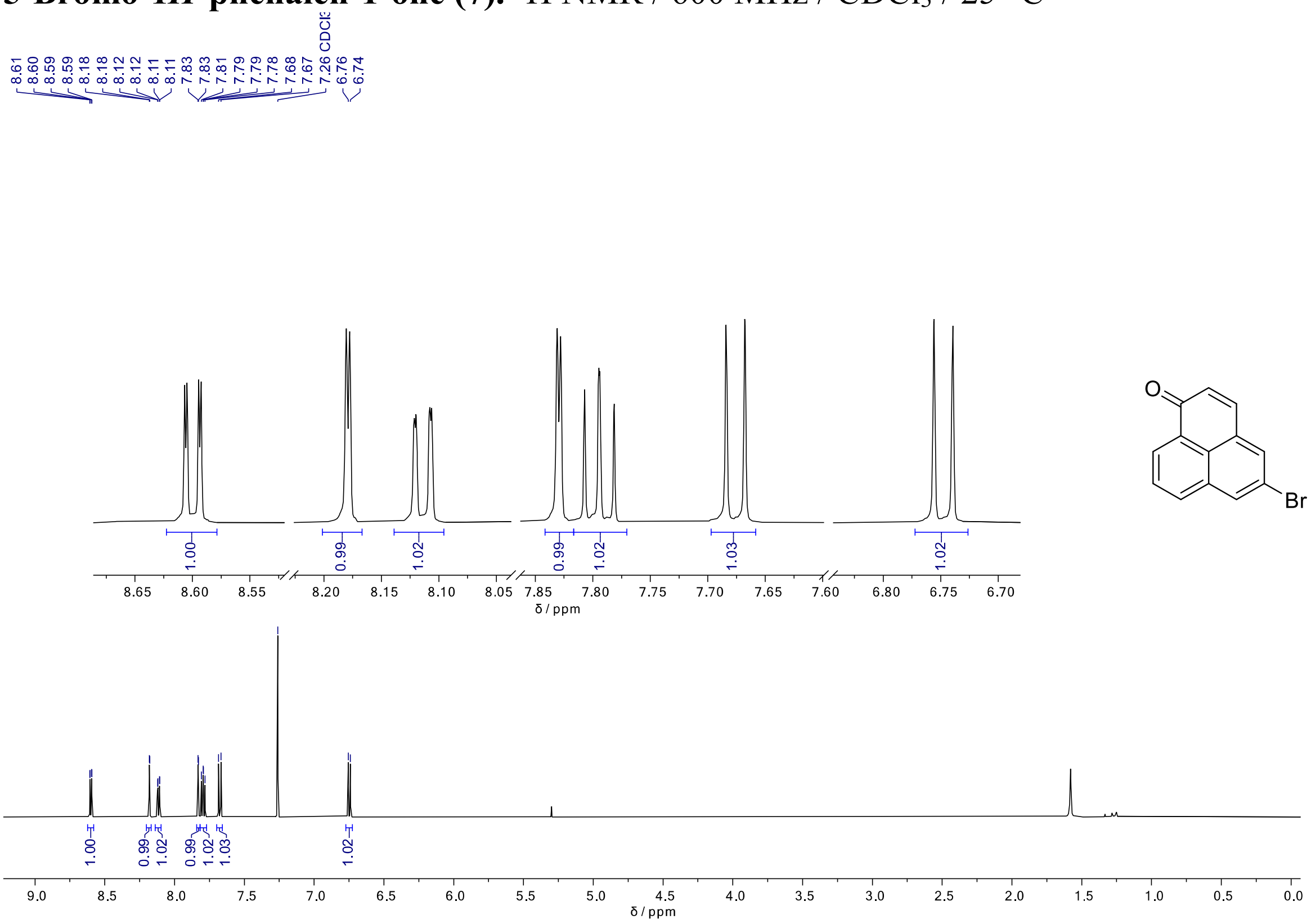

$^{13}$C NMR / 151 MHz / $CDCl_3$ / 25 °C

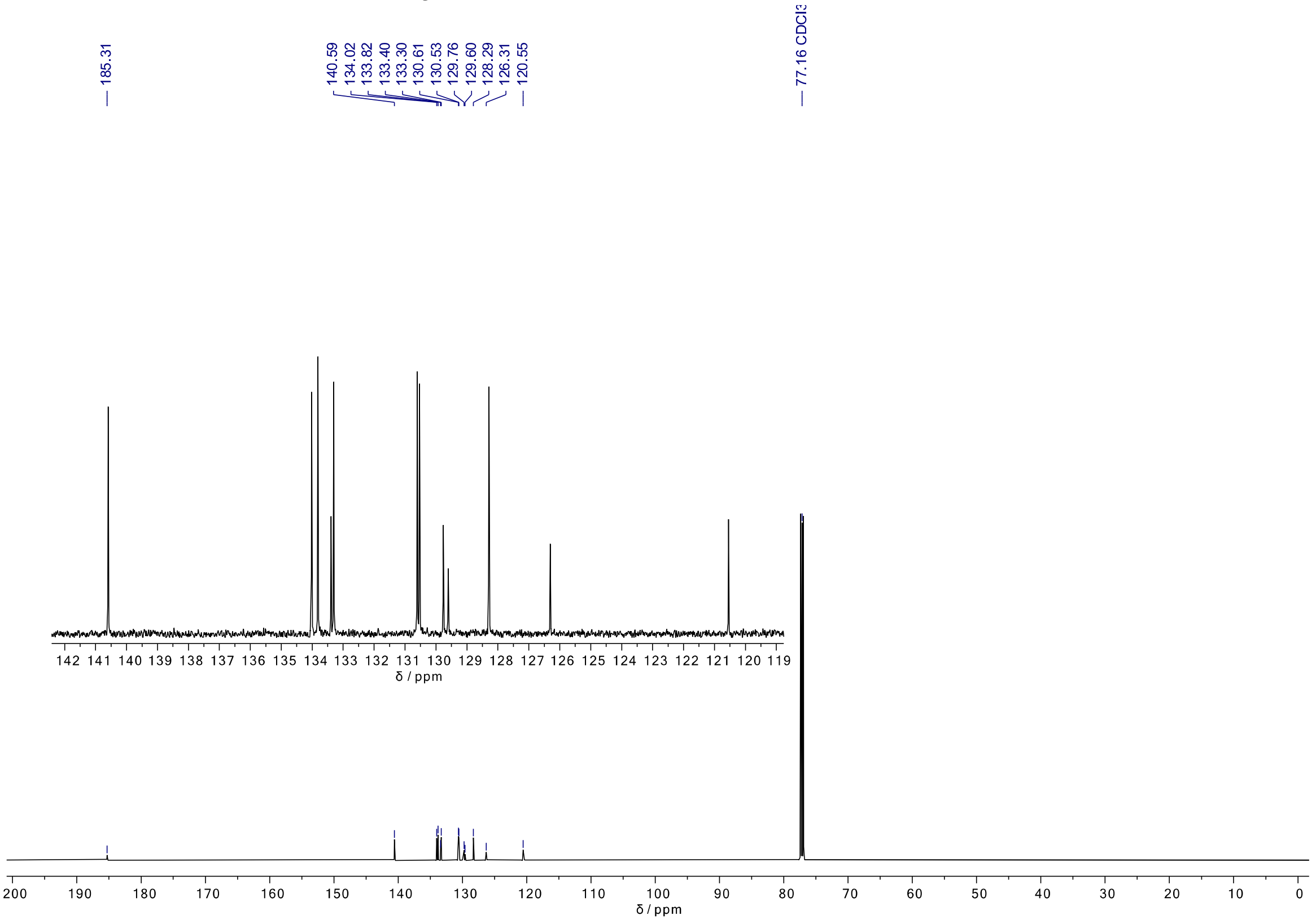

**5-(4,4,5,5-Tetramethyl-1,3,2-dioxaborolan-2-yl)-1*H*-phenalen-1-one (8).**

$^{1}$H NMR / 600 MHz / $CDCl_3$ / 25 °C

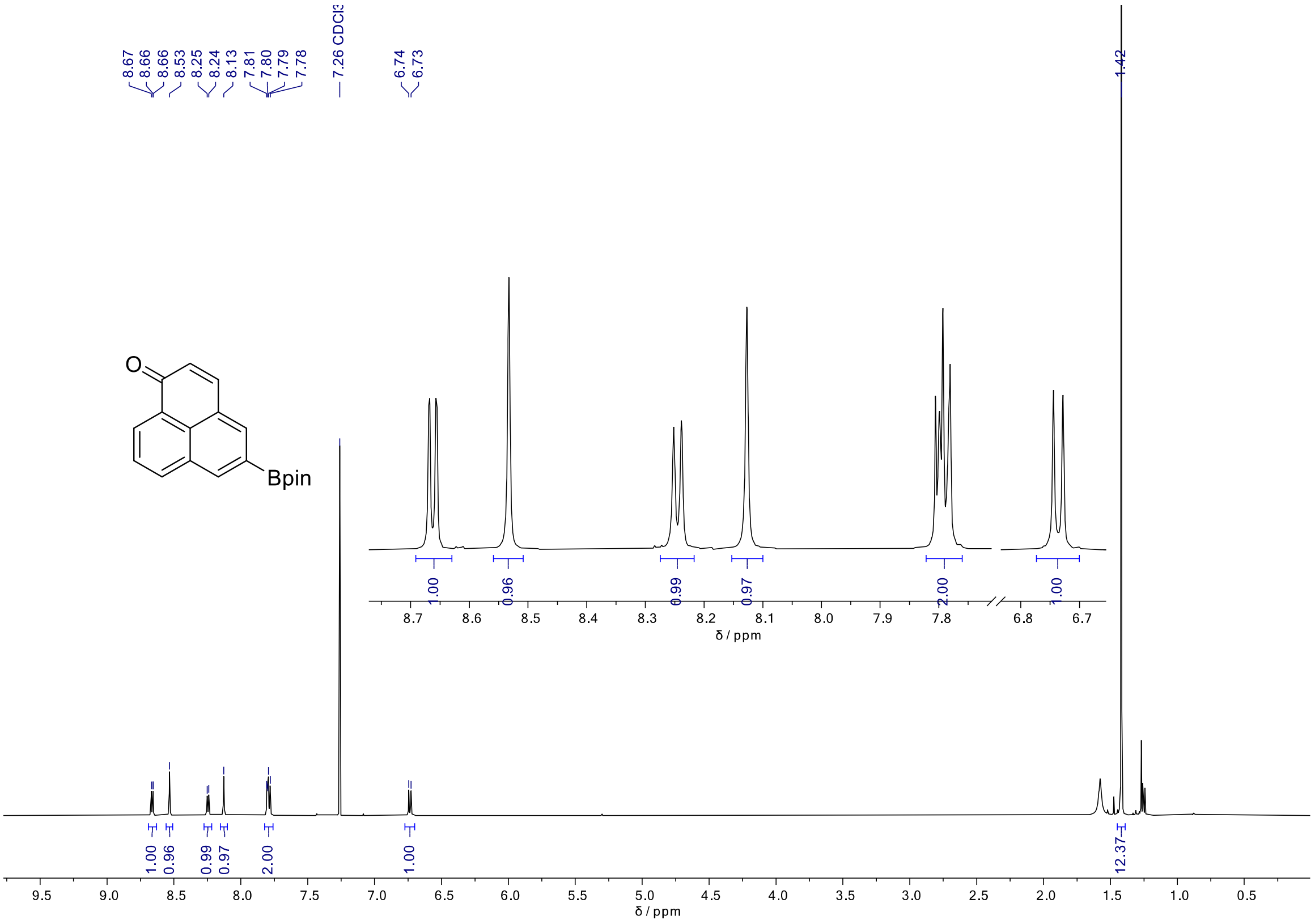

$^{13}$C NMR / 151 MHz / $CDCl_3$ / 25 °C

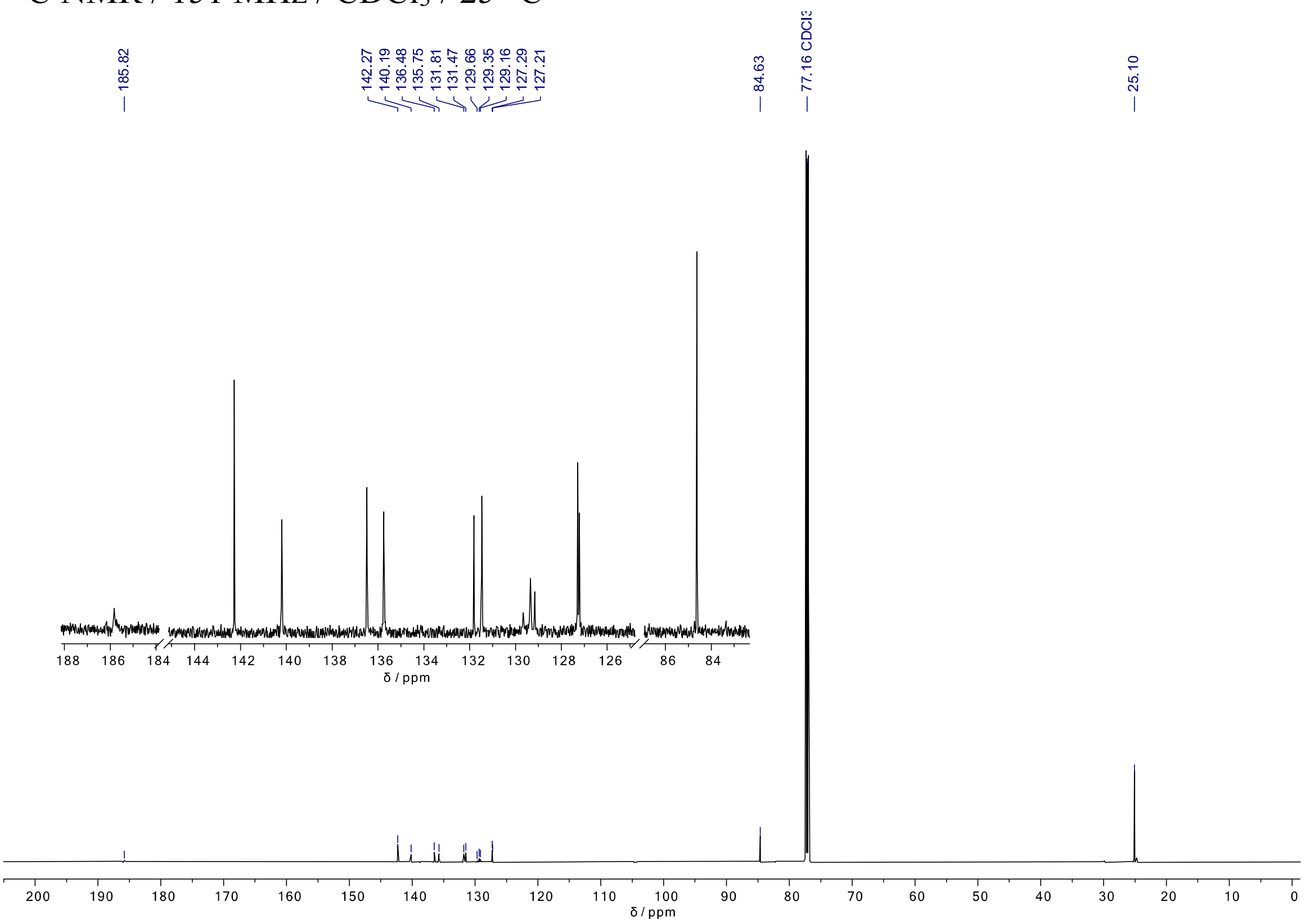

$^1H$–$^1H$ COSY NMR / 600 MHz / $CDCl_3$ / 25 °C

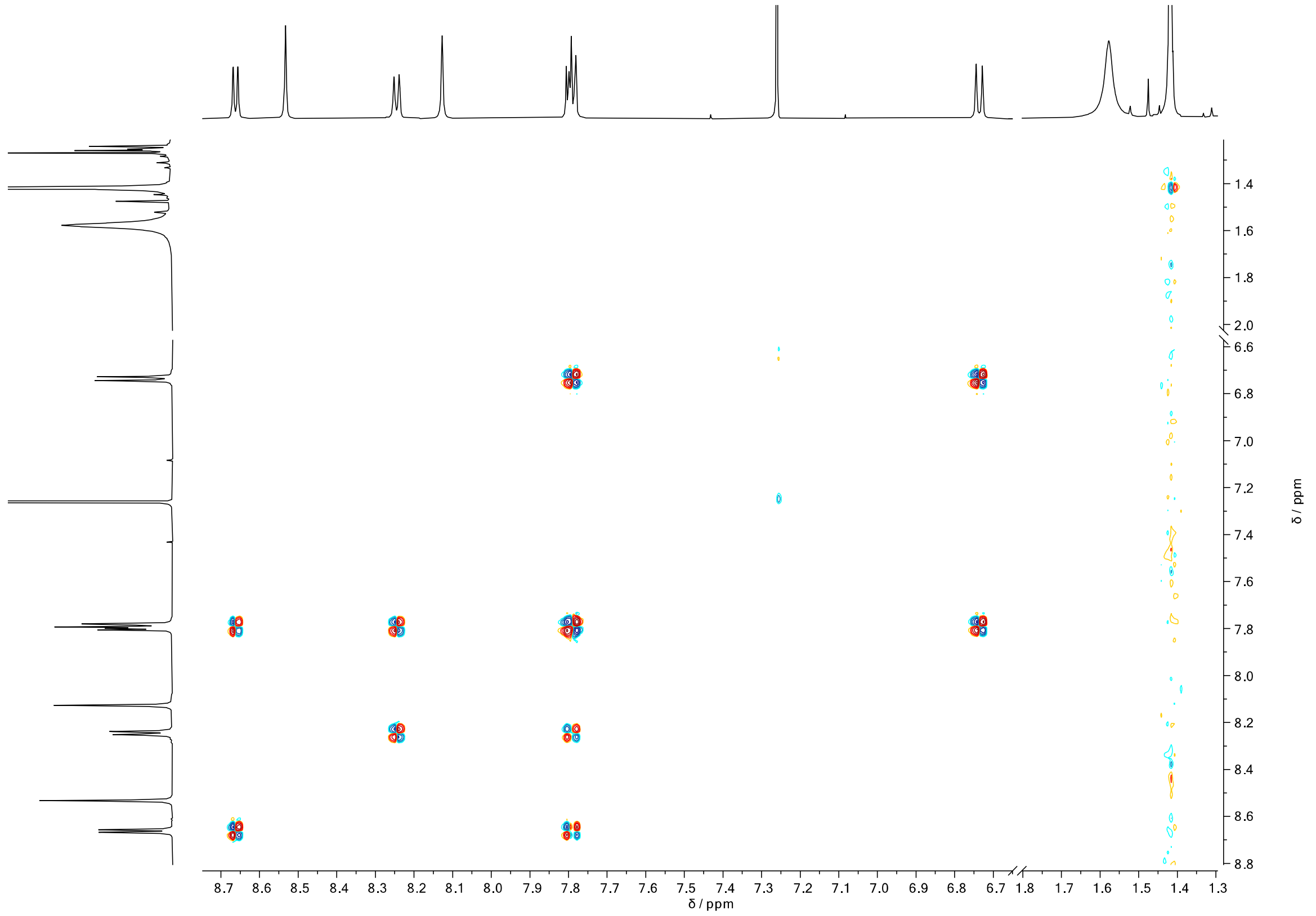

$^1H$–$^{13}C$ HSQC NMR / 600 MHz / $CDCl_3$ / 25 °C

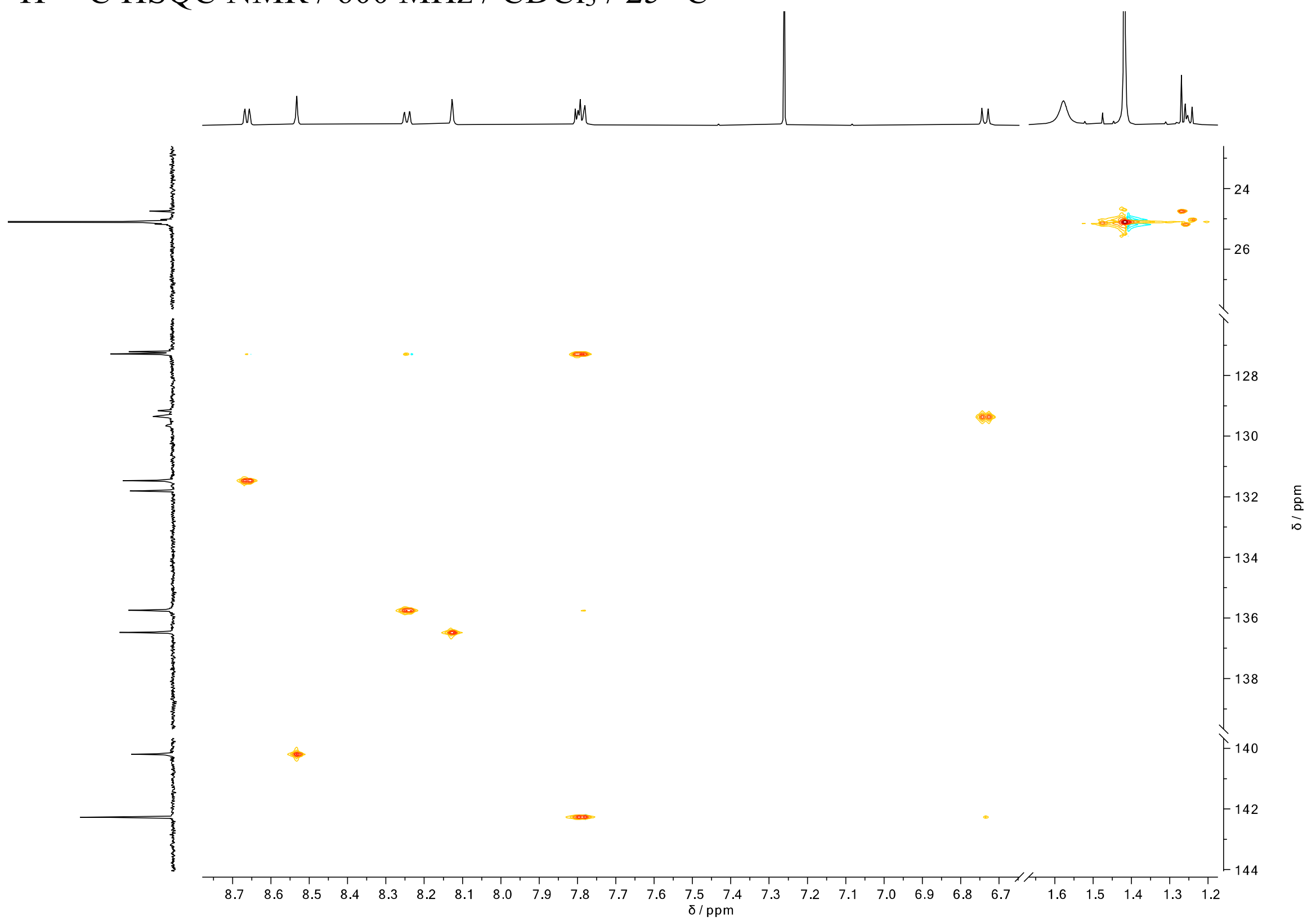

$^{1}H$–$^{13}C$ HMBC NMR / 600 MHz / $CDCl_3$ / 25 °C

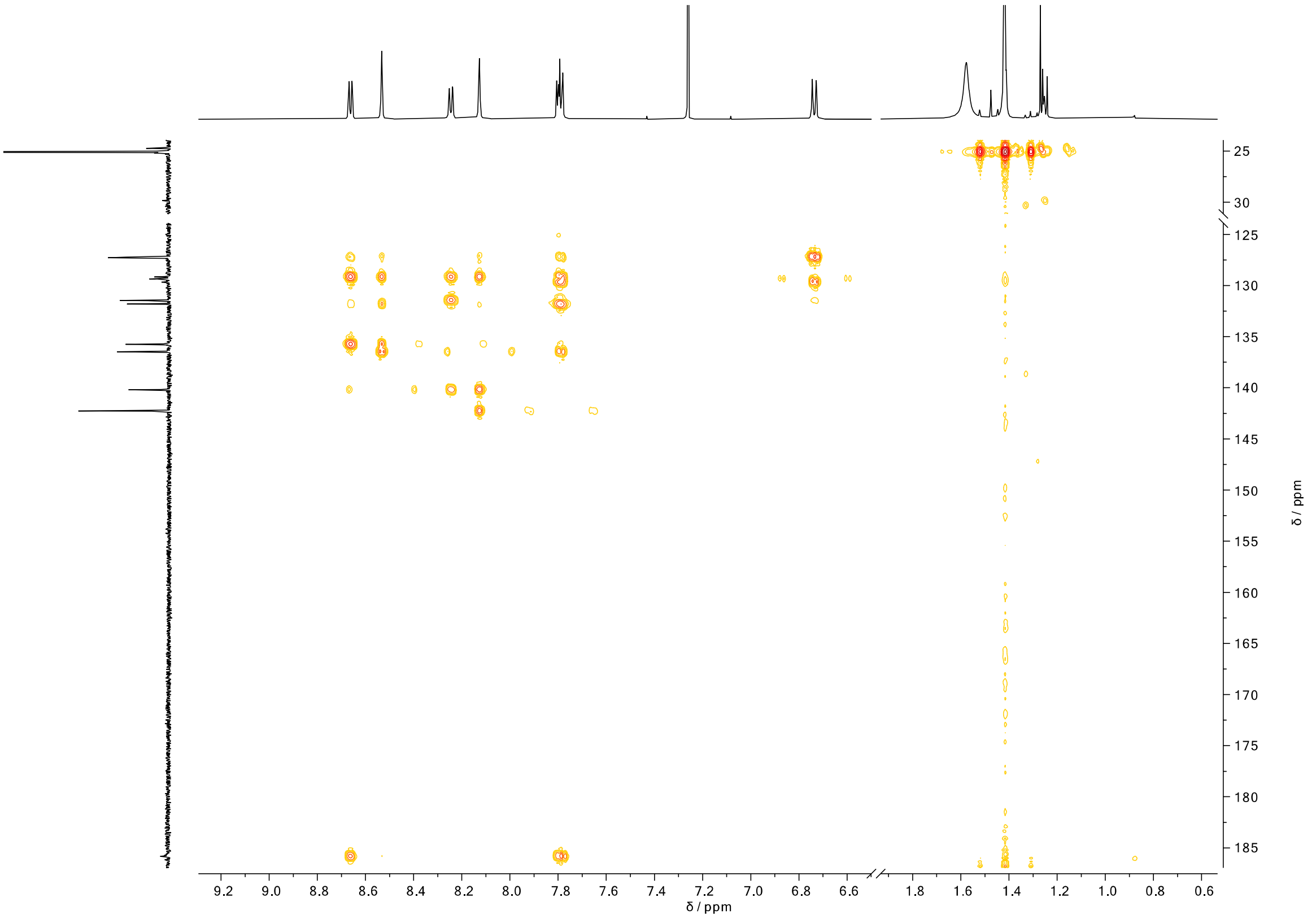

Assignment of proton and carbon resonances

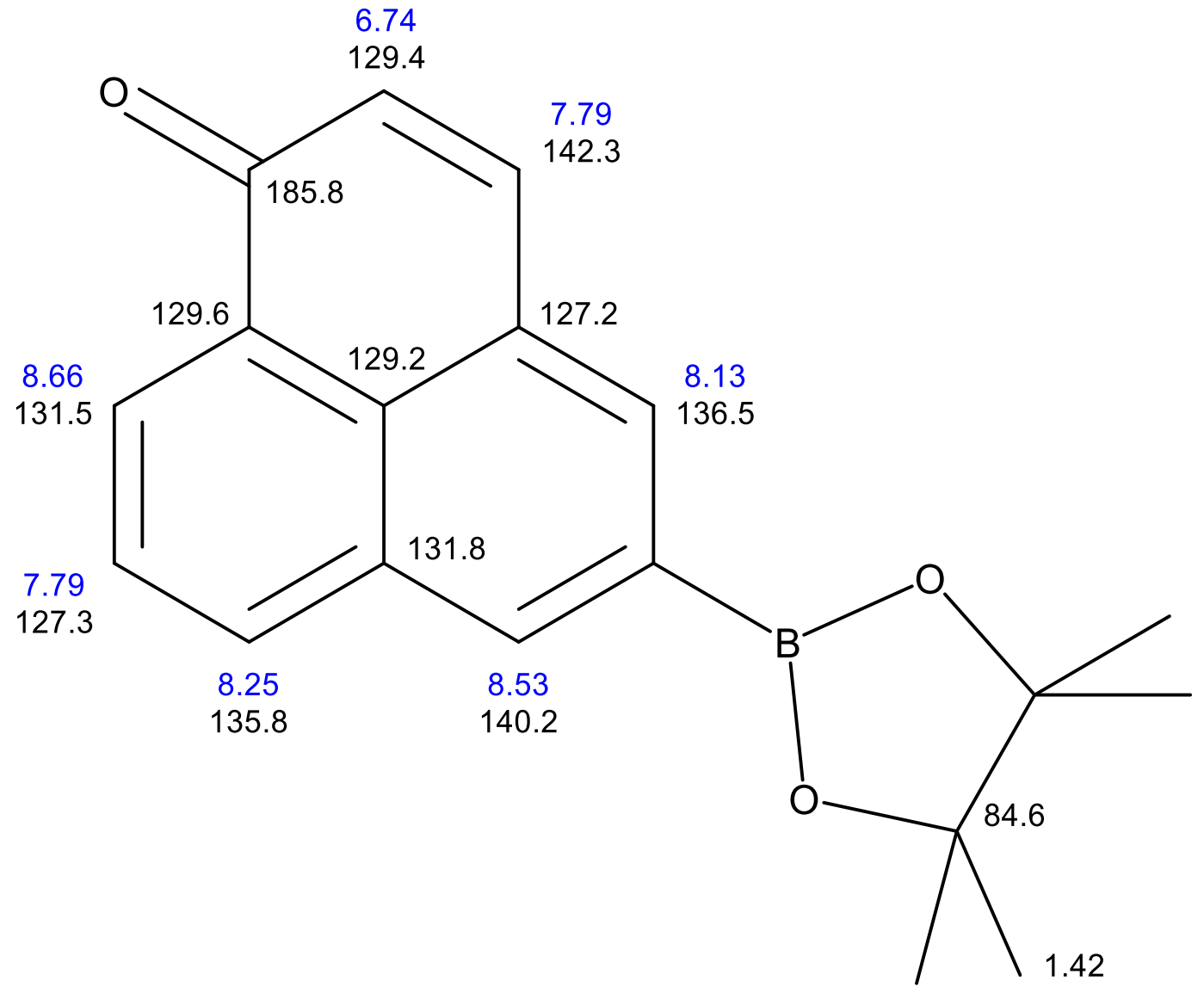

HRMS (EI)

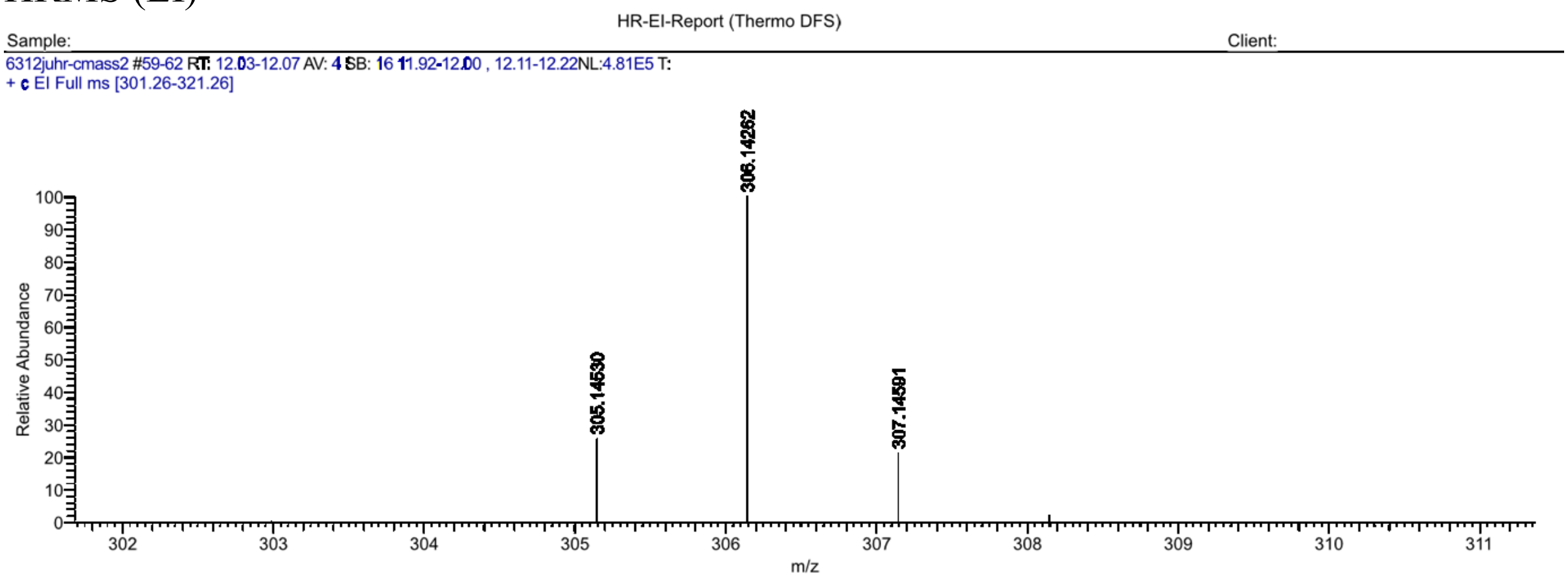

6312juhr-cmass2#59-62 RT: 12.03-12.07 AV: 4 T: + c
EI Full ms [301.26-321.26]
m/z= 306.04750-306.27642

| m/z | Intensity | Relative | Theo. Mass | Delta (ppm) | Composition |
|---|---|---|---|---|---|
| 306.14261 | 482130.8 | 100.00 | 306.14218 | 1.43 | $C_{19}H_{19}O_3B$ |

**6*H*,6'*H*-[2,2'-Biphenalene]-6,6'-dione (9).** $^{1}$H NMR / 600 MHz / $CDCl_3$ / 25 °C

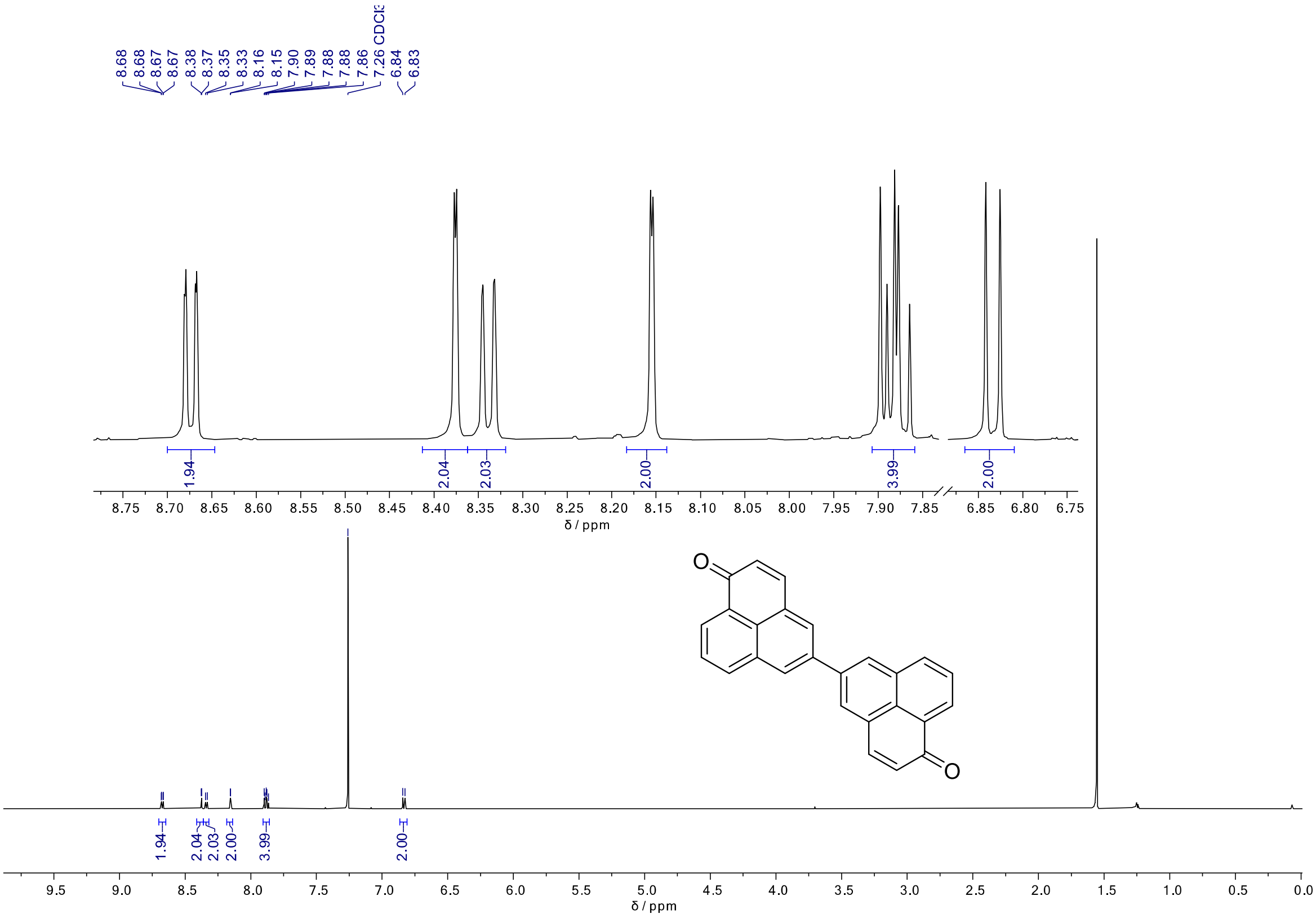

$^{13}$C NMR / 151 MHz / $CDCl_3$ / 25 °C

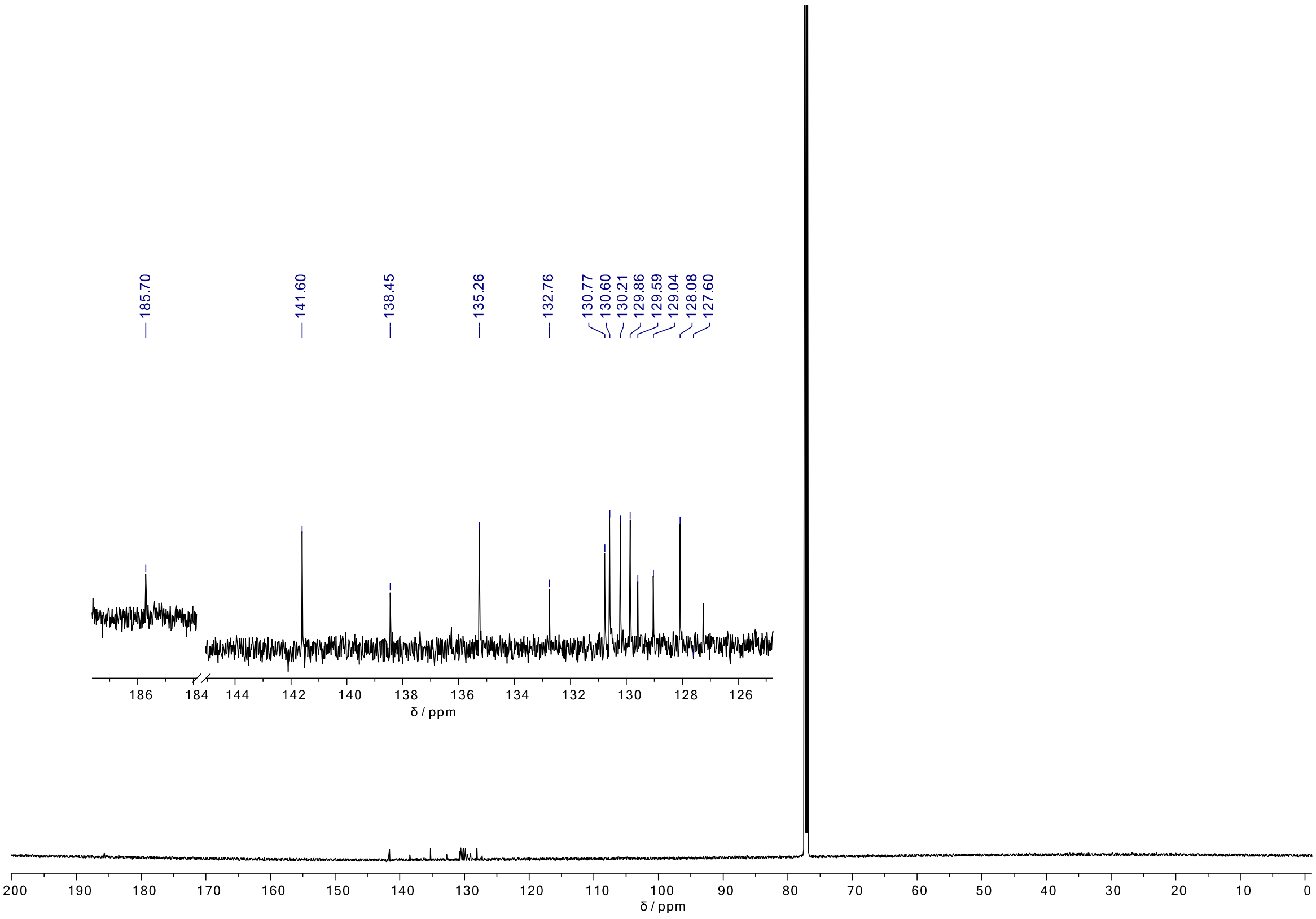

$^{1}$H–$^{1}$H COSY NMR / 600 MHz / $CDCl_3$ / 25 °C

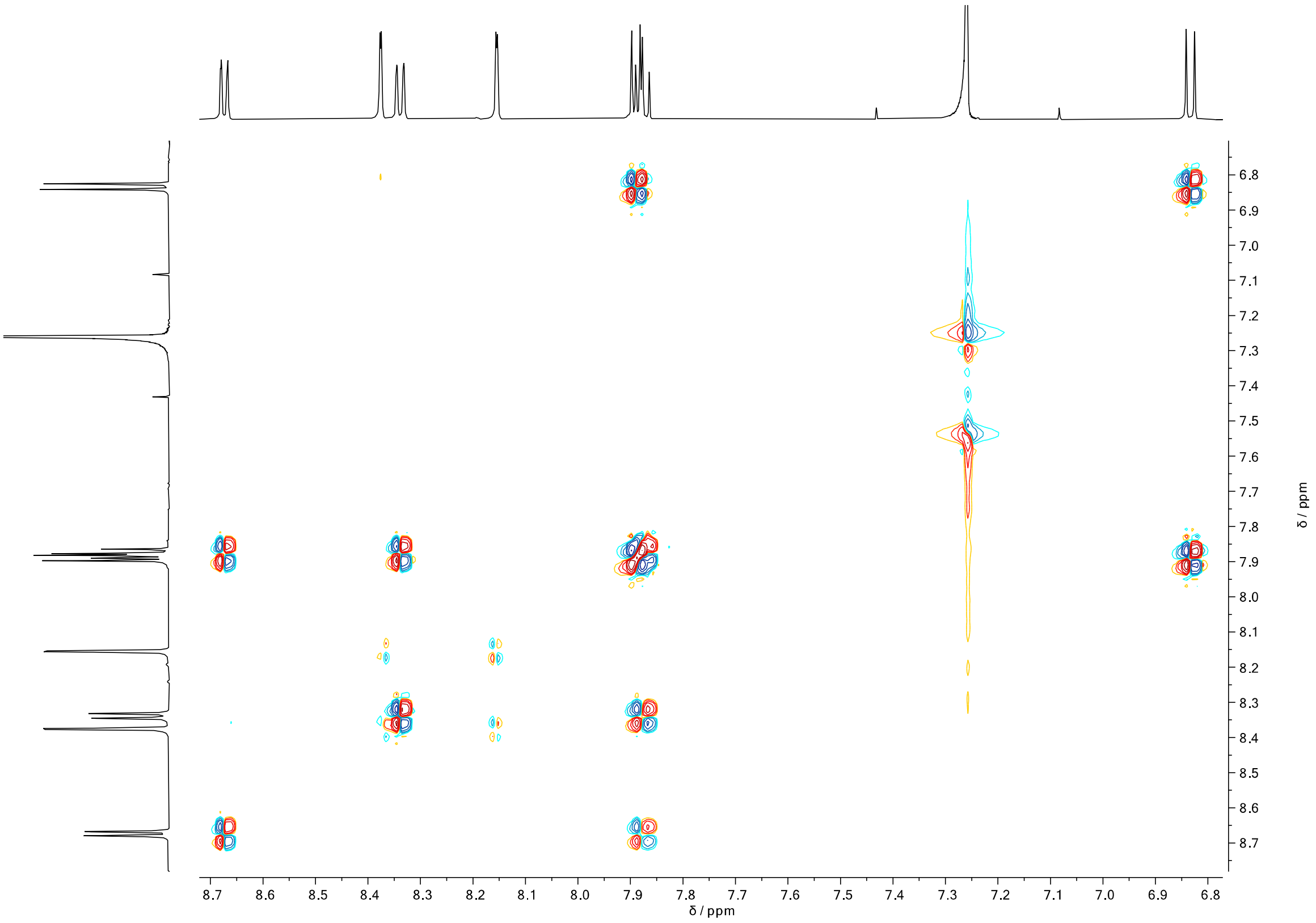

$^{1}$H–$^{13}$C HSQC NMR / 600 MHz / $CDCl_3$ / 25 °C

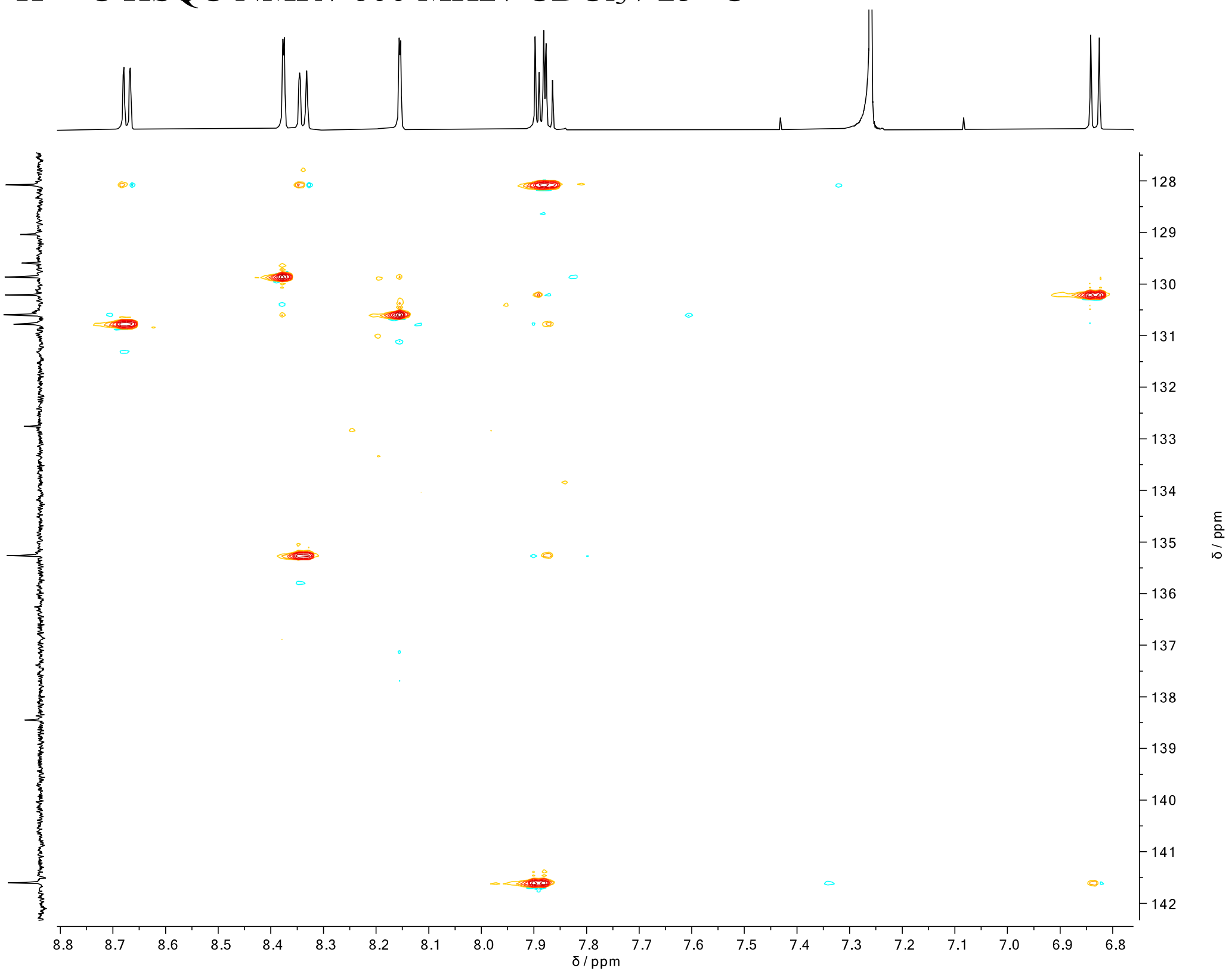

$^{1}H$–$^{13}C$ HMBC NMR / 600 MHz / $CDCl_3$ / 25 °C

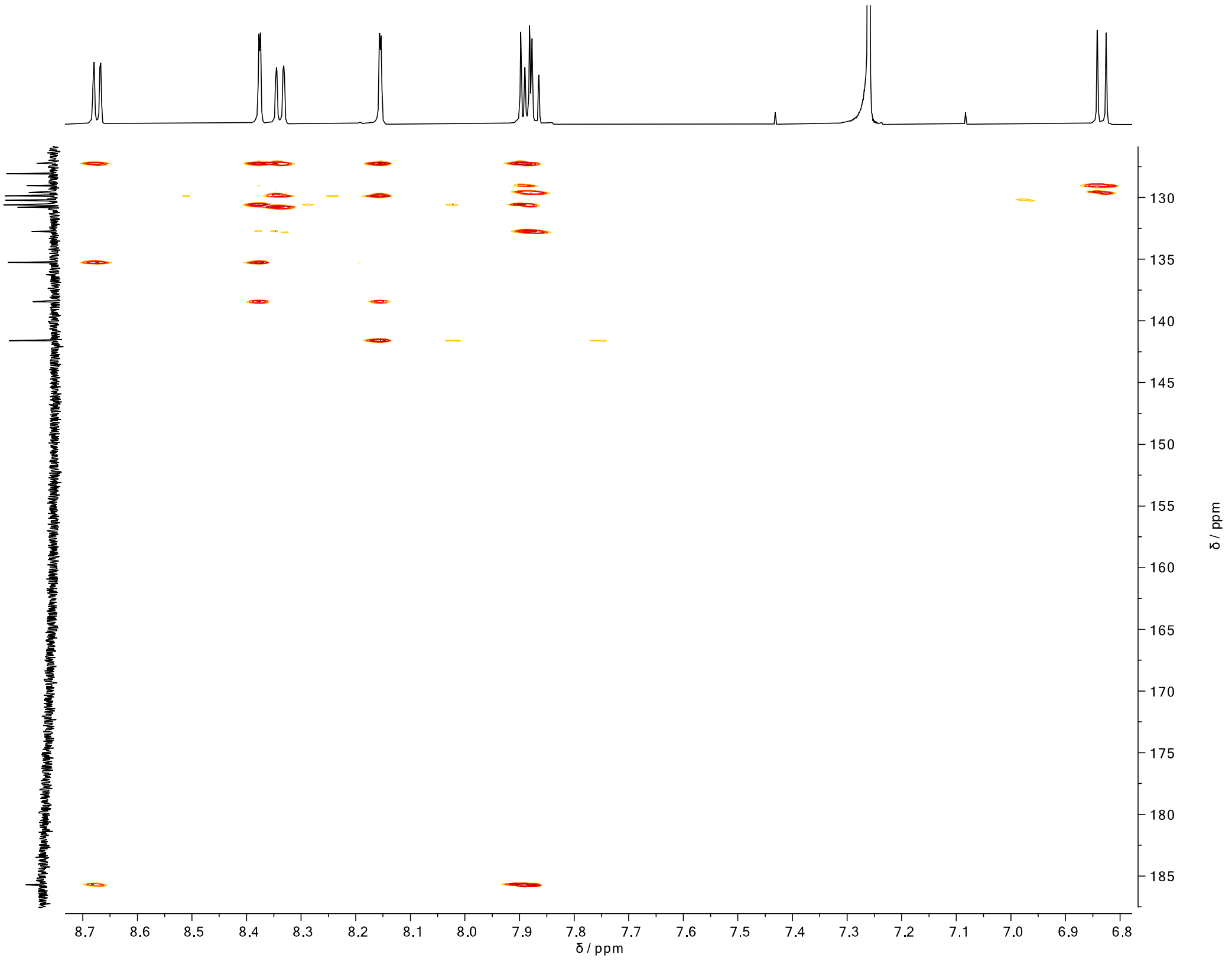

Assignment of proton and carbon resonances

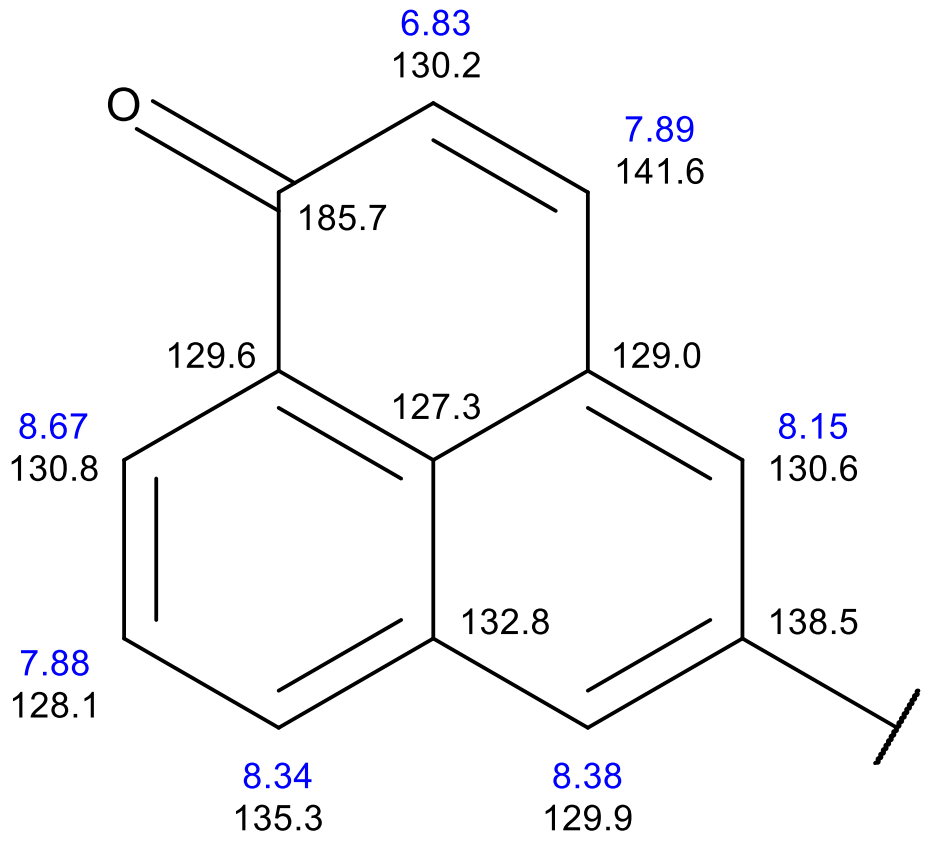

HRMS (EI)

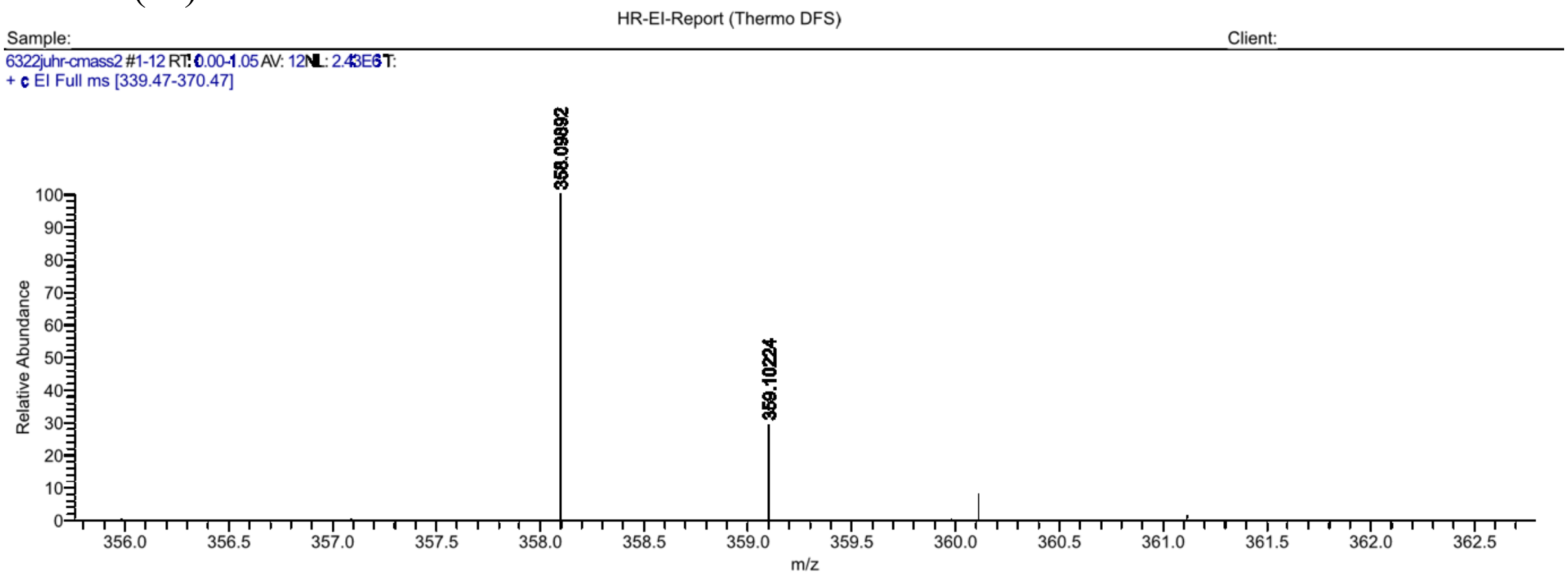

6322juhr-cmass2 #1-12 RT: 0.00-1.05 AV: 12 T: +
c EI Full ms [339.47-370.47]
m/z= 337.47200-372.47200

| m/z | Intensity | Relative | Theo. Mass | Delta (ppm) | Composition |
|---|---|---|---|---|---|
| 358.09892 | 2430546.0 | 100.00 | 358.09883 | 0.26 | $C_{25}H_{14}O_2$ |

**6*H*,6'*H*-2,2'-Biphenalene (10, 2*H*-diphenalenyl).** $^1$H NMR / 400 MHz / $CD_2Cl_2$ / 25 °C

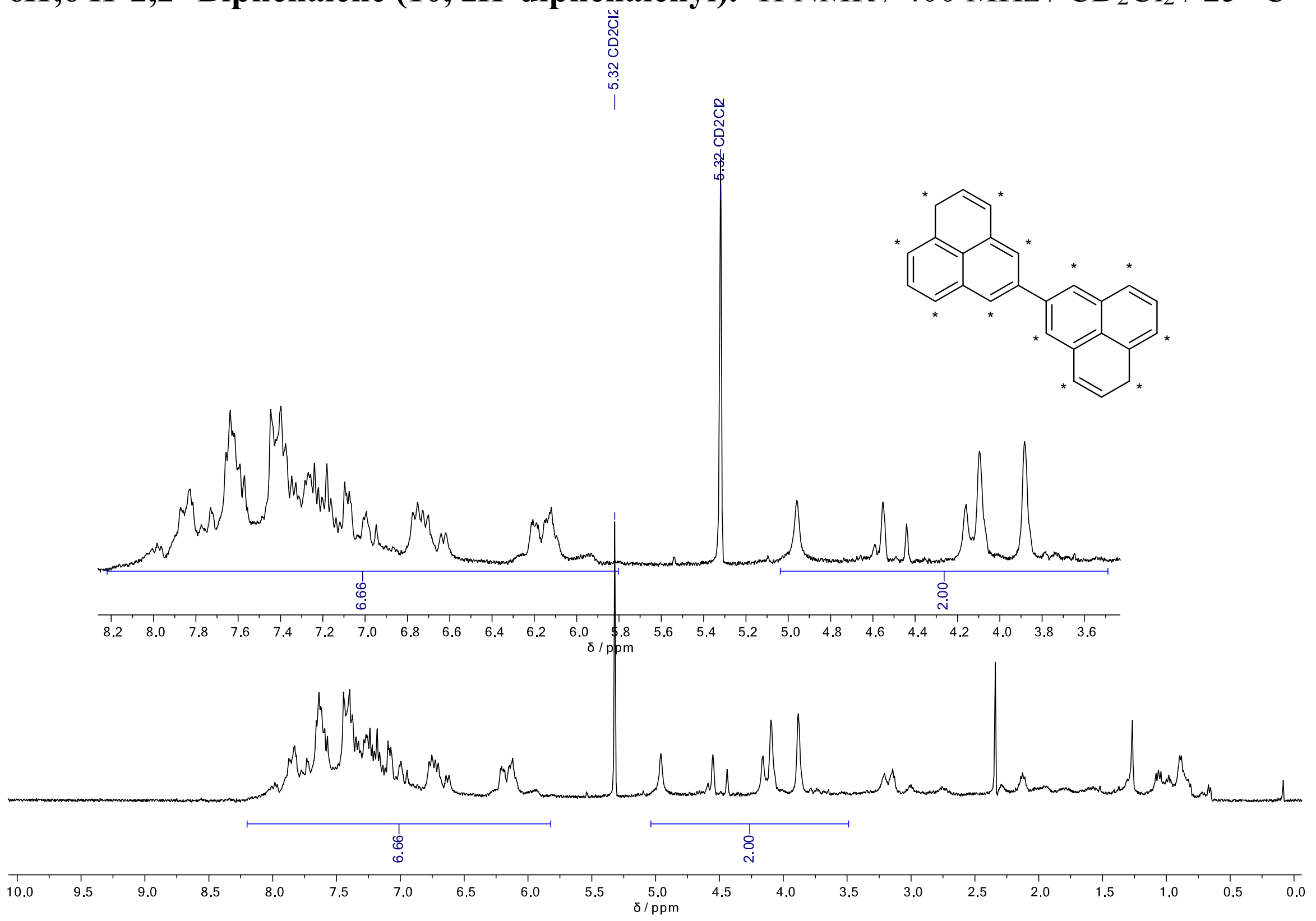

HRMS (EI)

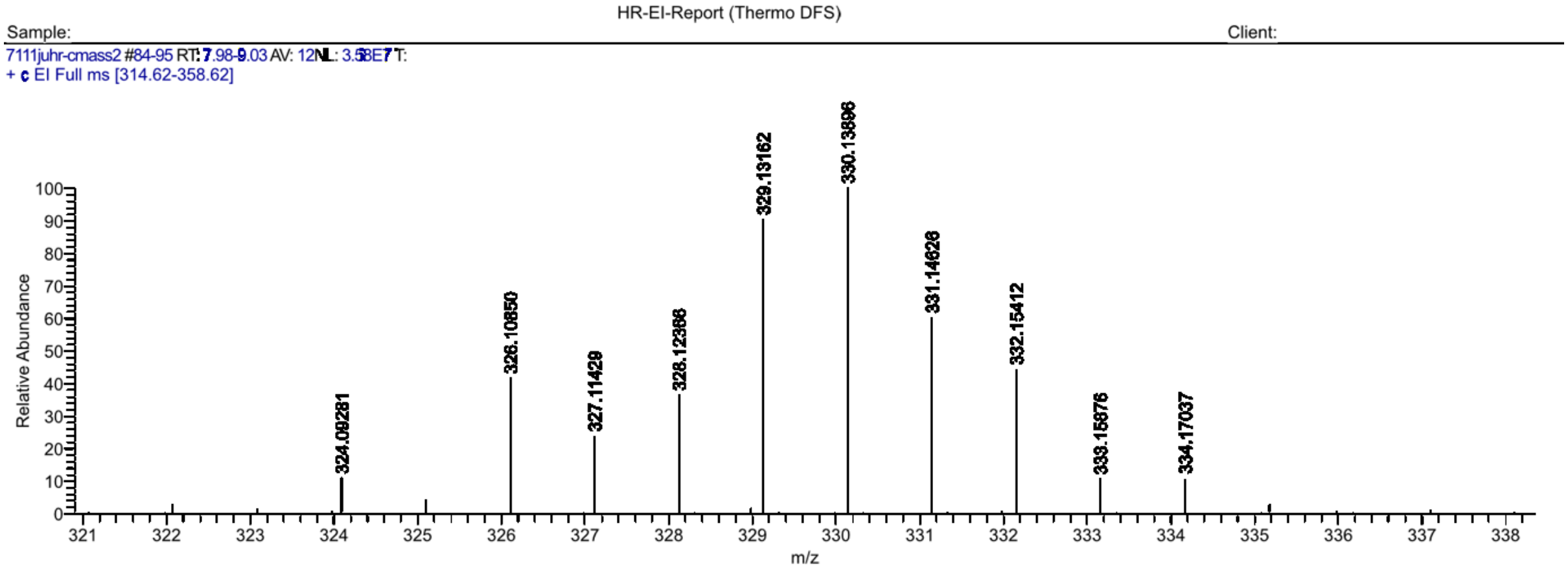

**7111juhr-cmass2#84-95 RT: 7.98-9.03 AV: 12 T: +**
**c EI Full ms [314.62-358.62]**
**m/z= 322.70692-330.65784**

| m/z | Intensity | Relative | Theo. Mass | Delta (ppm) | Composition |
|---|---|---|---|---|---|
| 324.09281 | 3921854.1 | 10.94 | 324.09335 | -1.67 | C 26 H 12 |
| 325.09758 | 1515706.3 | 4.23 | 325.10118 | -11.07 | C26 H13 |
| 326.10850 | 14899328.0 | 41.57 | 326.10900 | -1.53 | C26 H14 |
| 327.11429 | 8530389.1 | 23.80 | 327.11683 | -7.76 | C26 H15 |
| 328.12366 | 13025813.3 | 36.34 | 328.12465 | -3.03 | C26 H16 |
| 329.13162 | 32387328.0 | 90.36 | 329.13248 | -2.59 | C26 H17 |
| 330.13896 | 35843413.3 | 100.00 | 330.14030 | -4.07 | C26 H18 |